\documentclass[aps, prb, twocolumn, groupedaddress, showpacs, floatfix,superscriptaddress]{revtex4-1}
\usepackage{amsmath}
\usepackage{amssymb}
\usepackage{graphicx}
\usepackage{xr-hyper}
\usepackage{bm}

\begin{document}

\title{Quasiparticle energy bands and Fermi surfaces of monolayer $\textrm{NbSe}_{2}$}

\author{Sejoong Kim}
\email[Email: ]{sejoong@ust.ac.kr}
\affiliation{Korea Institute for Advanced Study, Hoegiro 85, Seoul 02455, Korea}
\affiliation{University of Science and Technology, Gajeong-ro 217, Daejeon 34113, Korea}
\author{Young-Woo Son}
\email[Email: ]{hand@kias.re.kr}
\affiliation{Korea Institute for Advanced Study, Hoegiro 85, Seoul 02455, Korea}

\date{\today}

\begin{abstract}
A quasiparticle band structure of a single layer 2$H$-$\mathrm{NbSe}_{2}$ 
is reported by using first-principles $GW$ calculation. 
We show that a self-energy correction increases 
the width of a partially occupied band
and alters its Fermi surface shape when comparing those using conventional mean-field calculation methods. 
Owing to a broken inversion symmetry in the trigonal prismatic single layer structure, 
the spin-orbit interaction is included
and its impact on the Fermi surface and quasiparticle energy bands are discussed. 
We also calculate the doping dependent static susceptibilities 
from the band structures obtained by the mean-field calculation as well as 
$GW$ calculation with and without spin-orbit interactions.
A complete tight-binding model is constructed within the three-band third nearest neighbour hoppings
and is shown to reproduce our $GW$ quasiparticle energy bands and Fermi surface very well. 
Considering variations of the Fermi surface shapes depending on self-energy corrections 
and spin-orbit interactions, 
we discuss the formations of charge density wave (CDW) with different dielectric environments
and their implications on recent controversial experimental results on CDW transition temperatures. 
\end{abstract}



\maketitle

\section{Introduction}

After the first experimental success in isolating various single layers from layered materials 
such as graphite, hexagonal boron nitrides, transition metal dichalcogenides (TMD)
and high-temperature cuprate superconductors~\cite{PNAS2005Novoselov}, 
there have been tremendous efforts to understand physical properties of 
two-dimensional (2D) crystals~\cite{RMP2009Neto,RMP2011Sarma,
NatNano2012Wang,RMP2012Kotov,Nature2013Geim}. 
Notably, owing to the reduced spatial dimension compared to their bulk counterparts, 
the Coulomb interaction in the 2D crystals as well as screening of substrates on which they are placed 
have played very important roles in modifying their electronic structures~\cite{RMP2011Sarma, RMP2012Kotov,PRB2012Komsa,PRB2012Ashwin,PRB2012Cheiwchanchamnagij,PRL2013Qiu,NatMater2014Ugeda}. 
For example, it is now well established that many-body interactions alter low-energy bands 
in graphene significantly~\cite{PRL2008Trevisanutto,PRL2009Park,Science2010Bostwick,NatPhys2011Elias,RMP2012Kotov,PRL2013Siegel,PRL2013Lischner}. 
Moreover, the substrate screening also changes the nature of Coulomb interactions 
in graphene on top of either dielectric materials or metals and hence modifies the bands further~\cite{PRL2009Park,SciRep2012Hwang}. 

Besides graphene, several recent experiments have revealed new interesting physical properties 
in a mono- and a few-layer TMDs such as series of phase transitions and novel superconductivity, 
being different from those shown in their bulk forms~\cite{Science2012Ye,NatNano2015Yu,ACSNano2015Sugawara,NatComm2015Chen,PRB2015Peng,Science2015Saito,NatPhys2016Saito,Nature2016Li,NatPhys2016Xi,NatPhys2016Tsen,NanoLett2015Cao,NatNano2015Xi,NatPhys2016Ugeda}.  
Formation of charge density waves (CDWs) 
in three-dimensional metallic TMDs has attracted interests for the last couple of decades and origins
of CDWs in some materials are still not settled yet~\cite{AdvPhys1975Wilson,JPCM2011Rossnagel}. 
Therefore, the current efforts in investigating physical properties of thin flakes of TMDs
may shed light on the origin of CDW phase in three dimensional TMDs and open a way
to find the characteristic new collective phenomena in 2D crystals~\cite{Science2015Saito,NatPhys2016Saito,Nature2016Li,NatPhys2016Xi,NatPhys2016Tsen}.

Among metallic TMDs, the stacked trigonal prismatic structure of niobium diselenide ($2H$-$\textrm{NbSe}_{2}$) 
is one of the most studied materials and an ideal system 
to study phase transitions as functions of temperature and dopings. 
It has been known for a long time~\cite{AdvPhys1975Wilson,JPCM2011Rossnagel} 
that three-dimensional stacking structure of $2H$-$\textrm{NbSe}_{2}$ is metallic at room temperature
and undergoes a CDW transition at 33 K before becoming a superconductor~\cite{SSC1963Revolinsky,JPCS1965Revolinsky}  at 7.2 K
although there has been the controversy regarding on the origin of CDW 
and the competition between CDW and superconducting (SC) 
states~\cite{AdvPhys1975Wilson,JPCM2011Rossnagel,PRL1974Wilson,PRL1999Straub,PRL2008Shen,PRL2009Borisenko,PRL1983Varma,PRL2004Valla,PRL2011Weber,PRB2012Rahn,PNAS2013Soumyanarayana,PRL2015Arguello}.
After a few earlier attempts to investigate physical properties of its thin flakes~\cite{PNAS2005Novoselov,PRL1972Frindt,PRB2009Staley}, 
a couple of recent works have reported successful isolations of its single layer form on top of various substrates
and measure their CDW and SC phase transitions~\cite{NatPhys2016Xi,NatPhys2016Tsen,NanoLett2015Cao,NatNano2015Xi,NatPhys2016Ugeda}. 

While all experiments~\cite{PRL1972Frindt,PRB2009Staley,NanoLett2015Cao,NatNano2015Xi,NatPhys2016Ugeda} hitherto have shown 
that the superconducting transition temperature decreases but does not diminish completely 
when the thickness of $2H$-$\textrm{NbSe}_{2}$ decreases to a single layer limit, 
the transition temperature ($T_\text{CDW}$) from metal to CDW phase differs from each other significantly~\cite{NatNano2015Xi,NatPhys2016Ugeda}.
The work by Xi {\it et al.}~\cite{NatNano2015Xi} measured $T_\text{CDW}$ of 145 K, more than four times larger than the bulk $T_\text{CDW}$ of 33 K 
whereas the work by Ugeda {\it et al.} reported that $T_\text{CDW}$ is similar to or less than that of the bulk.
Moreover, the former attribute the strong coupling mechanism to the formation of CDW at the very high temperature 
while the latter measured a small CDW energy gap of 4 meV together with CDW modulation under high biases
pointing to the puzzling dual nature (strong and weak) of CDW formation. 
It is noticeable that the former measured the transition in a sample on top of silicon substrate 
while the latter on top of epitaxial bilayer graphene (BLG) grown on 6$H$-SiC(0001) surface.

Previous theoretical studies have shown that the bulk CDW phase of $2H$-$\textrm{NbSe}_{2}$ does not have a prominent sharp peak at
the specific CDW wave vector, rather showing a broad peak in the real part of the bare susceptibility and its imaginary part does not
peak at the CDW wave vector at all~\cite{PRB2008Johaness,JPhysC1978Doran,PRB2006Johannes}. 
These suggest a possible strong electron-phonon coupling mechanism to form the CDW phase~\cite{PRL1983Varma,PRL2004Valla,PRL2011Weber,PRB2012Rahn,PNAS2013Soumyanarayana,PRL2015Arguello}, 
ruling out the Fermi surface nesting mechanism~\cite{PRL1974Wilson,PRL1999Straub,PRL2008Shen,PRL2009Borisenko}, 
or the saddle-point singularity driven CDW phase~\cite{PRL1975Rice,NatPhys2007Kiss}. 
A recent first-principles calculation using a semi-local exchange-correlation functional 
predicts the CDW instability in the single layer structure  with an enhancement of the electron-phonon interaction
at a specific CDW wave vector differing from that of the bulk~\cite{PRB2009Calandra}. 
However, the predicted 4$\times$1 CDW state is inconsistent with the recent measurement showing 3$\times$3 CDW state
by Ugeda and coworkers~\cite{NatPhys2016Ugeda}.
Another recent calculation using the similar method also suggests a good metallic behavior 
for a monolayer undistorted structure~\cite{PRB2009Lebegue} while  
a transport measurement shows a semimetallic nature~\cite{PNAS2005Novoselov}.

Considering significant changes in the low energy bands of the semimetallic and semiconducting 
2D crystals with a proper inclusion of electronic self-energy correction
as well as its modification by the substrate 
screening~\cite{PRL2008Trevisanutto,PRL2009Park,Science2010Bostwick,NatPhys2011Elias,RMP2012Kotov,PRL2013Siegel,PRL2013Lischner,SciRep2012Hwang,PRB2012Komsa,PRL2013Qiu},
a plain mean field calculation is not sufficient and 
it is necessary to investigate effects of suitable corrections from the many-body Coulomb interactions on the low energy bands of monolayer $\textrm{NbSe}_2$.
Although the CDW formation is quite sensitive to a shape of Fermi surface, its variation with the self energy corrections and alternations by the substrate screening 
have not been investigated yet fully. 
In this work, motivated by recent rapid progress in this field and experiments reporting different CDW formations~\cite{NatNano2015Xi,NatPhys2016Ugeda},
we report first-principles density functional calculation and $GW$ approximation results 
of single layer $\textrm{NbSe}_2$ in its normal metallic state
which provide comprehensive pictures of the low energy electronic structures, a prerequiste to understand the controversal CDW features. 

This paper is organized as follows. In Sec.~\ref{calculation}, we introduce a model system and calculation details. 
In Sec.~\ref{results}, our calculation results for the low energy band structures based on mean-field calculation methods 
and $GW$ approximation with and without spin-orbit interactions are presented. 
The doping dependent variations of Fermi surfaces are also discussed within various levels of approximations.  
The bare susceptibilities with calculated energy band structures are also presented in this section. 
Conclusions and discussion on CDW formation mechanism in the single layer are in Sec.~\ref{final}. In addition, the effects of including semi core orbitals, detalied derivation of atomic SOCs in this system, constructions of a tight binding Hamiltonian 
within the three-band third nearest neighbor hopping between $d$-oribital of niobiums fit for the DFT-GGA and $GW$ energy bands, respectively, and 
discussions on $GW$ energy bands of the bulk 2$H$-NbSe$_2$ are presented in Appendix. 

\section{\label{calculation}Systems and Calculation Details}

\begin{figure}[t]
\begin{center}
\includegraphics[width=1.0\columnwidth, clip=true]{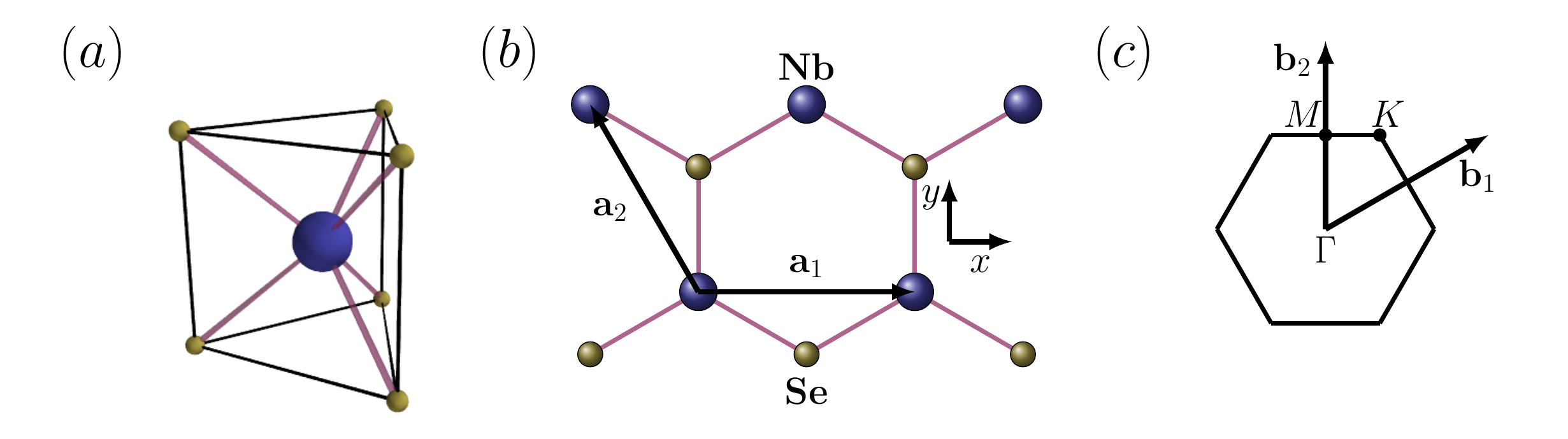} 
\end{center}
\caption{\label{crystal} (Color online) Crystal structure and Brillouin zone of a single layer of 2H-$\mathrm{NbSe}_{2}$. 
(a) 2H structure of $\mathrm{NbSe}_2$. A transition metal atom (Nb, blue sphere) are surrounded by six chalcogens (Se, yellow sphere). 
Three of them lie on the upper plane, and the other three on the lower one. 
(b) Crystal structure from a top view. Primitive lattice vectors $\mathbf{a}_{1}$ and $\mathbf{a}_{2}$ are denoted. 
(c) The first Brillouin zone and reciprocal lattice vectors $\mathbf{b}_{1}$ and $\mathbf{b}_{2}$.}
\end{figure}

Figure~\ref{crystal} displays the crystal structure of a single layer of 2$H$-$\mathrm{NbSe}_{2}$. 
The single layer is composed of three sub-layers: a triangular lattice of Nb at the center, 
which is sandwiched by two outermost layers of Se triangular lattice, forming a trigonal prismatic structure.
The two Se sub-layers of the 2$H$ structure are mirror symmetric to each other 
with respect to the plane of the Nb layer as shown in Fig.~\ref{crystal}(a). 
From a top view, the single layer of 2$H$-$\mathrm{NbSe}_{2}$ looks like a hexagonal lattice. 

We simulate the single layer of 2$H$-$\mathrm{NbSe}_{2}$ by 
taking a vacuum layer of $12$ $\textrm{\AA}$. 
We increase the vacuum layer up to 20$\textrm{\AA}$ and find no changes in calculation results.
Our relaxation calculation gives that the lattice constant of the layer is $a=3.45$ $\textrm{\AA}$ 
and the distance between Nb and Se sublayers is $1.68$ $\textrm{\AA}$, 
agreeing well with the previous studies~\cite{PRB2009Calandra,PRB2009Lebegue}.

We performed density functional theory (DFT) calculation 
by adopting the PBE generalized gradient approximation (GGA)~\cite{PRL1996Perdew} 
for the exchange-correlation functional and the norm-conserving pseudopotential~\cite{PRB1991Troullier} 
with a nonlinear core correction~\cite{PRB1982Louie}. 
Considering a crucial role of core levels in estimating self energy~\cite{PRL1995Rohlfing, PRL2002Marini},
we treated the semicore $4s^2 4p^6$ electrons of Nb atoms as valence electrons and discuss its impact on 
the quasiparticle energy bands in Fig.~\ref{SemiCore} in Appendix~\ref{AppendixA}. 
We used the plane-wave DFT code \textsc{Quantum-Espresso}~\cite{JPhys2009Giannozzi} 
with a cutoff of $55$ Ry, a $40 \times 40 \times 1$ $k$-point grid, 
and a smearing temperature $k_{B} \tau=0.005$ Ry. 
Quasiparticle energies were calculated within the level of $G_0 W_0$ approximation~\cite{PR1965Hedin, PRB1986Hybertsen} 
implemented in the \textsc{BerkeleyGW} code~\cite{CPC2012Deslippe} 
(hereafter, we will call the quasiparticle bands from $G_0 W_0$ approximation as $GW$ energy bands for convenience). 
The slab truncation scheme was used to treat the Coulomb interaction for the single layer geometry~\cite{CPC2012Deslippe}.
We used unoccupied bands up to $5$ Ry above the Fermi energy for the dielectric function calculation
to achieve the convergence of calculations.

\section{\label{results}Results}

\subsection{Band Structure}
\begin{figure}[t]
\begin{center}
\includegraphics[width=0.9\columnwidth, clip=true]{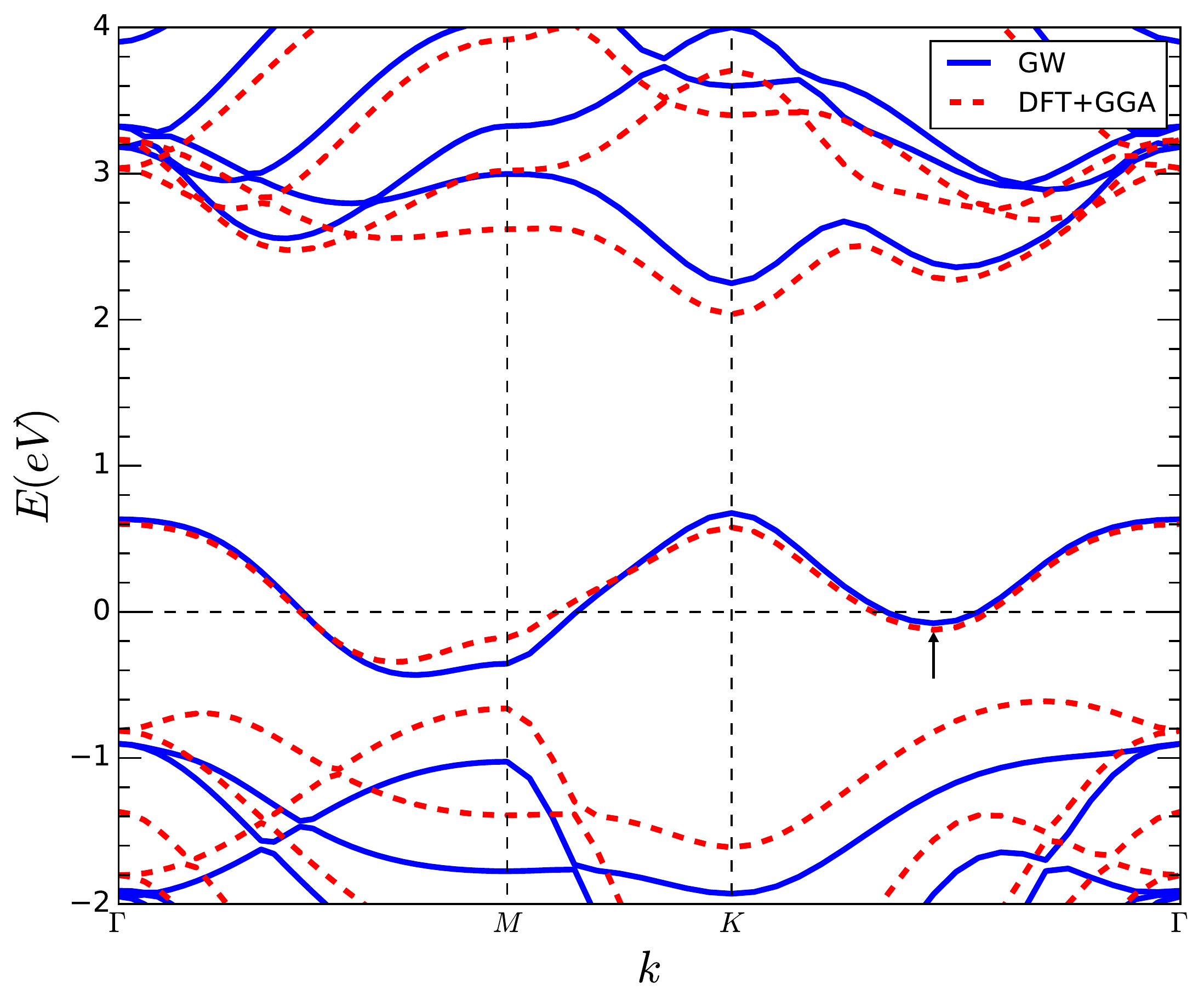} 
\end{center}
\caption{\label{bands} (Color online) Electronic band structures of monolayer $\textrm{NbSe}_{2}$ along the IBZ boundary. 
Red solid line and blue dashed one represent DFT-GGA and $GW$ band structures, respectively. 
The Fermi energy is set to be zero. Black arrow indicates the saddle point of the partially occupied band.}
\end{figure}

Figure~\ref{bands} shows electronic band structures calculated by DFT-GGA (red dashed line) 
and $GW$ approximation (blue solid line) respectively along symmetric lines of the irreducible Brillouin zone (IBZ). 
Our DFT-GGA band structure shows a quite good agreement with previous studies~\cite{PRB2009Calandra,PRB2009Lebegue}.
We note that there are noticeable differences in a partially occupied band (up to spin degeneracy) 
around the Fermi level in the two calculation schemes. 
First, the energy band minimum, which is located in the middle of the $\Gamma M$ line in the DFT-GGA calculation, moves toward the $M$ point
and becomes to be lowered, 
when the $GW$ correction is included. 
Second, the saddle point of the $GW$ band on the $\Gamma K$ line (indicated by black arrow in Fig.~\ref{bands}) 
is closer to the Fermi level than that of DFT-GGA. It is still slightly lower ($-77$ meV) than the Fermi level ($E_F$)
while the DFT-GGA value for the point is much lower ($-123$ meV) from  the $E_F$.
This close proximity of the saddle point to the charge neutral point is important in discussing CDW phase later.
The shape of unoccupied DFT-GGA band structure and their $GW$ corrections shown in Fig.~\ref{bands} 
looks quite similar to those of $\textrm{MoS}_2$~\cite{PRL2013Qiu} although
the semiconducting $\textrm{MoS}_2$ has a significant band gap enhancement from the self-energy corrections~\cite{PRL2013Qiu}.
Here, the spectral gap between the partially occupied band and unoccupied bands does not increases with the $GW$ correction
but the band width of partially occupied band crossing the $E_F$ is enlarged by 17~\% with $GW$ approximations.

\subsection{Fermi Surface}

\begin{figure}[t]
\begin{center}
\includegraphics[width=1.0\columnwidth, clip=true]{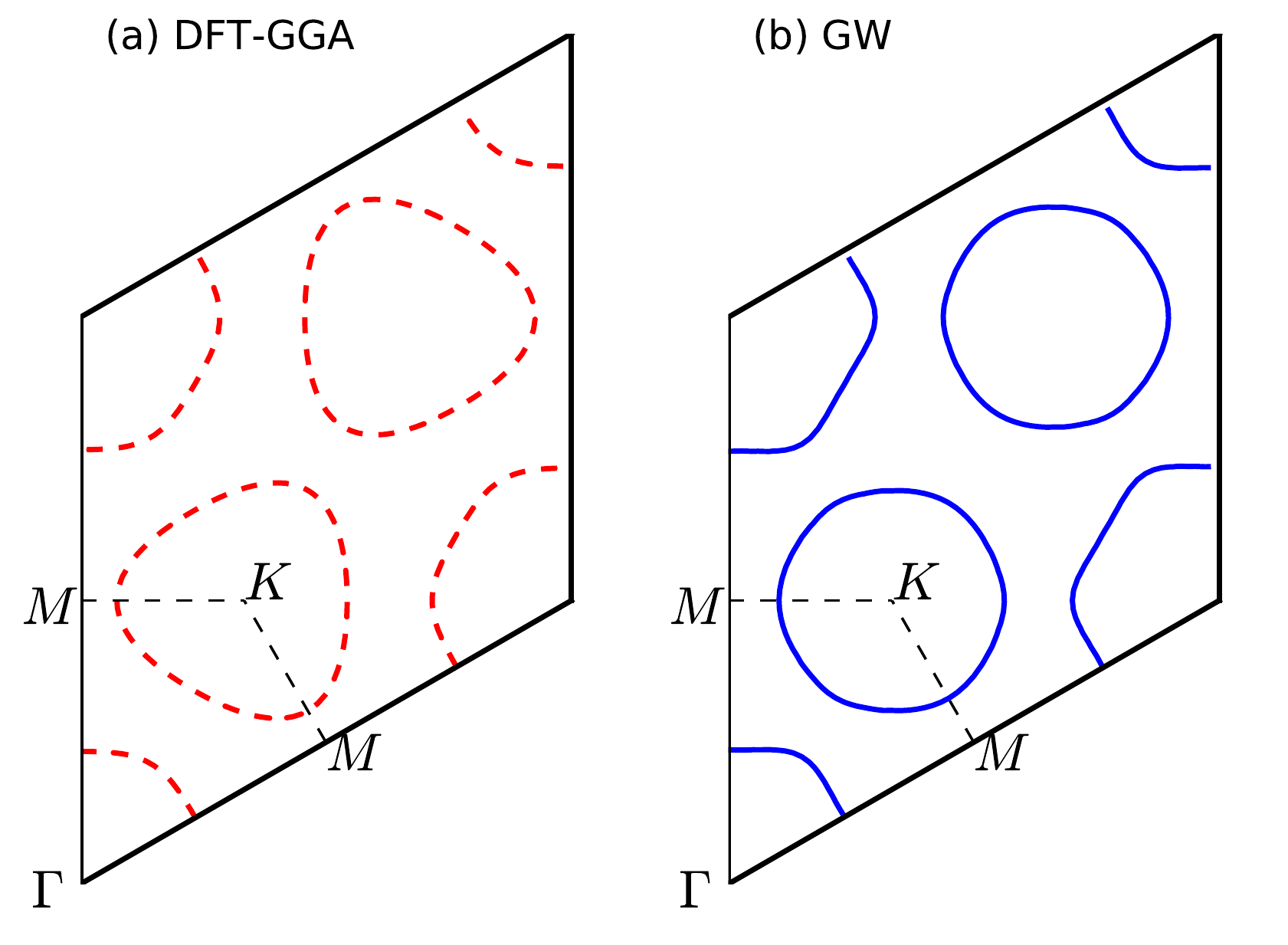} 
\end{center}
\caption{\label{fermi_surface} (Color online) Fermi surfaces of DFT-GGA (red solid line) (a) and $GW$ (blue dashed line) energy bands (b).}
\end{figure}

The Fermi surface also changes when the $GW$ correction is added to the DFT-GGA result. 
As shown in Fig.~\ref{fermi_surface}, calculated Fermi surfaces using the two methods basically 
contain two distinct hole pockets at the $K$ and $\Gamma$-points respectively. 
In the DFT-GGA Fermi surface shown in Fig.~\ref{fermi_surface} (a), a hexagonal hole pocket is at the $\Gamma$-point
and a rounded triangular hole pocket at the $K$-point agreeing well with previous calculations~\cite{PRB2008Johaness,JPhysC1978Doran,PRB2006Johannes,PRB2009Calandra,PRB2009Lebegue}.
In contrast, the triangular hole pockets in the DFT-GGA band becomes to be a rounded 
hexagonal pocket in the $GW$ band as shown in Fig.~\ref{fermi_surface} (b). 
So, flat sides of triangular pockets facing corners of hexagonal ones in the DFT-GGA Fermi surfaces
protrude toward the $\Gamma$ point with $GW$ corrections. 
Then, the distance between the two band crossing points 
along the $\Gamma K$ line decreases with $GW$ corrections.
This corresponds to the fact that the saddle point of the $GW$ band in the $MK$ line shifts up, 
as seen in Fig.~\ref{bands}, thus enabling one approach to the saddle point easier with slight hole doping.

\subsection{Noninteracting Susceptibility}

In order to investigate the implication of the change in the electronic structure 
on a formation of the charge density wave, we have calculated the real part 
of the noninteracting static susceptibility 
$\chi^{\prime}_{0}(\mathbf{q})$~\cite{PRB1974Cooke, PRB1975Rath, PRB1976Gupta, PRB1975Myron, PRB1977Myron, PRB2006Johannes, PRB2009Calandra},

\begin{equation}
\label{chiqr}
\chi^{\prime}_{0}(\mathbf{q}) = \sum_{n,n^\prime} \sum_{\mathbf{k}} 
\frac{f(\epsilon_{n\mathbf{k}})-f(\epsilon_{n^\prime \mathbf{k}+\mathbf{q}})}{\epsilon_{n\mathbf{k}} - \epsilon_{n^\prime \mathbf{k}+\mathbf{q}}} \left|\langle n\mathbf{k}| e^{i\mathbf{q}{r}} |n^\prime \mathbf{k}+\mathbf{q} \rangle \right|^2,
\end{equation}
where $\epsilon_{n \mathbf{k}}$ and $|n\mathbf{k}\rangle$ are the energy of the $n$-th band 
at the crystal momentum $\mathbf{k}$ and its corresponding Bloch state, 
and  $f(\epsilon_{n\mathbf{k}})$ is the Fermi-Dirac distribution function. 
Other quantum numbers such as spin are implicitly included in $n$. 
In principle, the matrix element~\cite{PRB1976Gupta, PRB1975Myron, PRB1977Myron, PRB2006Johannes} 
is needed to calculate 
but it is known that the constant matrix approximation~\cite{PRB1975Myron, PRB1977Myron, PRB2006Johannes, PRB2009Calandra}, 
in which the matrix element is set to be unity, is good 
for the transition metal dichalcogenide~\cite{PRB1977Myron}. 
The $k$-point grid used for the calculation is $300\times300\times1$. 

\begin{figure}[t]
\begin{center}
\includegraphics[width=1.0\columnwidth, clip=true]{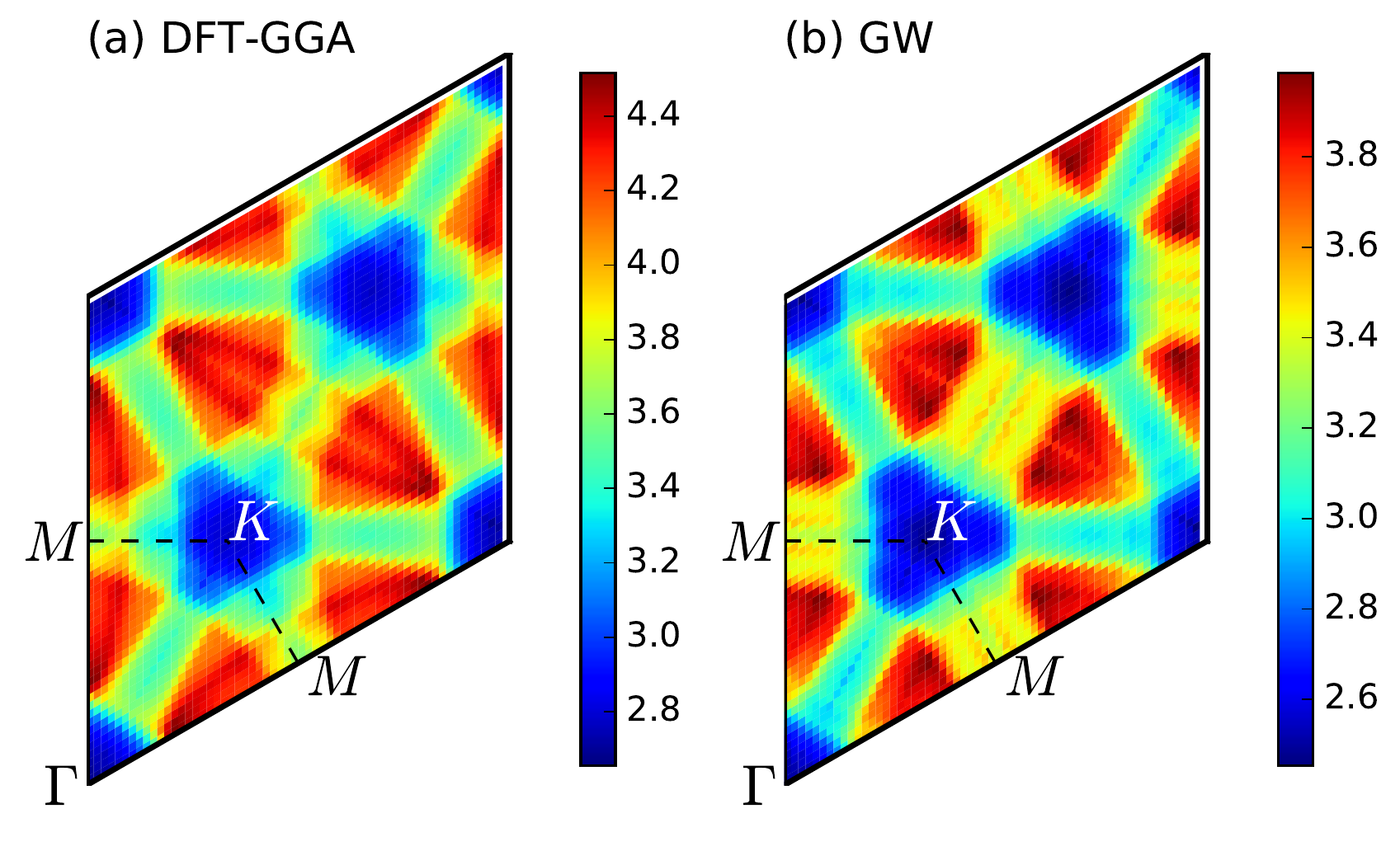} 
\end{center}
\caption{\label{chi-BZ} (Color online) The real part of the noninteracting susceptibility $\chi_{0}^{\prime}(q)$ of (a) DFT-GGA 
and (b) $GW$ bands within the constant matrix approximation (in arbitrary units). 
The susceptibility is calculated at $k_{B}T=10 \mathrm{meV}$.}
\end{figure}

\begin{figure}[b]
\begin{center}
\includegraphics[width=0.9\columnwidth, clip=true]{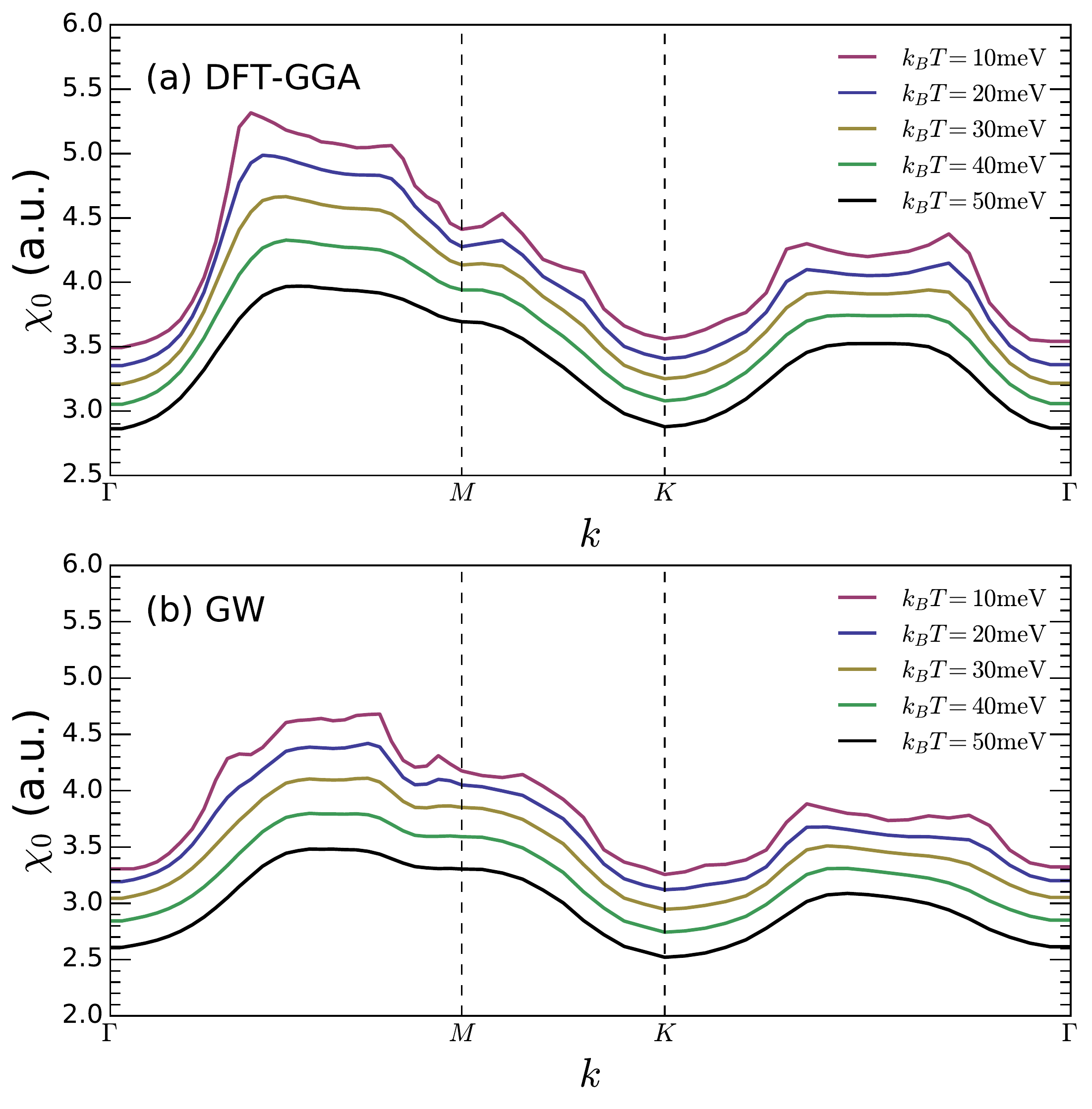} 
\end{center}
\caption{\label{chi-temp} (Color online) The real part of the noninteracting susceptibility $\chi_{0}(q)$ of 
(a) DFT-GGA and (b) $GW$ bands along the IBZ boundary ($\Gamma M K$) 
for several temperatures [$k_{B}T=10$ (purple), $20$ (blue), $30$ (yellow), $40$ (green), and $50$ (black) meV]. 
For clarify, the curves are shifted vertically by proper amounts.}
\end{figure}

Figure~\ref{chi-BZ} shows the real part of the bare static susceptibility $\chi_{0}^{\prime}(\mathbf{q})$ of 
(a) DFT-GGA  and (b) $GW$ calculations. 
We have checked that $\chi_{0}^{\prime}(\mathbf{q})$ of DFT-GGA has a broad maxima 
extending approximately from $2/5\Gamma M$ to $4/5\Gamma M$. 
Our calculation agrees well with other previous studies on the bulk structure~\cite{PRB2008Johaness,JPhysC1978Doran,PRB2006Johannes}
as well as the single layer~\cite{PRB2009Calandra}.
With $GW$ approximation, a general shape of $\chi_{0}^{\prime}(\mathbf{q})$ is quite similar to that with DFT-GGA [Fig.~\ref{chi-BZ} (b)]
in spite of apparent differences between Fermi surfaces from the two calculation methods. 
We note that a peak is at $\left( \frac{1}{3}\vec{a}_{1} + \frac{1}{15} \vec{a}_{2}\right)$ point of IBZ not along the high symmetric line unlike the case with DFT-GGA.

 In order to see the temperature dependence of $\chi_{0}^{\prime}(\mathbf{q})$, 
 we have repeated $\chi_{0}^{\prime}(\mathbf{q})$ calculations 
 by changing temperature $k_{B} T$. Figure~\ref{chi-temp} shows the bare susceptibility $\chi_{0}^{\prime}(\mathbf{q})$ 
 along the IBZ boundary at several temperatures. 
 At $k_{B}T\leq 10$ $\mathrm{meV}$, the maximum plateau of the DFT-GGA susceptibility 
 has a subpeak at $\mathbf{q}\approx 2/5\Gamma M$. 
 As temperature increases, this subpeak disappears and the broad plateau between $2/5\Gamma M$ and $4/5\Gamma M$ remains. 
 For the $GW$ case, the trend is quite similar to those with DFT-GGA along the high symmetric lines of BZ.

\subsection{Doping Effects}

\begin{figure}[t]
\begin{center}
\includegraphics[width=1.0\columnwidth, clip=true]{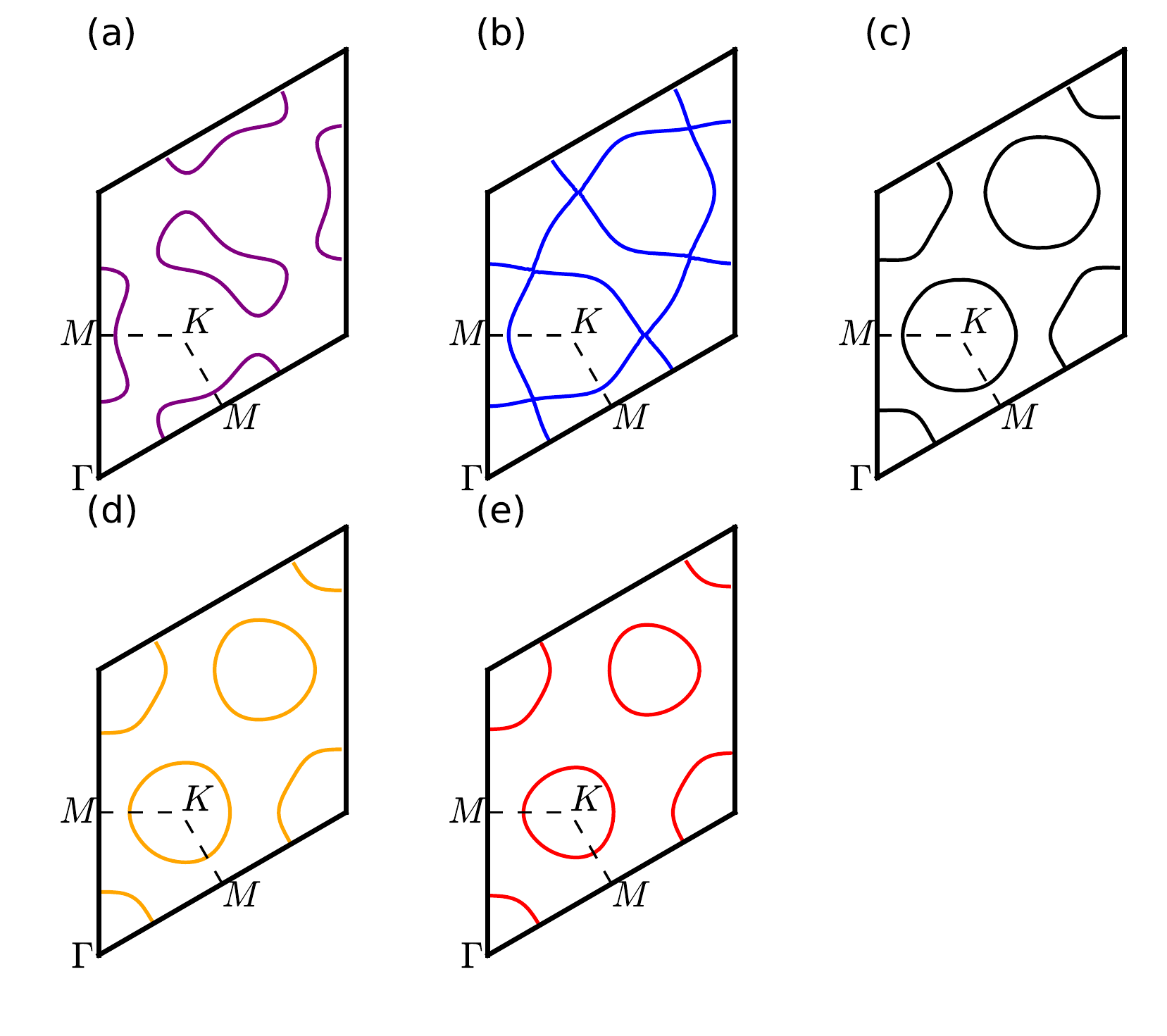} 
\end{center}
\caption{\label{doped_fermi_surface} (Color online) Fermi Surfaces (blue solid lines for the hole doped case and red for electron) 
of the single layer where the Fermi energy is rigidly shifted by $\delta \epsilon_{F}$ due to doping from the charge neutral one (black): 
(a) $\delta \epsilon_{F}=-150~\mathrm{meV}$, (b) $\delta \epsilon_{F}=-77~\mathrm{meV}$, 
(c) $\delta \epsilon_{F}= 0~\mathrm{meV}$, (d) $\delta \epsilon_{F}= 80~\mathrm{meV}$, and (e) $\delta \epsilon_{F}=150~\mathrm{meV}$.}
\end{figure}


Next we consider the doping effect on the bare static susceptibility. 
The doping effect might be taken into account by rigidly shifting the Fermi energy. 
$\delta \epsilon_{F}$ denotes the Fermi energy shift with respect to the charge neutral level 
due to the doping effect. 
Figure~\ref{doped_fermi_surface} shows Fermi surfaces for two $p$-doped cases
[(a) $\delta \epsilon_{F}=-150$ $\mathrm{meV}$, (b) $\delta \epsilon_{F}=-77$ $\mathrm{meV}$], 
the undoped case [(c) $\delta \epsilon_{F}=0$ $\mathrm{meV}$], and two $n$-doped cases 
[(d) $\delta \epsilon_{F}=80$ $\mathrm{meV}$, (e) $\delta \epsilon_{F}=150$ $\mathrm{meV}$]. 
First we consider $p$-doped cases [Fig.~\ref{doped_fermi_surface}(a) and (b)]. As seen in Fig.~\ref{bands}, 
the saddle point becomes closer to the Fermi level as the system is more $p$-doped. 
Thus, one can expect that the distance between the two hole pockets gets shorter. 
In particular, Fig.~\ref{doped_fermi_surface}(b) displays the special case 
where the Fermi level touches the saddle point on the $\Gamma K$ line. 
In this case, two hole pockets (triangle and hexagon) are connected to each other.
The Fermi surface in Fig.~\ref{doped_fermi_surface}(b) is exactly the same one discussed 
in Ref.~\onlinecite{PRL1975Rice}.
So, a slight hole doping enable the system undergo CDW phase transition through 
the logarithmic divergence of the susceptibility at the CDW wave vector connecting the saddle points~\cite{PRL1975Rice}.
Being more $p$-doped, the topology of the Fermi surface is totally different from that of the undoped one. 
As shown in Fig.~\ref{doped_fermi_surface}(a), triangular and hexagonal hole pockets are no longer observed, 
but there is only one rounded rectangular electron pocket, thus changing its carrier type.

In contrast, as shown in Figs.~\ref{doped_fermi_surface}(d) and (e), 
the topology of the $n$-doped system does not change 
compared with the undoped case, 
except for the fact that the size of the two pockets is reduced.

\begin{figure}[t]
\begin{center}
\includegraphics[width=1.0\columnwidth, clip=true]{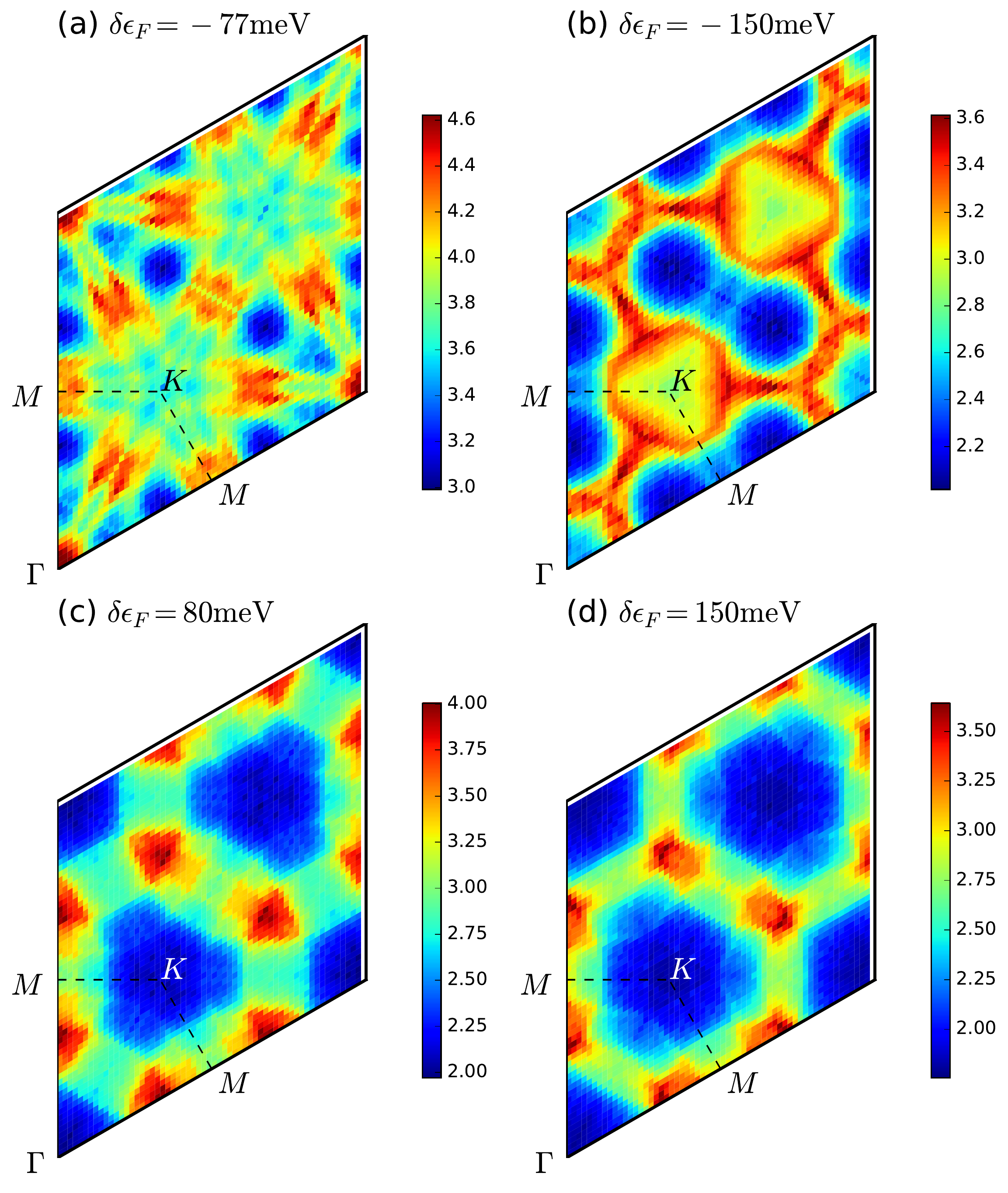} 
\end{center}
\caption{\label{chi-ndoped} (Color online)$\chi_{0}^{\prime}(\mathbf{q})$ of $p$-doped cases for $GW$ bands: 
(a) $\delta \epsilon_{F}= -77\mathrm{meV}$ and (b) $\delta \epsilon_{F}=-150\mathrm{meV}$.  
$\chi_{0}^{\prime}(\mathbf{q})$ of $n$-doped cases for $GW$ bands: 
(c) $\delta \epsilon_{F}= 80\mathrm{meV}$ and (d) $\delta \epsilon_{F}=150\mathrm{meV}$.}
\end{figure}

Such a difference between Fermi surfaces of $p$-doped and $n$-doped cases 
leads to a qualitative change in the bare static susceptibility $\chi_{0}^{\prime}(\mathbf{q})$. 
For the $n$-doped cases where the topology of the Fermi surface does not change, 
the landscape of the noninteracting susceptibility $\chi_{0}^{\prime}(\mathbf{q})$ resembles 
that of the undoped system. 
See Fig.~\ref{chi-ndoped}(c) and (d) for $\chi_{0}^{\prime}(\mathbf{q})$ 
of $\delta \epsilon_{F}=80$ $\mathrm{meV}$ and $150$ $\mathrm{meV}$, respectively. 
Compared with the undoped case, the minor difference is that the peak moves slightly toward $\Gamma M$ line, 
and the broad maximum from $3/5 \Gamma M$ to $4/5 \Gamma M$ is enhanced. 

On the contrary, for the $p$-doped cases where the relative position of the saddle point is critical, 
the topological change in the Fermi surface gives rise to a qualitatively different susceptibility pattern. 
For low doping, i.e., $\delta \epsilon_{F} > -77$ $\mathrm{meV}$, 
the susceptibility pattern does not greatly deviate from that of the undoped system [See Fig.~\ref{chi-ndoped}(a)]. 
However, when the system is more $p$-doped such that the Fermi energy is below the saddle point, 
quite a different pattern emerges. For example, see Fig.~\ref{chi-ndoped}(b) where $\delta \epsilon_{F} = -150$ $\mathrm{meV}$.
As shown in the Fermi surface for the case of $\delta \epsilon_{F} = -77$ $\mathrm{meV}$, 
the peak of $\chi_{0}^{\prime}(\mathbf{q})$ for this case  is at the $M$-point indicating its logarithmic divergence
shown in Fig.~\ref{chi-ndoped} (a).
One noticeable change for the case of further $p$-doping 
is that the largest peak lies neither at the $M$ point nor on the $\Gamma M$ line [Fig.~\ref{chi-ndoped} (b)]. 
Rather, the main peak is found on the $\Gamma K$ line for the case of $\delta \epsilon_{F} = -150$ $\mathrm{meV}$ . 

\subsection{Effect of Spin-Orbit Coupling}

So far the effect of spin-orbit coupling (SOC) has not been taken into account. 
Unlike the bulk 2$H$ structure where the combination of time-reversal and inversion symmetries 
prohibits band splitting due to SOC~\cite{PRB2012Ge}, 
a single layer of the 2$H$ structure can be significantly affected by SOC, 
since there is no center of inversion symmetry. 
One might expect that a new feature can emerge on the Fermi surface, 
depending on whether the band splitting due to SOC is comparable 
to the energy of the saddle point with respect to the Fermi level. 

\begin{figure}[t]
\begin{center}
\includegraphics[width=1.0\columnwidth, clip=true]{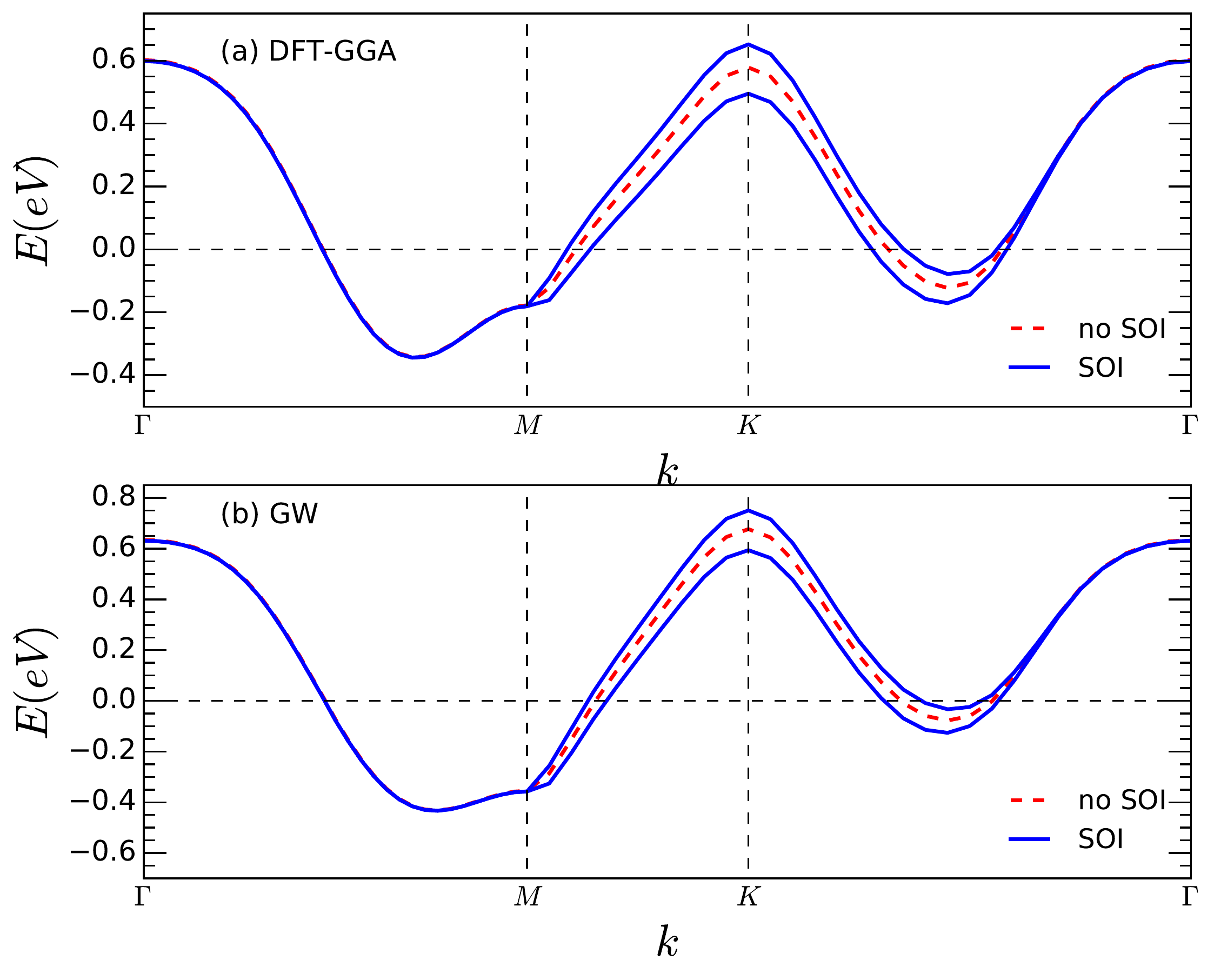} 
\end{center}
\caption{\label{soi-bands} (Color online) Partially occupied bands splitted by SOC 
along the IBZ boundary: (a) DFT-GGA and (b) $GW$ bands. 
Red dashed lines and blue solid ones indicate electronic bands without and with SOC, respectively.}
\end{figure}

For DFT calculations fully relativisitic pseudopotentials are used. 
SOC are included on the final stage of the $GW$ approximation
by adding DFT-SOC results since the effect of SOC to $GW$ results
is marginal as shown in Ref.~\onlinecite{PRB2012Yazyev}. 
We also note that a similar calculation method for the SOC effects has been used for explaining optical spectrum of 
a single layer MoS$_2$ successfully~\cite{PRL2013Qiu, NatMater2014Ugeda}.
Figure~\ref{soi-bands} shows the electronic band structure corrected by SOC 
around the Fermi energy from (a) DFT-GGA and (b) $GW$ schemes. 
As expected, the single band around the Fermi level is splitted to two bands for both cases. 
Note that the electronic band on the $\Gamma M$ line is not affected by SOC for the both cases. 
This can be understood by the fact that there is a mirror symmetry 
with respect to a plane perpendicular to the lattice vector $\mathbf{a}_{1}$ in Fig.~\ref{crystal}(b). 
This mirror symmetry together with time-reversal symmetry prohibits band splitting due to SOC along the $\Gamma M$ line.

\begin{figure}[t]
\begin{center}
\includegraphics[width=1.0\columnwidth, clip=true]{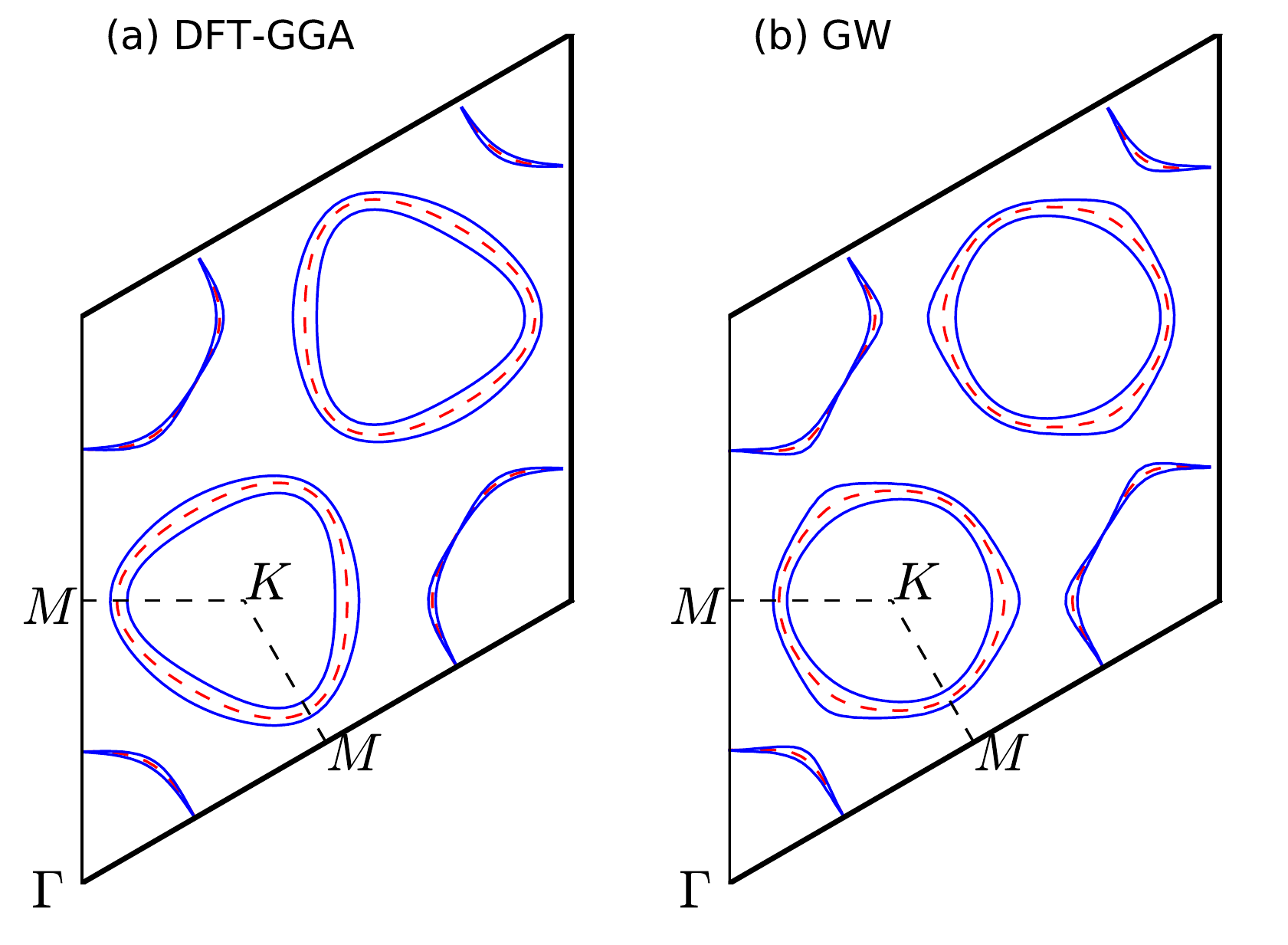} 
\end{center}
\caption{\label{soi_fermi_surface} (Color online) Fermi surfaces of electronic bands corrected by SOC: 
(a) DFT-GGA and (b) $GW$ bands. 
Red dashed lines and blue solid ones indicate Fermi surfaces without and with SOC, respectively.}
\end{figure}

With SOC corrections, we found that the general shape of energy bands are quite similar for the both calculation methods
except the local Fermi surfaces shape near the saddle point.
As shown in Fig.~\ref{soi_fermi_surface}(a), SOC introduces two rounded triangular pockets around $K$ 
and two hexagons around $\Gamma$ to the Fermi surface of DFT-GGA, 
which are just two copies of the Fermi surface without SOC. 
In contrast, with $GW$ approximation, 
the two pockets along the $\Gamma K$ line protrude toward each other further reducing the distance between the two pockets.
So, from the $GW$ Fermi surface with SOC [Fig.~\ref{soi_fermi_surface} (b)], we expect that 
the small amount of $p$ doping can push the Fermi energy down to $34$ meV below the 
charge neutral point connecting two rectangular pockets through saddle points between them. 
This also ensures the logarithmic divergence of the susceptibility making CDW phase a possibility~\cite{PRL1975Rice}.

We also emphasize that the Fermi surface of DFT-GGA with SOC [Fig.~\ref{soi_fermi_surface}(a)] is 
not a general result for a single layer of other metallic 2$H$-TMDs. 
For example, a single layer of 2$H$-$\textrm{TaSe}_{2}$ shows the Fermi surface similar to Fig.~\ref{soi_fermi_surface}(b) 
except for that the orientation of the triangular pocket is inverted~\cite{PRB2012Ge}. 
Thus, the Fermi surface of a single layer 2$H$-TMDs is determined 
by competition between the energy of the saddle point 
measured from the Fermi level and SOC-induced energy correction, 
which varies material by material. 

We next consider the bare static susceptibility $\chi_{0}^{\prime}(\mathbf{q})$ 
for bands corrected by SOC. 
Special cares are needed when the constant matrix approximation is applied 
to Eq.~(\ref{chiqr}) for the SOC-corrected band structure. 
For spin-degenerate bands without SOC, the oscillator strength matrix element 
between different spin components is obviously zero. 
Thus, the spin degree of freedom, which is implicit in Eq.~(\ref{chiqr}), 
just gives the factor two to $\chi_{0}^{\prime}(\mathbf{q})$. 
In contrast, the electronic bands corrected by SOC cannot be labelled 
by a definite spin quantum number over $\mathbf{k}$ space, 
but generally have a spin texture varying on $\mathbf{k}$ space. 
Thus it is likely that the interband matrix element of the oscillator strength 
might vanish for some points or regions of $\mathbf{k}$ space. 
It might not be correct to replace the interband matrix element 
of the oscillator strength just as unity for the whole $\mathbf{k}$ space without justification.

For the single layer of metallic TMDs, the bare susceptibility is mainly determined by two bands
splitted by SOC around the Fermi level. It is known that these bands are composed primarily of 
$d_{z^2}$, $d_{xy}$, and $d_{x^2-y^2}$ orbitals of transition metal~\cite{PR1957Miasek,PhysSS1968Egorov, PhysSS1978Kuznetsov,PRL1973Mattheiss, PRB2013Liu}.
In this effective subspace $\{d_{z^2}, d_{xy}, d_{x^2-y^2}\}$, 
the SOC term does not mix spin components. 
In other words, the two bands splitted by SOC have the opposite spin over $\mathbf{k}$ space. 
Therefore, interband components of the oscillator strength between the two bands might vanish, 
and the constant matrix approximation can be applied to intraband components. 
Note that this is the zeroth approximation, and this argument might be modified 
when $p$ orbitals of chalcogens and their SOC are involved. 
For detailed discussions on the effective tight-binding model of $d$ orbitals and the atomic SOC, 
see Appendix~\ref{AppendixB} and \ref{AppendixC}.

\begin{figure}[t]
\begin{center}
\includegraphics[width=1.0\columnwidth, clip=true]{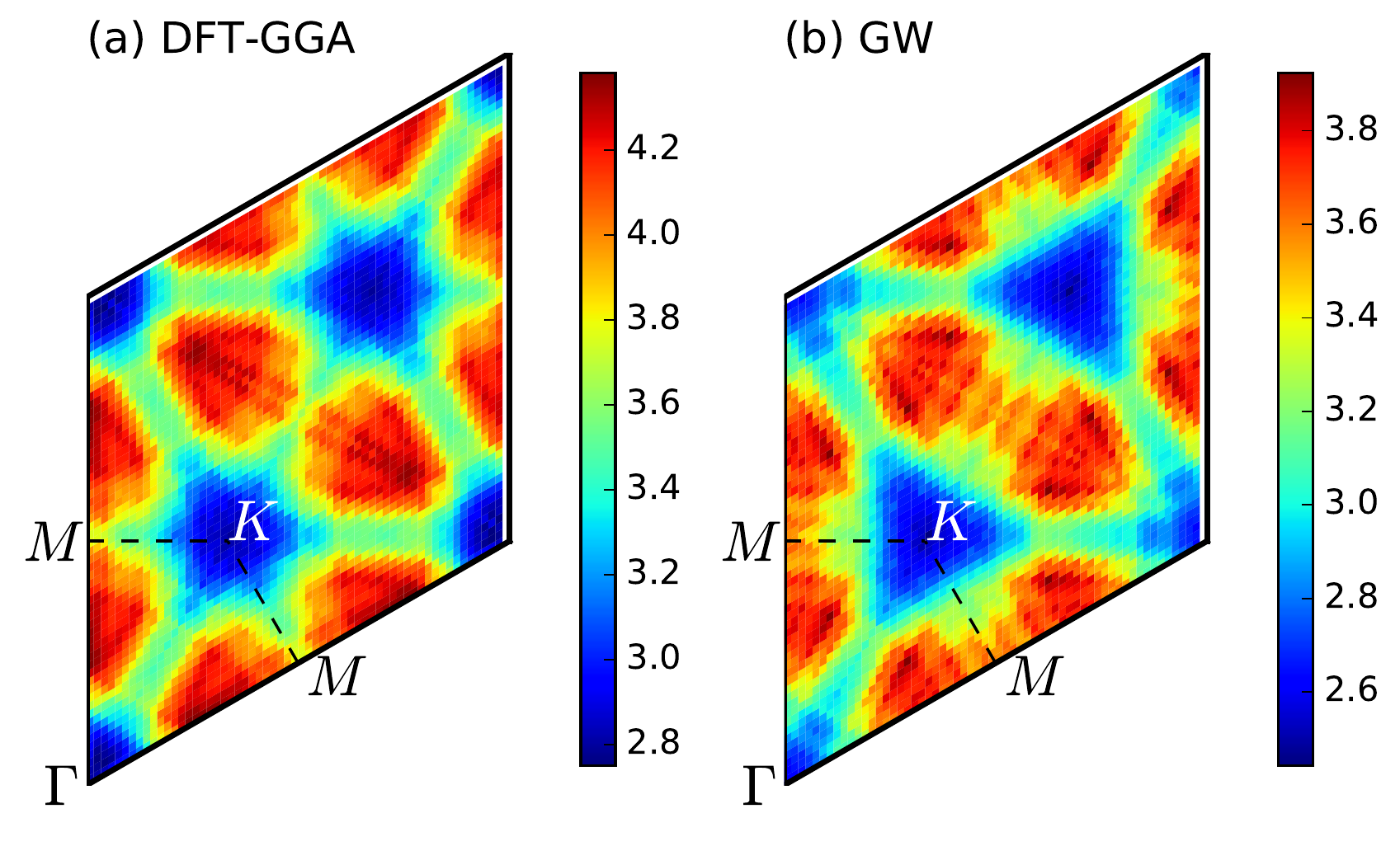} 
\end{center}
\caption{\label{soi-chi-BZ} (Color online) 
The real part of the non-interacting susceptibility $\chi_{0}^\prime(q)$ of 
SOC-corrected electronic bands from (a) DFT-GGA and (b) $GW$ methods.}
\end{figure}

Figure~\ref{soi-chi-BZ} shows the bare susceptibility $\chi_{0}^{\prime}(\mathbf{q})$ calculated 
from DFT-GGA and $GW$ methods based on the above argument. 
The bare susceptibility of SOC-corrected bands looks similar to 
that of electronic bands without SOC in Fig.~\ref{chi-BZ} 
except for that there are more substructures for the SOC-corrected bands. 
For example, the maximum plateau of DFT-GGA bands on the $\Gamma M$ 
line has more steps as shown in Fig.~\ref{soi-chi-BZ}(a). 
For $GW$ bands, the bare susceptibility follows a similar variation with the case of DFT-GGA [Fig.~\ref{soi-chi-BZ}(b)].

\section{\label{final}Discussion and Conclusion}

Having established the quasiparticle energy bands and the Fermi surface of a monolayer ${\textrm{NbSe}}_2$
using $GW$ approximation, now we discuss their implications on a possible CDW phase 
in the single layer on top of various substrates. 
Following previous discussions about effects of Coulomb interactions 
on the low energy bands of graphene on top of substrates~\cite{PRL2009Park,SciRep2012Hwang}, 
we consider alternations of the static screened Coulomb interaction 
in the momentum space~\cite{PR1965Hedin, PRB1986Hybertsen,CPC2012Deslippe}, 
$W_{\mathbf{GG'}}=\epsilon_{\mathbf{GG'}}^{-1} (\mathbf{q})v(\mathbf{q}+\mathbf{G'})$
by approximately including environmental static dielectric screening ($\epsilon_{\textrm{env}}$) 
in the dielectric matrix ($\epsilon_{\mathbf{GG'}}$) within the random phase approximation (RPA) such as, 
\begin{equation}
\label{dielectric}
\epsilon_{\mathbf{GG'}} (\mathbf{q}) \simeq \epsilon_{\textrm{env}}\delta_{\mathbf{GG'}}-v(\mathbf{q}+\mathbf{G})\chi_{\mathbf{GG'}}(\mathbf{q}),
\end{equation}
where
$v(\mathbf{q}+\mathbf{G})={4\pi}/{|\mathbf{q}+\mathbf{G}|^2}$ is the bare Coulomb interactions, 
$\chi_{\mathbf{GG'}}$ is the non-interacting polarizability, 
$\mathbf{q}$ is a 2D vector in the IBZ, 
and $\mathbf{G}$ is a 2D reciprocal-lattice vector. 
Here, $\epsilon_{\textrm{env}}\equiv(\epsilon_{\textrm{vac}}+\epsilon_{\textrm{sub}})/2$ 
where $\epsilon_{\textrm{vac}}=1$ is a dielectric constant of vacuum and 
$\epsilon_{\textrm{sub}}$ is a static dielectric constant of substrate~\cite{PRL2009Park,SciRep2012Hwang}.
We consider two different substrates, the silicon oxide substrate used in Ref.~\onlinecite{NatNano2015Xi} 
and the BLG/6$H$-SiC substrate in Ref.~\onlinecite{NatPhys2016Ugeda}.
For the former, $\epsilon_{\textrm{env}}=2.45$ where $\epsilon_{\textrm{sub}}=3.9$ for the silicon oxide~\cite{Gray}
so that, like graphene on top of BN, SiC or quartz substrates~\cite{PRL2009Park,SciRep2012Hwang},
the weak substrate screening would not change the $GW$ band structure.
Therefore, we expect that the proximity of saddle points near the Fermi level 
in our $GW$-SOC calculations may introduce the new type of CDW phase.
For the latter, there are some ambiguities in using the approximation of Eq.~\ref{dielectric} because
the dielectric function of bilayer graphene has a strong ${\mathbf q}$ dependence~\cite{RMP2011Sarma}
and because BLG on 6$H$-SiC(0001) surface is usually $n$-doped 
with electric field perpendicular to the plane~\cite{PRL2007Otha,NatMat2007Zhou,SSC2013Kim}. 
Considering a large lattice mismatch between ${\textrm{NbSe}}_2$ and BLG
and rotational disorders between them~\cite{NatPhys2016Ugeda}, we can assume no strong interlayer coupling
so that the ${\mathbf{q}}$-dependence of BLG dielectric function may be neglected. 
Taking into account of the doped BLG on top of 6$H$-SiC(0001) surface~\cite{PRL2007Otha,NatMat2007Zhou,SSC2013Kim}, 
$\epsilon_{\textrm{sub}}$ becomes large for the in-plane polarization.
Such a large substrate screening indicates that
the Fermi surface and quasiparticles bands of the system follow those obtained from DFT-GGA calculations, 
thus favoring the bulk-like (strong electron-phonon coupling induced) CDW phase transition~\cite{PRB2009Calandra}.
We also note that a similar situation can occur in the bulk NbSe$_2$. 
As shown in Fig.~\ref{Bulk-NbSe2}(e) of Appendix~\ref{AppendixD},  
our $GW$ bands calculation of bulk $2H$-NbSe$_2$ shows a quite similar Fermi surface shape 
near around $A$- and $H$-points compared with those from a single layer DFT-GGA calculation.
Such a similarity of Fermi surface shape is due to the relatively large screening effects of adjacent layers in the bulk.
Although a large screening of the substrate could reduce the effects of strong Coulomb interaction, 
there may be a still considerable mismatch between the dispersion with a relatively large substrate screening
and the DFT-GGA energy bands as shown in a previous study 
on the low energy quasiparticle energy bands of graphene~\cite{SciRep2012Hwang}. 
So, we expect that the Fermi surface of the latter system could not
entirely follow those from the mean-field calculation but that it could be a mixture between those two calculation schemes.
So, subtle changes in the Fermi surface from the weak correction of $GW$ approximation could change the CDW periodicity
different from the recent DFT-GGA calculation~\cite{PRB2009Calandra}.

The Fermi surface variation within the $GW$ approximation alone cannot explain the large discrepancy of $T_\textrm{CDW}$
between the two experiments~\cite{NatNano2015Xi,NatPhys2016Ugeda}.
Since a giant phonon softening should occur in the vicinity of $M$-point 
when the saddle points touch the Fermi energy~\cite{PRL1975Rice},
this will enhance the transition temperature beyond the simple mean-field estimation 
but the more comprehensive studies
to understand electron-phonon interaction with strong Coulomb interaction
are need to understand this phenomena, that is beyond the scope of current work. 
A recent detailed calculation~\cite{Canadell} for the single layer on top of graphene 
indicates a negligible interlayer interactionbetween them. 
Contrary to the experiment interpretation~\cite{NatPhys2016Ugeda}, 
the detailed density of state analysis using DFT-GGA calculation~\cite{Canadell}
suggests a strong coupling mechanism for the CDW in the single layer limit but the origin of small energy gap 
at the Fermi energy is still not clear yet.

In conclusion, we have calculated quasiparticle band structures of a single layer 2$H$-$\mathrm{NbSe}_{2}$ 
by using first-principles $GW$ calculation. 
We found that the width of a partially occupied band increases
and its Fermi surface shape changes significantly
compared with those obtained using DFT-GGA calculation method.
The SOC changes the Fermi surfaces further and the resulting energy bands
have the singular saddle points very close to the Fermi surface, 
enabling the system undergo CDW phase transition. 
Considering a relatively easy control of charge doping in 2D crystals, 
we think that the present system is particularly interesting in realizing doping-dependent phase transition. 
We also provide a simple tight-binding model with a basis of three $d$-orbitals of niobium
that can be useful for many-body calculation 
of large hybrid systems involving $\mathrm{NbSe}_{2}$ and other layered systems.

\begin{acknowledgements}
We thank Antonio Castro Neto, Cheol-Hwan Park, Ting Cao and Steven Louie for discussion. Y.-W.S. was supported by the National Research Foundation of the Ministry of Science, ICT and Future Planning Grants 
of Korean government (QMMRC, No. R11-2008-053-01002-0).
Computations were supported by the Center for Advanced Computation of Korea Insitute for Advanced Study.
\end{acknowledgements}

\appendix

\section{\label{AppendixA}Effects of semicore states on $GW$ approximations}
Here we compare band calculation results of the $GW$ approximation with and without including Nb semicore states in the pseudopotential. It has been investigated that $GW$ energy bands can be significantly changed by the inclusion of core states of the transition metal, for example, in II-VI semiconductors like CdS~\cite{PRL1995Rohlfing} or bulk copper~\cite{PRL2002Marini}. In these materials, usual DFT band structures rarely change by including core states whose binding energies are well separated from those of valence states. Despite a distinct separation between core and valence states, when core states have a large spatial overlap with valence states, core states can strongly interact with valence states when the exchange part of the $GW$ self-energy is calculated. This strong interaction between core and valence states indeed leads to a significant modification on the $GW$ band structure~\cite{PRL1995Rohlfing, PRL2002Marini}. 

\begin{figure}[h!]
\begin{center}
\includegraphics[width=1.0\columnwidth, clip=true]{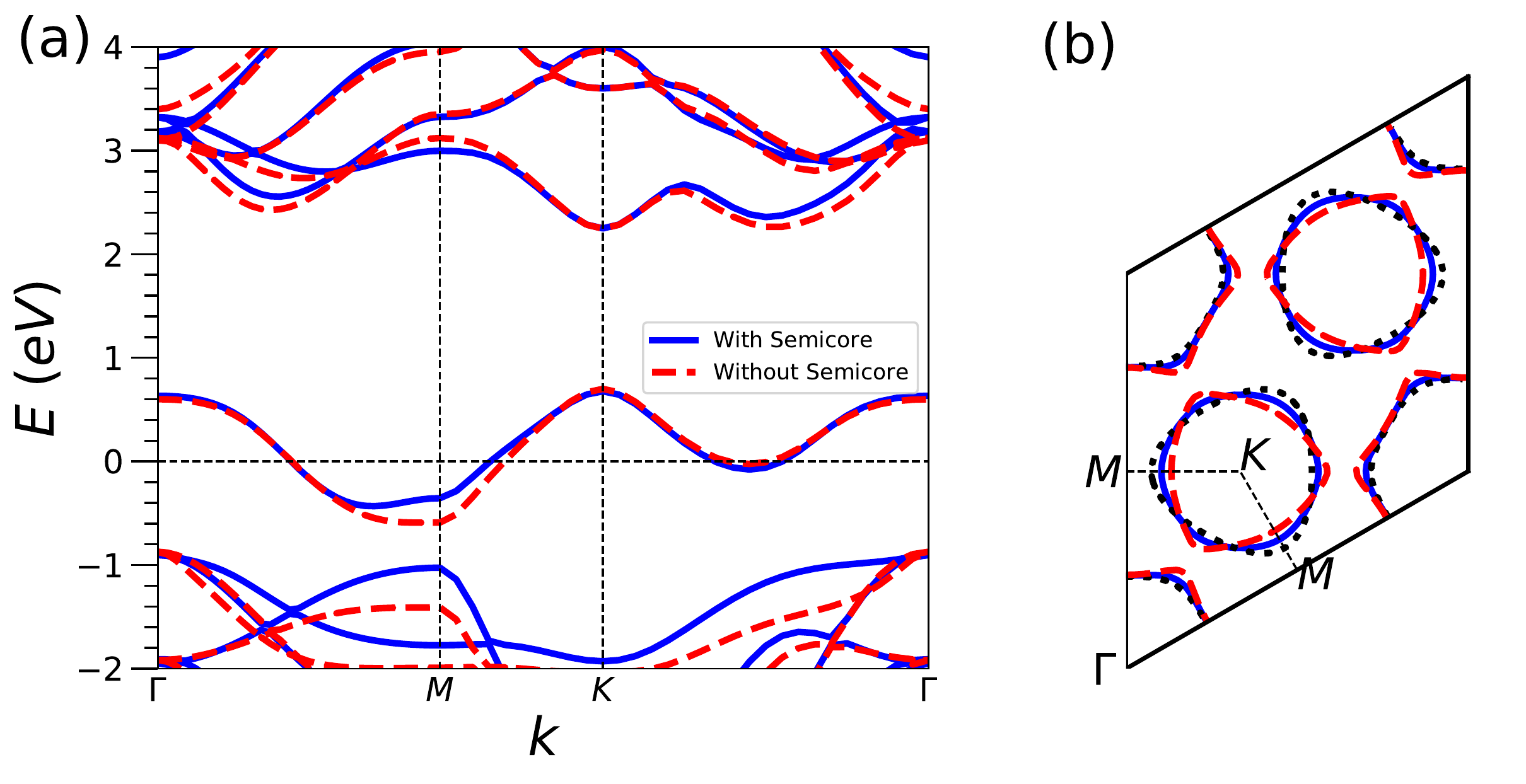} 
\end{center}
\caption{\label{SemiCore} (Color online) (a) $GW$ bands for a single layer NbSe$_2$ with and without semicore levels of $4s^2 4p^6$ of Nb atom, which are denoted by blue solid lines and red dashed ones. (b) Fermi surfaces of $GW$ bands with semicore states (blue solid lines), $GW$ without semicore states (red dashed lines), and DFT-GGA bands (black dotted lines)}  
\end{figure}

Regarding this effect, we have generated pseudo-potentials of Nb atoms with the semicore $4s^2 4p^6$ electrons as valence ones and without them, and we have applied the $GW$ approximation to the two cases. Almost flat bands originating from $4s$ and $4p$ states are located at about $54.03$ eV and $30.39$ eV, respectively, below the Fermi level, which are separated from $4d$ bands around the Fermi level. All-electron calculation of the Nb atom reveals that radial wave functions of $4s$, $4p$, and $4d$ states mostly spread over the similar radial range, thereby implying that there is a strong interaction between core ($4s$ and $4p$) and valence ($4d)$ states in the exchange interaction of the $GW$ approximation. 

Figure~\ref{SemiCore}(a) shows the resulting $GW$ band structures of the NbSe$_2$ monolayer by using pseudopotentials with and without $4s$ and $4p$ semicore states. One distinguished feature induced by the inclusion of semicore states is the shape of the partially occupied band in the neighborhood of the symmetric point $M$. Without semicore states the minimum of the partially occupied band is located at $M$. It is in a sharp contrast to the fact that the partially occupied band has the energy minimum at the middle of the $\Gamma M$ line in the $GW$ approximation with semicore states. The energy difference of the two cases at $M$ is approximately $233$ meV. Fermi surfaces of $GW$ bands with and without semicore states are shown in Fig.~\ref{SemiCore}(b) together with the Fermi surface of DFT-GGA bands. Due to the energy shift around $M$ caused by semicore states, the crossing point of this band with the Fermi level on the $MK$ line moves toward the $M$ point, resulting in a rounded hexagonal Fermi surface around $M$. Without semicore states, the Fermi surface around $M$ looks like a triangular shape whose flat sides face the $M$ point, so the triangular pocket is distinguished from that of the DFT-GGA calculation in terms of the orientation. Except for the energy correction around $M$, energies and shapes of partially occupied bands from the two calculations are almost the same. 

When it comes to energy bands far from the Fermi level, it is first found that four lowest unoccupied bands originating from $4d$ orbitals of Nb do not significantly change. In contrast higher unoccupied bands and fully occupied ones below the Fermi energy are shifted upwards overall.

\section{\label{AppendixB}Atomic Spin-Orbit Coupling in monolayer $\textrm{NbSe}_2$}
We here discuss the effect of SOC and the spin texture of the partially occupied band. Here we consider SOC as the atomic one that gives only on-site terms, 
\begin{equation}
\mathcal{H}_{\mathrm{so}} = \lambda_{\mathrm{TM}} \hat{\mathbf{S}}\cdot \sum_{m\in \mathrm{Nb}} \hat{\mathbf{L}}_{m} +  \lambda_{\mathrm{CH}} \hat{\mathbf{S}}\cdot \sum_{m\in \mathrm{Se}} \hat{\mathbf{L}}_{m},
\end{equation}
where $\lambda_{\mathrm{TM}}$ and $\lambda_{\mathrm{CH}}$ are SOC constants for transition metal and chalcogen respectively, $\hat{\mathbf{S}}$ is the spin $1/2$ operator, and $\hat{\mathbf{L}}_{m}$ is the angular momentum operator of an atom $m$. 
First of all, for the $d$-orbital space, the SOC term at transition metal atom $m$ reads
\begin{widetext}
\begin{equation}
\label{SOC_d}
\lambda_{\mathrm{TM}}\hat{\mathbf{S}}\cdot \hat{\mathbf{L}}_{m} \dot{=} \lambda_{\mathrm{TM}}
\left[\begin{array}{cccccccccc}
0 & 0 & 0 & 0 & 0 & 0 & 0 & 0 & i\frac{\sqrt{3}}{2} & -\frac{\sqrt{3}}{2} \\
0 & 0 & i & 0 & 0 & 0 & 0 & 0 & \frac{1}{2} & -\frac{i}{2}                \\
0 & -i & 0 & 0 & 0 & 0 & 0 & 0 & \frac{i}{2} & \frac{1}{2}                \\
0 & 0 & 0 & 0 & \frac{i}{2} & -i\frac{\sqrt{3}}{2} & -\frac{1}{2} & -\frac{i}{2} & 0 & 0 \\
0 & 0 & 0 & -\frac{i}{2} & 0 & \frac{\sqrt{3}}{2} & \frac{i}{2} & -\frac{1}{2} & 0 & 0 \\
0 & 0 & 0 & i\frac{\sqrt{3}}{2} & \frac{\sqrt{3}}{2} & 0 & 0 & 0 & 0 & 0 \\
0 & 0 & 0 & -\frac{1}{2} & -\frac{i}{2} & 0 & 0 & -i & 0 & 0 \\
0 & 0 & 0 & \frac{i}{2} & -\frac{1}{2} & 0 & i & 0 & 0 & 0 \\
-i\frac{\sqrt{3}}{2} & \frac{1}{2} & -\frac{i}{2} & 0 & 0 & 0 & 0 & 0 & 0 & -\frac{i}{2} \\
-\frac{\sqrt{3}}{2} & \frac{i}{2} & \frac{1}{2} & 0 & 0 & 0 & 0 & 0 & \frac{i}{2} & 0 \\
\end{array}\right],
\end{equation}
with the basis $\{\left|d_{z^2},\uparrow\right\rangle, \left|d_{xy},\uparrow\right\rangle, \left|d_{x^2-y^2},\uparrow\right\rangle, \left|d_{yz},\uparrow\right\rangle, \left|d_{xz},\uparrow\right\rangle, \left|d_{z^2},\downarrow\right\rangle, \left|d_{xy},\downarrow\right\rangle, \left|d_{x^2-y^2},\downarrow\right\rangle, \left|d_{yz}\downarrow\right\rangle, \left|d_{xz},\downarrow\right\rangle \}$.
Here $\left|\uparrow\right\rangle$ and $\left|\downarrow\right\rangle$ are spin eigenstates of the $\hat{S}_{z}$ operator. 
\end{widetext}
Without SOC, it is shown that two subspaces of $d$ orbitals $\{d_{z^2}, d_{xy}, d_{x^2-y^2} \}$ and $\{d_{yz}, d_{xz}\}$ are not coupled in the group-theoretical construction of the $d$-orbital TB model~\cite{PR1957Miasek, PhysSS1968Egorov}. 
Equation~(\ref{SOC_d}) tells us that SOC leads to mixing of the two subspaces $\{d_{z^2}, d_{xy}, d_{x^2-y^2} \}$ and $\{d_{yz}, d_{xz}\}$. Furthermore, $\left|\uparrow\right\rangle$ and $\left|\downarrow\right\rangle$ are no longer spin eigenstates, since $\langle \psi_{1},\uparrow(\downarrow)|\hat{\mathbf{S}}\cdot \hat{\mathbf{L}}_{m} | \psi_{2}, \downarrow (\uparrow) \rangle \neq 0$ in which $\psi_{1} \in \{d_{z^2}, d_{xy}, d_{x^2-y^2} \}$ and $\psi_{2} \in \{d_{yz}, d_{xz}\}$. 

When we consider the partially occupied band around the Fermi energy coming from $\{d_{z^2}, d_{xy}, d_{x^2-y^2} \}$, this band is well separated from the other four bands at least by $2$ eV. Therefore, from the viewpoint of the perturbation theory, the mixing term of Eq.~(\ref{SOC_d}) between $\{d_{z^2}, d_{xy}, d_{x^2-y^2} \}$ and $\{d_{yz}, d_{xz}\}$ might give a negligible contribution to the partially occupied band. In the subspace $\{d_{z^2}, d_{xy}, d_{x^2-y^2} \}$, the SOC term [Eq.~(\ref{SOC_d})] approximately reads
\begin{equation}
\label{SOC_d2}
\lambda_{\mathrm{TM}}\hat{\mathbf{S}}\cdot \hat{\mathbf{L}}_{m} \approx \lambda_{\mathrm{TM}}
\left[\begin{array}{cccccc}
0 &  0 & 0 & 0 & 0 &  0  \\
0 &  0 & i & 0 & 0 &  0  \\
0 & -i & 0 & 0 & 0 &  0 \\
0 &  0 & 0 & 0 & 0 &  0  \\
0 &  0 & 0 & 0 & 0 & -i \\
0 &  0 & 0 & 0 & i &  0  \\
\end{array}\right],
\end{equation}
which is written in the order of $\{\left|d_{z^2},\uparrow\right\rangle, \left|d_{xy},\uparrow\right\rangle, \left|d_{x^2-y^2},\uparrow\right\rangle, \left|d_{z^2},\downarrow\right\rangle, \left|d_{xy},\downarrow\right\rangle, \left|d_{x^2-y^2},\downarrow\right\rangle \}$.
Clearly, Eq.~(\ref{SOC_d2}) has no mixing term between $\left|\uparrow\right\rangle$ and $\left|\downarrow\right\rangle$. Thus $\left|\uparrow\right\rangle$ and $\left|\downarrow\right\rangle$ can be regarded as good spin eigenstates for the partially occupied band in the effective $d$-orbital TB model discussed in Sec.~\ref{AppendixC}.

However, spin eigenstates for the real system might be modified, since there is the SOC term of $p$-orbital from chalcogens. The partially occupied band of our interest is approximately described in terms of three $d$ orbitals $\{d_{z^2}, d_{xy}, d_{x^2-y^2} \}$. However, it is an effective description, but the real band involves contribution from $p$ orbitals of chalcogens. In fact, according to the orbital-projected density of states calculation, the wavefunction of the partially occupied band contains $p$ orbitals of chalcogens by about $20\%$.

The SOC term on a chalcogen $m$ is
\begin{equation}
\label{SOC_p}
\lambda_{\mathrm{CH}}\hat{\mathbf{S}}\cdot \hat{\mathbf{L}}_{m} \approx \lambda_{\mathrm{CH}}
\left[\begin{array}{cccccc}
0 & -\frac{i}{2} & 0 & 0 & 0 & \frac{1}{2}  \\
\frac{i}{2} & 0 & 0 & 0 & 0 & -\frac{i}{2}  \\
0 & 0 & 0 & -\frac{1}{2} & \frac{i}{2} & 0  \\
0 & 0 & -\frac{1}{2} & 0 & \frac{i}{2} & 0  \\
0 & 0 & -\frac{i}{2} & -\frac{i}{2} & 0 & 0 \\
\frac{1}{2} & \frac{i}{2} & 0 & 0 & 0 & 0   \\
\end{array}\right],
\end{equation}
which is ordered in the basis of $\{\left|p_{x},\uparrow\right\rangle, \left|p_{y},\uparrow\right\rangle, \left|p_{z},\uparrow\right\rangle, \left|p_{x},\downarrow\right\rangle, \left|p_{y},\downarrow\right\rangle, \left|p_{z},\downarrow\right\rangle \}$. Two subspaces of $p$ orbitals $\{p_{x}, p_{y}\}$ and $\{p_{z}\}$ are mixed under SOC. Further, $\left|\uparrow\right\rangle$ and $\left|\downarrow\right\rangle$ are not exact spin eigenstates. Eigenstates of the total angular momentum operator $\hat{J} = \hat{L} + \hat{S}$ could be true ones since orbital and spin angular momenta are correlated via SOC. Considering contributions of $p$ orbitals, it might be expected that there is relatively small mixing between $\left|\uparrow\right\rangle$ and $\left|\downarrow\right\rangle$ components for bands splitted by SOC. 

\section{\label{AppendixC}Three-Band Tight-Binding Model}


\begin{figure}[b]
\begin{center}
\includegraphics[width=1.0\columnwidth, clip=true]{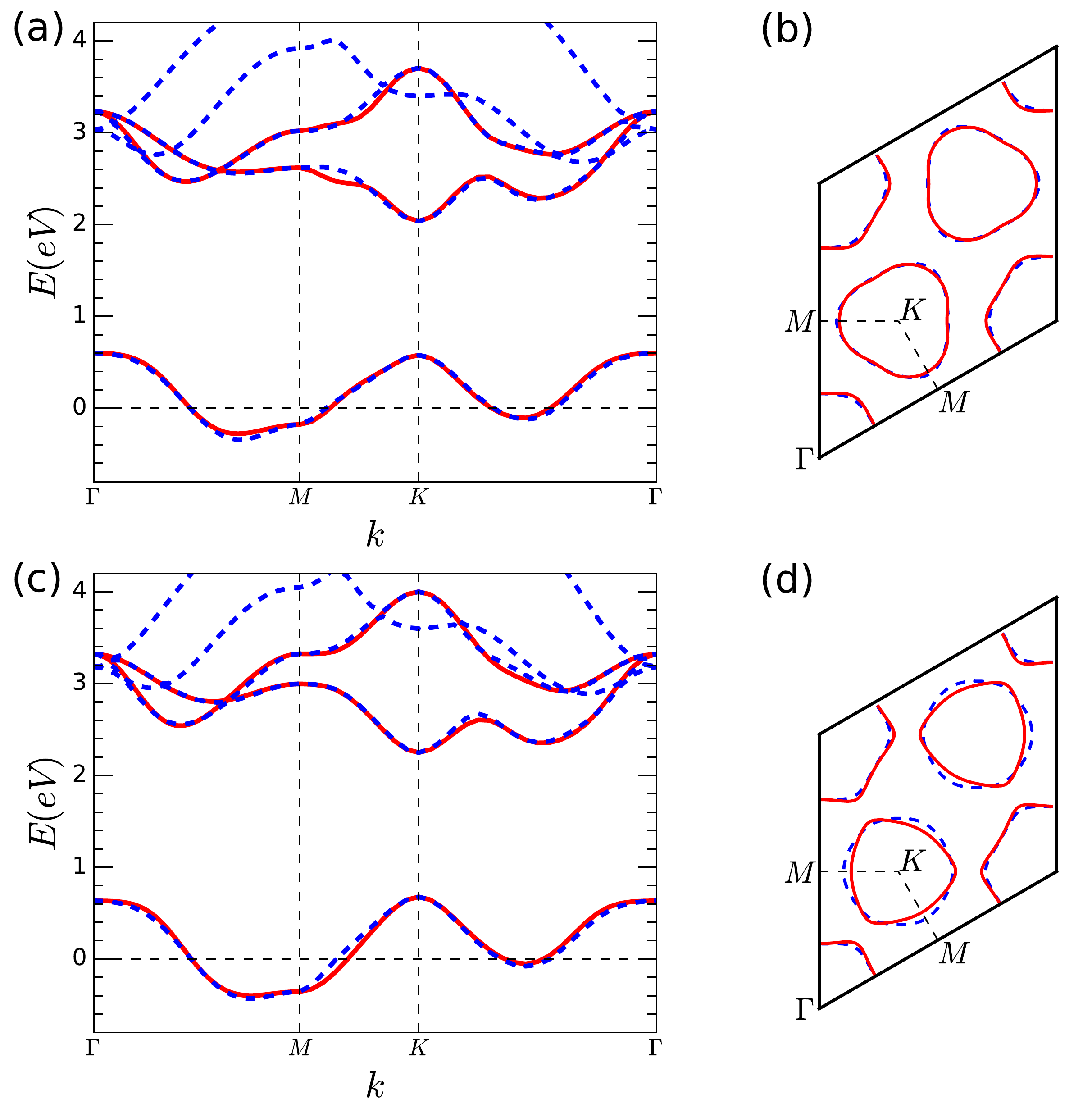} 
\end{center}
\caption{\label{NbSe2-TB} (Color online) (a) Electronic bands from the DFT-GGA calculation (blue dashed line) and ones from the three-band TNN TB model fit to the DFT-GGA bands (red solid line).
(b) Fermi surfaces of electronic bands from the DFT-GGA calculation (blue dashed line) and the TB model (red solid line).(c) Electronic bands from the $GW$ approximation (blue dashed line) and ones from the three-band TNN TB model fit to the $GW$ bands (red solid line). (d) Fermi surfaces of electronic bands from the $GW$ calculation (blue dashed line) and the TB model (red solid line).} 
\end{figure}

In this Section, we construct a tight-binding (TB) model which closely reproduces $GW$ bands, 
especially the partially occupied band around the Fermi energy. 
The partially occupied band comes mostly from $d_{z^2}$, $d_{xy}$, and $d_{x^2-y^2}$ 
orbitals of transition metal~\cite{PR1957Miasek,PhysSS1968Egorov, PhysSS1978Kuznetsov,PRL1973Mattheiss, PRB2013Liu}. 
$p$ orbitals from chalcogen atoms gives the next contribution to the band. 
Although it is more precise to construct the tight-binding model by including the $p$ orbitals, 
we would like to have a minimal model capturing the main physics 
so that it could be used for other theoretical investigations. 
For this purpose, we start with the three-orbital TB model including only $d_{z^2}$, $d_{xy}$, and $d_{x^2-y^2}$ orbitals. 

The form of the TB model and the number of independent parameters 
can be determined by the lattice symmetry. 
In particular, Ref.~\onlinecite{PR1957Miasek} provides a complete table of matrix components 
of the TB model of $d$ orbitals for hexagonal structure, 
which is relevant for TMDs. 
For detailed discussion on the group-theoretical construction of the TB model, 
see Refs.~\onlinecite{PhysSS1968Egorov, PhysSS1978Kuznetsov}. 

The effective TB model of $d$ orbitals has been applied to TMDs in the literature~\cite{PRL1973Mattheiss, PRB2013Liu}. 
Recently, this model has been extensively used in order to fit DFT band structures 
for monolayers of group-VIB TMDs~\cite{PRB2013Liu}. 
In Ref.~\onlinecite{PRB2013Liu}, the authors have extended the three $d$-orbital TB model 
up to the third nearest neighbor (TNN) hoppings. 
Using the TNN TB model, they have successfully reproduced DFT electronic bands. 

As mentioned in Refs.~\onlinecite{PRL1973Mattheiss, PRB2013Liu}, 
the $d$-$d$ interactions in the effective TB model contain the direct hoppings 
between $d$ orbitals of transition metal and the indirect $d$-$d$ hoppings via $p$ orbitals of chalcogens. 
However, within this effective model, one cannot know how much contribution $p$ orbitals give 
to the energy band of our interest, which might be important for some aspects, 
for example, the effect of SOC. 

\begin{figure}[t]
\begin{center}
\includegraphics[width=1.0\columnwidth, clip=true]{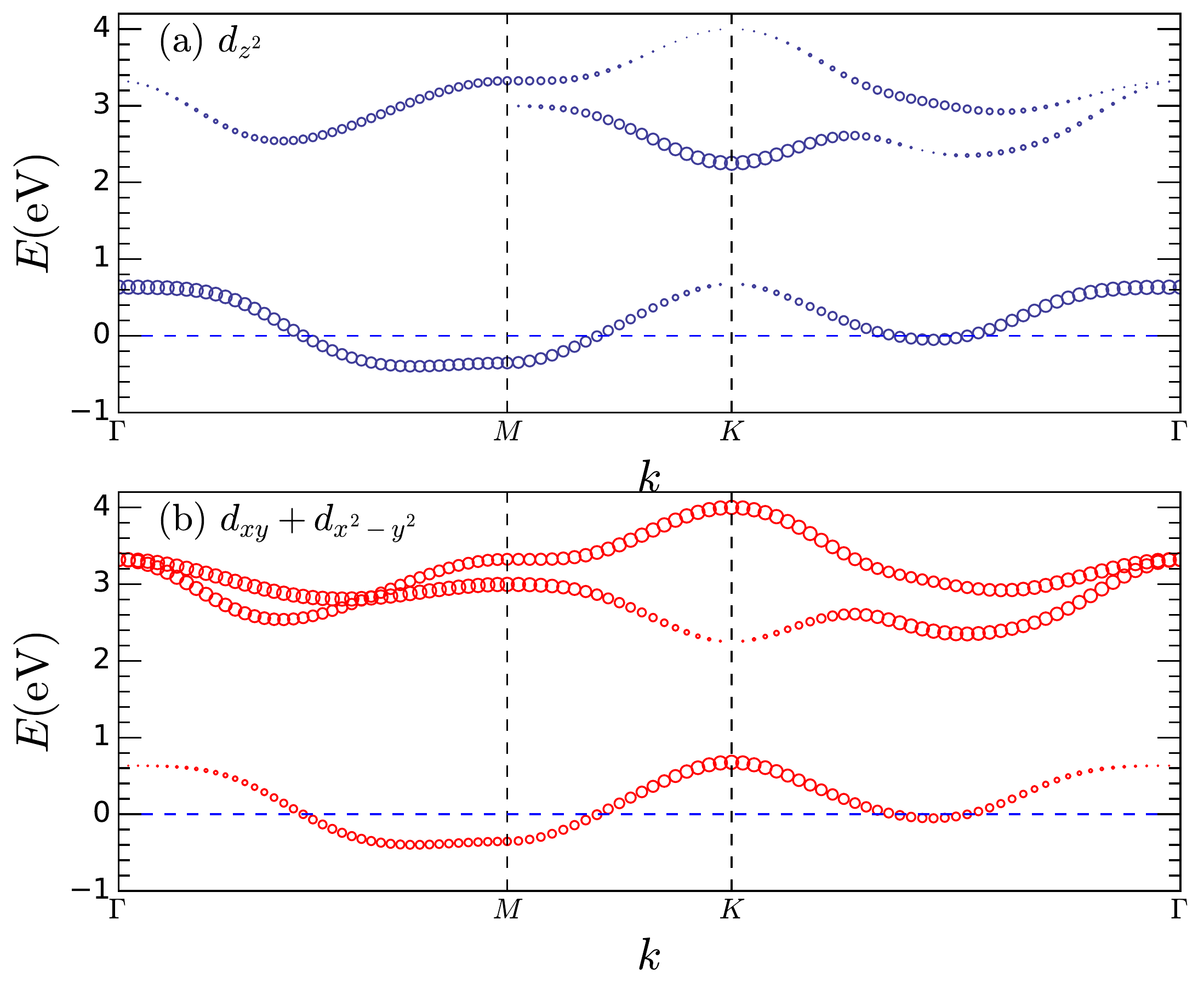} 
\end{center}
\caption{\label{NbSe2-TB-orbitals} (Color online) Orbital projected band structures 
from the three-band TNN TB model of the $GW$ bands: contributions from (a) the $d_{z^2}$ orbital (blue circle) 
and (b) $d_{xy}$ and $d_{x^2-y^2}$ orbitals (red circle). 
The size of open circle is proportional to contribution of the corresponding orbital. 
The Fermi energy (blue dashed line) is set to be zero.}
\end{figure}

Using the three-band TNN model of Ref.~\onlinecite{PRB2013Liu}, 
we fit DFT-GGA and $GW$ bands on the IBZ boundary. 
For this, we first adopt the least-square fitting procedure, and finely tune parameters 
in order to fit the saddle point of the $MK$ line. 
Figure~\ref{NbSe2-TB}(a) shows energy bands of the three-band TNN TB model together with DFT-GGA bands. 
Following notations of Ref.~\onlinecite{PRB2013Liu}, 
fitting parameters of the TB model for DFT-GGA bands are $\epsilon_{1}=1.408$, $\epsilon_{2}=2.048$, 
$t_{0}=-0.128$, $t_{1}=0.115$, $t_{2}=-0.466$, $t_{11}=0.115$, $t_{12}=0.122$, 
$t_{22}=0.036$, $r_{0}=0.025$, $r_{1}=0.194$, $r_{2}=-0.079$, 
$r_{11}=0.021$, $r_{12}=0.096$, $u_{0}=-0.031$, $u_{1}=-0.037$, 
$u_{2}=-0.002$, $u_{11}=0.258$, $u_{12}=-0.179$, and $u_{22}=-0.167$ in units of $\mathrm{eV}$. 
The energy bands of the TB model are in a very good agreement with DFT-GGA bands 
except for a small deviation at the energy minimum of the partially occupied band. As shown in Fig.~\ref{NbSe2-TB}(b), the fitted TB model well reproduces the Fermi surface of the DFT-GGA calculation.

We also construct the three-band TNN model of $GW$ bands, whose fitting parameters are 
$\epsilon_{1}=1.148$, $\epsilon_{2}=2.379$, 
$t_{0}=-0.118$, $t_{1}=-0.386$, $t_{2}=-0.366$, $t_{11}=0.167$, $t_{12}=0.243$, 
$t_{22}=-0.075$, $r_{0}=0.094$, $r_{1}=0.043$, $r_{2}=-0.152$, 
$r_{11}=0.055$, $r_{12}=-0.012$, $u_{0}=-0.061$, $u_{1}=-0.010$, 
$u_{2}=0.002$, $u_{11}=0.140$, $u_{12}=-0.077$, and $u_{22}=-0.014$ in units of $\mathrm{eV}$. Energy bands reproduced by the TNN tight-binding model and $GW$ bands are shown in Fig.~\ref{NbSe2-TB}(c). The energy bands of the TB model agree well with those of the $GW$ calculation, but there is some deviation of the partially occupied band on the $MK$ line. While hexagonal pockets around $\Gamma$ nicely match, the triangular pocket of the TB model is less rounded than that of $GW$ bands. This is due to the small mismatch of the two bands on the $MK$ line in Fig.~\ref{NbSe2-TB}(c). 

Figure~\ref{NbSe2-TB-orbitals} displays orbital-projected bands calculated from the three-orbital TNN TB model of the $GW$ bands. $d_{z^2}$ orbital is dominant for the lowest band around the Fermi level, while $d_{xy}$ and $d_{x^2-y^2}$ orbitals give main contributions to the other two unoccupied bands. 


\section{\label{AppendixD}$GW$ bands of bulk $2H$-NbSe$_2$}


\begin{figure}[t]
\begin{center}
\includegraphics[width=1.0\columnwidth, clip=true]{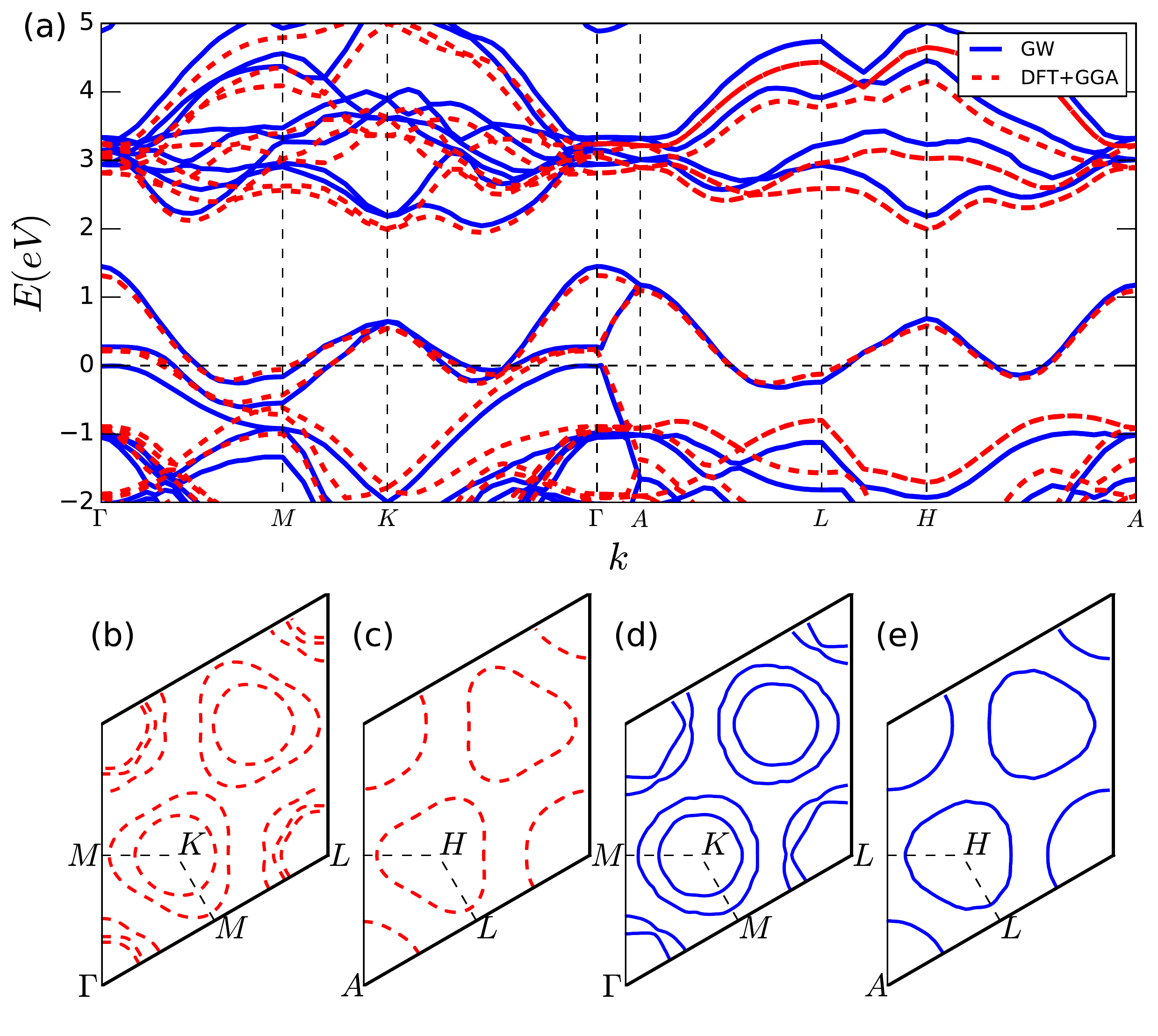} 
\end{center}
\caption{\label{Bulk-NbSe2} (Color online) (a) Energy bands of the bulk $2H$-NbSe$_2$ along symmetric lines of the IBZ. Red dashed lines and blue solid lines correspond to DFT-GGA and $GW$ electronic bands, respectively. $(b)$ and $(c)$ are Fermi surfaces of DFT-GGA bands on $\Gamma M K$ and $ALH$ planes of the IBZ respectively. Similarly $(d)$ and $(e)$ shows Fermi surfaces of $GW$ bands on $\Gamma M K$ and $ALH$ planes, respectively.}
\end{figure}

We have extended DFT-GGA and $GW$ calculations to energy bands of the bulk $2H$-NbSe$_2$ by using the same pseudo-potential including semi-core states. For detailed calculations we have used the plane-wave basis set with a cutoff of 55 Ry, a $20 \times 20 \times 5$ k-point grid, and a smearing temperature $k_{B} \tau=0.005 $ Ry~\cite{JPhys2009Giannozzi}. The $GW$ bands of the bulk $2H$-NbSe$_2$ is calculated within the level of the $G_0 W_0$ approximation~\cite{CPC2012Deslippe} by including about $400$ unoccupied bands, which are up to 10 Ry above the Fermi energy. 

Figure~\ref{Bulk-NbSe2}(a) shows DFT-GGA and $GW$ electronic bands of the bulk $2H$-NbSe$_2$ along symmetric lines of the IBZ, which are denoted by red dashed lines and blue solid ones respectively. Their corresponding Fermi surfaces are depicted in Fig.~\ref{Bulk-NbSe2}(b)-(d). Figure~\ref{Bulk-NbSe2}(b) and (c) show Fermi surfaces of DFT-GGA bands on the $\Gamma MK$ and $ALH$ planes, respectively, while Fig.~\ref{Bulk-NbSe2}~(d) and (e) are Fermi surfaces of $GW$ bands on $\Gamma MK$ and $ALH$ planes respectively. 
As discussed in Sec.~\ref{final}, the Fermi surface of the bulk $2H$-NbSe$_2$ $GW$ bands on the $ALH$ plane is similar to that of the single-layer DFT-GGA bands on the $\Gamma MK$ plane. In particular the Fermi surface around $H$ is a rounded triangular pocket whose flat lines face the symmetric point $A$, but not a rounded hexagonal pocket as seen in the Fermi surface of the monolayer $GW$ calculation. The relative large screening effect of neighboring layers could lead to this similarity between the Fermi surface of bulk $GW$ bands and that of monolayer DFT-GGA bands. 

\bibliography{NbSe2-GW}

\begin{thebibliography}{80}%
\makeatletter
\providecommand \@ifxundefined [1]{%
 \@ifx{#1\undefined}
}%
\providecommand \@ifnum [1]{%
 \ifnum #1\expandafter \@firstoftwo
 \else \expandafter \@secondoftwo
 \fi
}%
\providecommand \@ifx [1]{%
 \ifx #1\expandafter \@firstoftwo
 \else \expandafter \@secondoftwo
 \fi
}%
\providecommand \natexlab [1]{#1}%
\providecommand \enquote  [1]{``#1''}%
\providecommand \bibnamefont  [1]{#1}%
\providecommand \bibfnamefont [1]{#1}%
\providecommand \citenamefont [1]{#1}%
\providecommand \href@noop [0]{\@secondoftwo}%
\providecommand \href [0]{\begingroup \@sanitize@url \@href}%
\providecommand \@href[1]{\@@startlink{#1}\@@href}%
\providecommand \@@href[1]{\endgroup#1\@@endlink}%
\providecommand \@sanitize@url [0]{\catcode `\\12\catcode `\$12\catcode
  `\&12\catcode `\#12\catcode `\^12\catcode `\_12\catcode `\%12\relax}%
\providecommand \@@startlink[1]{}%
\providecommand \@@endlink[0]{}%
\providecommand \url  [0]{\begingroup\@sanitize@url \@url }%
\providecommand \@url [1]{\endgroup\@href {#1}{\urlprefix }}%
\providecommand \urlprefix  [0]{URL }%
\providecommand \Eprint [0]{\href }%
\providecommand \doibase [0]{http://dx.doi.org/}%
\providecommand \selectlanguage [0]{\@gobble}%
\providecommand \bibinfo  [0]{\@secondoftwo}%
\providecommand \bibfield  [0]{\@secondoftwo}%
\providecommand \translation [1]{[#1]}%
\providecommand \BibitemOpen [0]{}%
\providecommand \bibitemStop [0]{}%
\providecommand \bibitemNoStop [0]{.\EOS\space}%
\providecommand \EOS [0]{\spacefactor3000\relax}%
\providecommand \BibitemShut  [1]{\csname bibitem#1\endcsname}%
\let\auto@bib@innerbib\@empty
\bibitem [{\citenamefont {Novoselov}\ \emph {et~al.}(2005)\citenamefont
  {Novoselov}, \citenamefont {Jiang}, \citenamefont {Schedin}, \citenamefont
  {Booth}, \citenamefont {Khotkevich}, \citenamefont {Morozov},\ and\
  \citenamefont {Geim}}]{PNAS2005Novoselov}%
  \BibitemOpen
  \bibfield  {author} {\bibinfo {author} {\bibfnamefont {K.~S.}\ \bibnamefont
  {Novoselov}}, \bibinfo {author} {\bibfnamefont {D.}~\bibnamefont {Jiang}},
  \bibinfo {author} {\bibfnamefont {F.}~\bibnamefont {Schedin}}, \bibinfo
  {author} {\bibfnamefont {T.~J.}\ \bibnamefont {Booth}}, \bibinfo {author}
  {\bibfnamefont {V.~V.}\ \bibnamefont {Khotkevich}}, \bibinfo {author}
  {\bibfnamefont {S.~V.}\ \bibnamefont {Morozov}}, \ and\ \bibinfo {author}
  {\bibfnamefont {A.~K.}\ \bibnamefont {Geim}},\ }\href@noop {} {\bibfield
  {journal} {\bibinfo  {journal} {Proc. Natl. Acad. Sci. USA}\ }\textbf
  {\bibinfo {volume} {102}},\ \bibinfo {pages} {10451} (\bibinfo {year}
  {2005})}\BibitemShut {NoStop}%
\bibitem [{\citenamefont {Castro~Neto}\ \emph {et~al.}(2009)\citenamefont
  {Castro~Neto}, \citenamefont {Guinea}, \citenamefont {Peres}, \citenamefont
  {Novoselov},\ and\ \citenamefont {Geim}}]{RMP2009Neto}%
  \BibitemOpen
  \bibfield  {author} {\bibinfo {author} {\bibfnamefont {A.~H.}\ \bibnamefont
  {Castro~Neto}}, \bibinfo {author} {\bibfnamefont {F.}~\bibnamefont {Guinea}},
  \bibinfo {author} {\bibfnamefont {N.~M.~R.}\ \bibnamefont {Peres}}, \bibinfo
  {author} {\bibfnamefont {K.~S.}\ \bibnamefont {Novoselov}}, \ and\ \bibinfo
  {author} {\bibfnamefont {A.~K.}\ \bibnamefont {Geim}},\ }\href@noop {}
  {\bibfield  {journal} {\bibinfo  {journal} {Rev. Mod. Phys.}\ }\textbf
  {\bibinfo {volume} {81}},\ \bibinfo {pages} {109} (\bibinfo {year}
  {2009})}\BibitemShut {NoStop}%
\bibitem [{\citenamefont {Das~Sarma}\ \emph {et~al.}(2011)\citenamefont
  {Das~Sarma}, \citenamefont {Adam}, \citenamefont {Hwang},\ and\ \citenamefont
  {Rossi}}]{RMP2011Sarma}%
  \BibitemOpen
  \bibfield  {author} {\bibinfo {author} {\bibfnamefont {S.}~\bibnamefont
  {Das~Sarma}}, \bibinfo {author} {\bibfnamefont {S.}~\bibnamefont {Adam}},
  \bibinfo {author} {\bibfnamefont {E.~H.}\ \bibnamefont {Hwang}}, \ and\
  \bibinfo {author} {\bibfnamefont {E.}~\bibnamefont {Rossi}},\ }\href@noop {}
  {\bibfield  {journal} {\bibinfo  {journal} {Rev. Mod. Phys.}\ }\textbf
  {\bibinfo {volume} {83}},\ \bibinfo {pages} {407} (\bibinfo {year}
  {2011})}\BibitemShut {NoStop}%
\bibitem [{\citenamefont {Wang}\ \emph {et~al.}(2012)\citenamefont {Wang},
  \citenamefont {Kalantar-Zadeh}, \citenamefont {Kis}, \citenamefont
  {Coleman},\ and\ \citenamefont {Strano}}]{NatNano2012Wang}%
  \BibitemOpen
  \bibfield  {author} {\bibinfo {author} {\bibfnamefont {Q.~H.}\ \bibnamefont
  {Wang}}, \bibinfo {author} {\bibfnamefont {K.}~\bibnamefont
  {Kalantar-Zadeh}}, \bibinfo {author} {\bibfnamefont {A.}~\bibnamefont {Kis}},
  \bibinfo {author} {\bibfnamefont {J.~N.}\ \bibnamefont {Coleman}}, \ and\
  \bibinfo {author} {\bibfnamefont {M.~S.}\ \bibnamefont {Strano}},\
  }\href@noop {} {\bibfield  {journal} {\bibinfo  {journal} {Nat. Nano.}\
  }\textbf {\bibinfo {volume} {7}},\ \bibinfo {pages} {699} (\bibinfo {year}
  {2012})}\BibitemShut {NoStop}%
\bibitem [{\citenamefont {Kotov}\ \emph {et~al.}(2012)\citenamefont {Kotov},
  \citenamefont {Uchoa}, \citenamefont {Pereira}, \citenamefont {Guinea},\ and\
  \citenamefont {Castro~Neto}}]{RMP2012Kotov}%
  \BibitemOpen
  \bibfield  {author} {\bibinfo {author} {\bibfnamefont {V.~N.}\ \bibnamefont
  {Kotov}}, \bibinfo {author} {\bibfnamefont {B.}~\bibnamefont {Uchoa}},
  \bibinfo {author} {\bibfnamefont {V.~M.}\ \bibnamefont {Pereira}}, \bibinfo
  {author} {\bibfnamefont {F.}~\bibnamefont {Guinea}}, \ and\ \bibinfo {author}
  {\bibfnamefont {A.~H.}\ \bibnamefont {Castro~Neto}},\ }\href@noop {}
  {\bibfield  {journal} {\bibinfo  {journal} {Rev. Mod. Phys.}\ }\textbf
  {\bibinfo {volume} {84}},\ \bibinfo {pages} {1067} (\bibinfo {year}
  {2012})}\BibitemShut {NoStop}%
\bibitem [{\citenamefont {Geim}\ and\ \citenamefont
  {Grigorieva}(2013)}]{Nature2013Geim}%
  \BibitemOpen
  \bibfield  {author} {\bibinfo {author} {\bibfnamefont {A.~K.}\ \bibnamefont
  {Geim}}\ and\ \bibinfo {author} {\bibfnamefont {I.~V.}\ \bibnamefont
  {Grigorieva}},\ }\href@noop {} {\bibfield  {journal} {\bibinfo  {journal}
  {Nature}\ }\textbf {\bibinfo {volume} {499}},\ \bibinfo {pages} {419}
  (\bibinfo {year} {2013})}\BibitemShut {NoStop}%
\bibitem [{\citenamefont {Komsa}\ and\ \citenamefont
  {Krasheninnikov}(2012)}]{PRB2012Komsa}%
  \BibitemOpen
  \bibfield  {author} {\bibinfo {author} {\bibfnamefont {H.-P.}\ \bibnamefont
  {Komsa}}\ and\ \bibinfo {author} {\bibfnamefont {A.~V.}\ \bibnamefont
  {Krasheninnikov}},\ }\href@noop {} {\bibfield  {journal} {\bibinfo  {journal}
  {Phys. Rev. B}\ }\textbf {\bibinfo {volume} {86}},\ \bibinfo {pages} {241201}
  (\bibinfo {year} {2012})}\BibitemShut {NoStop}%
\bibitem [{\citenamefont {Ramasubramaniam}(2012)}]{PRB2012Ashwin}%
  \BibitemOpen
  \bibfield  {author} {\bibinfo {author} {\bibfnamefont {A.}~\bibnamefont
  {Ramasubramaniam}},\ }\href@noop {} {\bibfield  {journal} {\bibinfo
  {journal} {Phys. Rev. B}\ }\textbf {\bibinfo {volume} {86}},\ \bibinfo
  {pages} {115409} (\bibinfo {year} {2012})}\BibitemShut {NoStop}%
\bibitem [{\citenamefont {Cheiwchanchamnangij}\ and\ \citenamefont
  {Lambrecht}(2012)}]{PRB2012Cheiwchanchamnagij}%
  \BibitemOpen
  \bibfield  {author} {\bibinfo {author} {\bibfnamefont {T.}~\bibnamefont
  {Cheiwchanchamnangij}}\ and\ \bibinfo {author} {\bibfnamefont {W.~R.~L.}\
  \bibnamefont {Lambrecht}},\ }\href@noop {} {\bibfield  {journal} {\bibinfo
  {journal} {Phys. Rev. B}\ }\textbf {\bibinfo {volume} {85}},\ \bibinfo
  {pages} {205302} (\bibinfo {year} {2012})}\BibitemShut {NoStop}%
\bibitem [{\citenamefont {Qiu}\ \emph {et~al.}(2013)\citenamefont {Qiu},
  \citenamefont {da~Jornada},\ and\ \citenamefont {Louie}}]{PRL2013Qiu}%
  \BibitemOpen
  \bibfield  {author} {\bibinfo {author} {\bibfnamefont {D.~Y.}\ \bibnamefont
  {Qiu}}, \bibinfo {author} {\bibfnamefont {F.~H.}\ \bibnamefont {da~Jornada}},
  \ and\ \bibinfo {author} {\bibfnamefont {S.~G.}\ \bibnamefont {Louie}},\
  }\href@noop {} {\bibfield  {journal} {\bibinfo  {journal} {Phys. Rev. Lett.}\
  }\textbf {\bibinfo {volume} {111}},\ \bibinfo {pages} {216805} (\bibinfo
  {year} {2013})}\BibitemShut {NoStop}%
\bibitem [{\citenamefont {Ugeda}\ \emph {et~al.}(2014)\citenamefont {Ugeda},
  \citenamefont {Bradley}, \citenamefont {Shi}, \citenamefont {da~Jornada},
  \citenamefont {Zhang}, \citenamefont {Qiu}, \citenamefont {Ruan},
  \citenamefont {Mo}, \citenamefont {Hussain}, \citenamefont {Shen},
  \citenamefont {Wang}, \citenamefont {Louie},\ and\ \citenamefont
  {Crommie}}]{NatMater2014Ugeda}%
  \BibitemOpen
  \bibfield  {author} {\bibinfo {author} {\bibfnamefont {M.~M.}\ \bibnamefont
  {Ugeda}}, \bibinfo {author} {\bibfnamefont {A.~J.}\ \bibnamefont {Bradley}},
  \bibinfo {author} {\bibfnamefont {S.-F.}\ \bibnamefont {Shi}}, \bibinfo
  {author} {\bibfnamefont {F.~H.}\ \bibnamefont {da~Jornada}}, \bibinfo
  {author} {\bibfnamefont {Y.}~\bibnamefont {Zhang}}, \bibinfo {author}
  {\bibfnamefont {D.~Y.}\ \bibnamefont {Qiu}}, \bibinfo {author} {\bibfnamefont
  {W.}~\bibnamefont {Ruan}}, \bibinfo {author} {\bibfnamefont {S.-K.}\
  \bibnamefont {Mo}}, \bibinfo {author} {\bibfnamefont {Z.}~\bibnamefont
  {Hussain}}, \bibinfo {author} {\bibfnamefont {Z.-X.}\ \bibnamefont {Shen}},
  \bibinfo {author} {\bibfnamefont {F.}~\bibnamefont {Wang}}, \bibinfo {author}
  {\bibfnamefont {S.~G.}\ \bibnamefont {Louie}}, \ and\ \bibinfo {author}
  {\bibfnamefont {M.~F.}\ \bibnamefont {Crommie}},\ }\href@noop {} {\bibfield
  {journal} {\bibinfo  {journal} {Nat. Mater.}\ }\textbf {\bibinfo {volume}
  {13}},\ \bibinfo {pages} {1091} (\bibinfo {year} {2014})}\BibitemShut
  {NoStop}%
\bibitem [{\citenamefont {Trevisanutto}\ \emph {et~al.}(2008)\citenamefont
  {Trevisanutto}, \citenamefont {Giorgetti}, \citenamefont {Reining},
  \citenamefont {Ladisa},\ and\ \citenamefont {Olevano}}]{PRL2008Trevisanutto}%
  \BibitemOpen
  \bibfield  {author} {\bibinfo {author} {\bibfnamefont {P.~E.}\ \bibnamefont
  {Trevisanutto}}, \bibinfo {author} {\bibfnamefont {C.}~\bibnamefont
  {Giorgetti}}, \bibinfo {author} {\bibfnamefont {L.}~\bibnamefont {Reining}},
  \bibinfo {author} {\bibfnamefont {M.}~\bibnamefont {Ladisa}}, \ and\ \bibinfo
  {author} {\bibfnamefont {V.}~\bibnamefont {Olevano}},\ }\href@noop {}
  {\bibfield  {journal} {\bibinfo  {journal} {Phys. Rev. Lett.}\ }\textbf
  {\bibinfo {volume} {101}},\ \bibinfo {pages} {226405} (\bibinfo {year}
  {2008})}\BibitemShut {NoStop}%
\bibitem [{\citenamefont {Park}\ \emph {et~al.}(2009)\citenamefont {Park},
  \citenamefont {Giustino}, \citenamefont {Spataru}, \citenamefont {Cohen},\
  and\ \citenamefont {Louie}}]{PRL2009Park}%
  \BibitemOpen
  \bibfield  {author} {\bibinfo {author} {\bibfnamefont {C.-H.}\ \bibnamefont
  {Park}}, \bibinfo {author} {\bibfnamefont {F.}~\bibnamefont {Giustino}},
  \bibinfo {author} {\bibfnamefont {C.~D.}\ \bibnamefont {Spataru}}, \bibinfo
  {author} {\bibfnamefont {M.~L.}\ \bibnamefont {Cohen}}, \ and\ \bibinfo
  {author} {\bibfnamefont {S.~G.}\ \bibnamefont {Louie}},\ }\href@noop {}
  {\bibfield  {journal} {\bibinfo  {journal} {Phys. Rev. Lett.}\ }\textbf
  {\bibinfo {volume} {102}},\ \bibinfo {pages} {076803} (\bibinfo {year}
  {2009})}\BibitemShut {NoStop}%
\bibitem [{\citenamefont {Bostwick}\ \emph {et~al.}(2010)\citenamefont
  {Bostwick}, \citenamefont {Speck}, \citenamefont {Seyller}, \citenamefont
  {Horn}, \citenamefont {Polini}, \citenamefont {Asgari}, \citenamefont
  {MacDonald},\ and\ \citenamefont {Rotenberg}}]{Science2010Bostwick}%
  \BibitemOpen
  \bibfield  {author} {\bibinfo {author} {\bibfnamefont {A.}~\bibnamefont
  {Bostwick}}, \bibinfo {author} {\bibfnamefont {F.}~\bibnamefont {Speck}},
  \bibinfo {author} {\bibfnamefont {T.}~\bibnamefont {Seyller}}, \bibinfo
  {author} {\bibfnamefont {K.}~\bibnamefont {Horn}}, \bibinfo {author}
  {\bibfnamefont {M.}~\bibnamefont {Polini}}, \bibinfo {author} {\bibfnamefont
  {R.}~\bibnamefont {Asgari}}, \bibinfo {author} {\bibfnamefont {A.~H.}\
  \bibnamefont {MacDonald}}, \ and\ \bibinfo {author} {\bibfnamefont
  {E.}~\bibnamefont {Rotenberg}},\ }\href@noop {} {\bibfield  {journal}
  {\bibinfo  {journal} {Science}\ }\textbf {\bibinfo {volume} {328}},\ \bibinfo
  {pages} {999} (\bibinfo {year} {2010})}\BibitemShut {NoStop}%
\bibitem [{\citenamefont {Elias}\ \emph {et~al.}(2011)\citenamefont {Elias},
  \citenamefont {Gorbachev}, \citenamefont {Mayorov}, \citenamefont {Morozov},
  \citenamefont {Zhukov}, \citenamefont {Blake}, \citenamefont {Ponomarenko},
  \citenamefont {Grigorieva}, \citenamefont {Novoselov}, \citenamefont
  {Guinea},\ and\ \citenamefont {Geim}}]{NatPhys2011Elias}%
  \BibitemOpen
  \bibfield  {author} {\bibinfo {author} {\bibfnamefont {D.~C.}\ \bibnamefont
  {Elias}}, \bibinfo {author} {\bibfnamefont {R.~V.}\ \bibnamefont
  {Gorbachev}}, \bibinfo {author} {\bibfnamefont {A.~S.}\ \bibnamefont
  {Mayorov}}, \bibinfo {author} {\bibfnamefont {S.~V.}\ \bibnamefont
  {Morozov}}, \bibinfo {author} {\bibfnamefont {A.~A.}\ \bibnamefont {Zhukov}},
  \bibinfo {author} {\bibfnamefont {P.}~\bibnamefont {Blake}}, \bibinfo
  {author} {\bibfnamefont {L.~A.}\ \bibnamefont {Ponomarenko}}, \bibinfo
  {author} {\bibfnamefont {I.~V.}\ \bibnamefont {Grigorieva}}, \bibinfo
  {author} {\bibfnamefont {K.~S.}\ \bibnamefont {Novoselov}}, \bibinfo {author}
  {\bibfnamefont {F.}~\bibnamefont {Guinea}}, \ and\ \bibinfo {author}
  {\bibfnamefont {A.~K.}\ \bibnamefont {Geim}},\ }\href@noop {} {\bibfield
  {journal} {\bibinfo  {journal} {Nat. Phys.}\ }\textbf {\bibinfo {volume}
  {7}},\ \bibinfo {pages} {701} (\bibinfo {year} {2011})}\BibitemShut {NoStop}%
\bibitem [{\citenamefont {Siegel}\ \emph {et~al.}(2013)\citenamefont {Siegel},
  \citenamefont {Regan}, \citenamefont {Fedorov}, \citenamefont {Zettl},\ and\
  \citenamefont {Lanzara}}]{PRL2013Siegel}%
  \BibitemOpen
  \bibfield  {author} {\bibinfo {author} {\bibfnamefont {D.~A.}\ \bibnamefont
  {Siegel}}, \bibinfo {author} {\bibfnamefont {W.}~\bibnamefont {Regan}},
  \bibinfo {author} {\bibfnamefont {A.~V.}\ \bibnamefont {Fedorov}}, \bibinfo
  {author} {\bibfnamefont {A.}~\bibnamefont {Zettl}}, \ and\ \bibinfo {author}
  {\bibfnamefont {A.}~\bibnamefont {Lanzara}},\ }\href@noop {} {\bibfield
  {journal} {\bibinfo  {journal} {Phys. Rev. Lett.}\ }\textbf {\bibinfo
  {volume} {110}},\ \bibinfo {pages} {146802} (\bibinfo {year}
  {2013})}\BibitemShut {NoStop}%
\bibitem [{\citenamefont {Lischner}\ \emph {et~al.}(2013)\citenamefont
  {Lischner}, \citenamefont {Vigil-Fowler},\ and\ \citenamefont
  {Louie}}]{PRL2013Lischner}%
  \BibitemOpen
  \bibfield  {author} {\bibinfo {author} {\bibfnamefont {J.}~\bibnamefont
  {Lischner}}, \bibinfo {author} {\bibfnamefont {D.}~\bibnamefont
  {Vigil-Fowler}}, \ and\ \bibinfo {author} {\bibfnamefont {S.~G.}\
  \bibnamefont {Louie}},\ }\href@noop {} {\bibfield  {journal} {\bibinfo
  {journal} {Phys. Rev. Lett.}\ }\textbf {\bibinfo {volume} {110}},\ \bibinfo
  {pages} {146801} (\bibinfo {year} {2013})}\BibitemShut {NoStop}%
\bibitem [{\citenamefont {Hwang}\ \emph {et~al.}(2012)\citenamefont {Hwang},
  \citenamefont {Siegel}, \citenamefont {Mo}, \citenamefont {Regan},
  \citenamefont {Ismach}, \citenamefont {Zhang}, \citenamefont {Zettl},\ and\
  \citenamefont {Lanzara}}]{SciRep2012Hwang}%
  \BibitemOpen
  \bibfield  {author} {\bibinfo {author} {\bibfnamefont {C.}~\bibnamefont
  {Hwang}}, \bibinfo {author} {\bibfnamefont {D.~A.}\ \bibnamefont {Siegel}},
  \bibinfo {author} {\bibfnamefont {S.-K.}\ \bibnamefont {Mo}}, \bibinfo
  {author} {\bibfnamefont {W.}~\bibnamefont {Regan}}, \bibinfo {author}
  {\bibfnamefont {A.}~\bibnamefont {Ismach}}, \bibinfo {author} {\bibfnamefont
  {Y.}~\bibnamefont {Zhang}}, \bibinfo {author} {\bibfnamefont
  {A.}~\bibnamefont {Zettl}}, \ and\ \bibinfo {author} {\bibfnamefont
  {A.}~\bibnamefont {Lanzara}},\ }\href@noop {} {\bibfield  {journal} {\bibinfo
   {journal} {Sci. Rep.}\ }\textbf {\bibinfo {volume} {2}},\ \bibinfo {pages}
  {590} (\bibinfo {year} {2012})}\BibitemShut {NoStop}%
\bibitem [{\citenamefont {Ye}\ \emph {et~al.}(2012)\citenamefont {Ye},
  \citenamefont {Zhang}, \citenamefont {Akashi}, \citenamefont {Bahramy},
  \citenamefont {Arita},\ and\ \citenamefont {Iwasa}}]{Science2012Ye}%
  \BibitemOpen
  \bibfield  {author} {\bibinfo {author} {\bibfnamefont {J.~T.}\ \bibnamefont
  {Ye}}, \bibinfo {author} {\bibfnamefont {Y.~J.}\ \bibnamefont {Zhang}},
  \bibinfo {author} {\bibfnamefont {R.}~\bibnamefont {Akashi}}, \bibinfo
  {author} {\bibfnamefont {M.~S.}\ \bibnamefont {Bahramy}}, \bibinfo {author}
  {\bibfnamefont {R.}~\bibnamefont {Arita}}, \ and\ \bibinfo {author}
  {\bibfnamefont {Y.}~\bibnamefont {Iwasa}},\ }\href@noop {} {\bibfield
  {journal} {\bibinfo  {journal} {Science}\ }\textbf {\bibinfo {volume}
  {338}},\ \bibinfo {pages} {1193} (\bibinfo {year} {2012})}\BibitemShut
  {NoStop}%
\bibitem [{\citenamefont {Yu}\ \emph {et~al.}(2015)\citenamefont {Yu},
  \citenamefont {Yang}, \citenamefont {Lu}, \citenamefont {Yan}, \citenamefont
  {Cho}, \citenamefont {Ma}, \citenamefont {Niu}, \citenamefont {Kim},
  \citenamefont {Son}, \citenamefont {Feng}, \citenamefont {Li}, \citenamefont
  {Cheong}, \citenamefont {Chen},\ and\ \citenamefont {Zhang}}]{NatNano2015Yu}%
  \BibitemOpen
  \bibfield  {author} {\bibinfo {author} {\bibfnamefont {Y.}~\bibnamefont
  {Yu}}, \bibinfo {author} {\bibfnamefont {F.}~\bibnamefont {Yang}}, \bibinfo
  {author} {\bibfnamefont {X.~F.}\ \bibnamefont {Lu}}, \bibinfo {author}
  {\bibfnamefont {Y.~J.}\ \bibnamefont {Yan}}, \bibinfo {author} {\bibfnamefont
  {Y.-H.}\ \bibnamefont {Cho}}, \bibinfo {author} {\bibfnamefont
  {L.}~\bibnamefont {Ma}}, \bibinfo {author} {\bibfnamefont {X.}~\bibnamefont
  {Niu}}, \bibinfo {author} {\bibfnamefont {S.}~\bibnamefont {Kim}}, \bibinfo
  {author} {\bibfnamefont {Y.-W.}\ \bibnamefont {Son}}, \bibinfo {author}
  {\bibfnamefont {D.}~\bibnamefont {Feng}}, \bibinfo {author} {\bibfnamefont
  {S.}~\bibnamefont {Li}}, \bibinfo {author} {\bibfnamefont {S.-W.}\
  \bibnamefont {Cheong}}, \bibinfo {author} {\bibfnamefont {X.~H.}\
  \bibnamefont {Chen}}, \ and\ \bibinfo {author} {\bibfnamefont
  {Y.}~\bibnamefont {Zhang}},\ }\href@noop {} {\bibfield  {journal} {\bibinfo
  {journal} {Nat. Nano.}\ }\textbf {\bibinfo {volume} {10}},\ \bibinfo {pages}
  {270} (\bibinfo {year} {2015})}\BibitemShut {NoStop}%
\bibitem [{\citenamefont {Sugawara}\ \emph {et~al.}(2015)\citenamefont
  {Sugawara}, \citenamefont {Nakata}, \citenamefont {Shimizu}, \citenamefont
  {Han}, \citenamefont {Hitosugi}, \citenamefont {Sato},\ and\ \citenamefont
  {Takahashi}}]{ACSNano2015Sugawara}%
  \BibitemOpen
  \bibfield  {author} {\bibinfo {author} {\bibfnamefont {K.}~\bibnamefont
  {Sugawara}}, \bibinfo {author} {\bibfnamefont {Y.}~\bibnamefont {Nakata}},
  \bibinfo {author} {\bibfnamefont {R.}~\bibnamefont {Shimizu}}, \bibinfo
  {author} {\bibfnamefont {P.}~\bibnamefont {Han}}, \bibinfo {author}
  {\bibfnamefont {T.}~\bibnamefont {Hitosugi}}, \bibinfo {author}
  {\bibfnamefont {T.}~\bibnamefont {Sato}}, \ and\ \bibinfo {author}
  {\bibfnamefont {T.}~\bibnamefont {Takahashi}},\ }\href@noop {} {\bibfield
  {journal} {\bibinfo  {journal} {ACS Nano}\ }\textbf {\bibinfo {volume}
  {10}},\ \bibinfo {pages} {1341} (\bibinfo {year} {2015})}\BibitemShut
  {NoStop}%
\bibitem [{\citenamefont {Chen}\ \emph {et~al.}(2015)\citenamefont {Chen},
  \citenamefont {Chan}, \citenamefont {Fang}, \citenamefont {Zhang},
  \citenamefont {Chou}, \citenamefont {Mo}, \citenamefont {Hussain},
  \citenamefont {Fedorov},\ and\ \citenamefont {Chiang}}]{NatComm2015Chen}%
  \BibitemOpen
  \bibfield  {author} {\bibinfo {author} {\bibfnamefont {P.}~\bibnamefont
  {Chen}}, \bibinfo {author} {\bibfnamefont {Y.-H.}\ \bibnamefont {Chan}},
  \bibinfo {author} {\bibfnamefont {X.-Y.}\ \bibnamefont {Fang}}, \bibinfo
  {author} {\bibfnamefont {Y.}~\bibnamefont {Zhang}}, \bibinfo {author}
  {\bibfnamefont {M.~Y.}\ \bibnamefont {Chou}}, \bibinfo {author}
  {\bibfnamefont {S.-K.}\ \bibnamefont {Mo}}, \bibinfo {author} {\bibfnamefont
  {Z.}~\bibnamefont {Hussain}}, \bibinfo {author} {\bibfnamefont {A.-V.}\
  \bibnamefont {Fedorov}}, \ and\ \bibinfo {author} {\bibfnamefont {T.-C.}\
  \bibnamefont {Chiang}},\ }\href@noop {} {\bibfield  {journal} {\bibinfo
  {journal} {Nat. Comm.}\ }\textbf {\bibinfo {volume} {6}},\ \bibinfo {pages}
  {8943} (\bibinfo {year} {2015})}\BibitemShut {NoStop}%
\bibitem [{\citenamefont {Peng}\ \emph {et~al.}(2015)\citenamefont {Peng},
  \citenamefont {Guan}, \citenamefont {Zhang}, \citenamefont {Song},
  \citenamefont {Wang}, \citenamefont {He}, \citenamefont {Xue},\ and\
  \citenamefont {Ma}}]{PRB2015Peng}%
  \BibitemOpen
  \bibfield  {author} {\bibinfo {author} {\bibfnamefont {J.-P.}\ \bibnamefont
  {Peng}}, \bibinfo {author} {\bibfnamefont {J.-Q.}\ \bibnamefont {Guan}},
  \bibinfo {author} {\bibfnamefont {H.-M.}\ \bibnamefont {Zhang}}, \bibinfo
  {author} {\bibfnamefont {C.-L.}\ \bibnamefont {Song}}, \bibinfo {author}
  {\bibfnamefont {L.}~\bibnamefont {Wang}}, \bibinfo {author} {\bibfnamefont
  {K.}~\bibnamefont {He}}, \bibinfo {author} {\bibfnamefont {Q.-K.}\
  \bibnamefont {Xue}}, \ and\ \bibinfo {author} {\bibfnamefont {X.-C.}\
  \bibnamefont {Ma}},\ }\href@noop {} {\bibfield  {journal} {\bibinfo
  {journal} {Phys. Rev. B}\ }\textbf {\bibinfo {volume} {91}},\ \bibinfo
  {pages} {121113} (\bibinfo {year} {2015})}\BibitemShut {NoStop}%
\bibitem [{\citenamefont {Saito}\ \emph {et~al.}(2015)\citenamefont {Saito},
  \citenamefont {Kasahara}, \citenamefont {Ye}, \citenamefont {Iwasa},\ and\
  \citenamefont {Nojima}}]{Science2015Saito}%
  \BibitemOpen
  \bibfield  {author} {\bibinfo {author} {\bibfnamefont {Y.}~\bibnamefont
  {Saito}}, \bibinfo {author} {\bibfnamefont {Y.}~\bibnamefont {Kasahara}},
  \bibinfo {author} {\bibfnamefont {J.~T.}\ \bibnamefont {Ye}}, \bibinfo
  {author} {\bibfnamefont {Y.}~\bibnamefont {Iwasa}}, \ and\ \bibinfo {author}
  {\bibfnamefont {T.}~\bibnamefont {Nojima}},\ }\href@noop {} {\bibfield
  {journal} {\bibinfo  {journal} {Science}\ }\textbf {\bibinfo {volume}
  {350}},\ \bibinfo {pages} {409} (\bibinfo {year} {2015})}\BibitemShut
  {NoStop}%
\bibitem [{\citenamefont {Saito}\ \emph {et~al.}(2016)\citenamefont {Saito},
  \citenamefont {Nakamura}, \citenamefont {Bahramy}, \citenamefont {Kohama},
  \citenamefont {Ye}, \citenamefont {Kasahara}, \citenamefont {Nakagawa},
  \citenamefont {Onga}, \citenamefont {Tokunaga}, \citenamefont {Nojima},\ and\
  \citenamefont {adn Y.~Iwasa}}]{NatPhys2016Saito}%
  \BibitemOpen
  \bibfield  {author} {\bibinfo {author} {\bibfnamefont {Y.}~\bibnamefont
  {Saito}}, \bibinfo {author} {\bibfnamefont {Y.}~\bibnamefont {Nakamura}},
  \bibinfo {author} {\bibfnamefont {M.~S.}\ \bibnamefont {Bahramy}}, \bibinfo
  {author} {\bibfnamefont {Y.}~\bibnamefont {Kohama}}, \bibinfo {author}
  {\bibfnamefont {J.~T.}\ \bibnamefont {Ye}}, \bibinfo {author} {\bibfnamefont
  {Y.}~\bibnamefont {Kasahara}}, \bibinfo {author} {\bibfnamefont
  {Y.}~\bibnamefont {Nakagawa}}, \bibinfo {author} {\bibfnamefont
  {M.}~\bibnamefont {Onga}}, \bibinfo {author} {\bibfnamefont {M.}~\bibnamefont
  {Tokunaga}}, \bibinfo {author} {\bibfnamefont {T.}~\bibnamefont {Nojima}}, \
  and\ \bibinfo {author} {\bibfnamefont {Y.~Y.}\ \bibnamefont {adn Y.~Iwasa}},\
  }\href@noop {} {\bibfield  {journal} {\bibinfo  {journal} {Nat. Phys.}\
  }\textbf {\bibinfo {volume} {12}},\ \bibinfo {pages} {144} (\bibinfo {year}
  {2016})}\BibitemShut {NoStop}%
\bibitem [{\citenamefont {Li}\ \emph {et~al.}(2016)\citenamefont {Li},
  \citenamefont {O'Farrell}, \citenamefont {Loh}, \citenamefont {Eda},
  \citenamefont {\"Ozyilmaz},\ and\ \citenamefont
  {Castro-Neto}}]{Nature2016Li}%
  \BibitemOpen
  \bibfield  {author} {\bibinfo {author} {\bibfnamefont {L.~J.}\ \bibnamefont
  {Li}}, \bibinfo {author} {\bibfnamefont {E.~C.~T.}\ \bibnamefont
  {O'Farrell}}, \bibinfo {author} {\bibfnamefont {K.~P.}\ \bibnamefont {Loh}},
  \bibinfo {author} {\bibfnamefont {G.}~\bibnamefont {Eda}}, \bibinfo {author}
  {\bibfnamefont {B.}~\bibnamefont {\"Ozyilmaz}}, \ and\ \bibinfo {author}
  {\bibfnamefont {A.~H.}\ \bibnamefont {Castro-Neto}},\ }\href@noop {}
  {\bibfield  {journal} {\bibinfo  {journal} {Nature}\ }\textbf {\bibinfo
  {volume} {529}},\ \bibinfo {pages} {185} (\bibinfo {year}
  {2016})}\BibitemShut {NoStop}%
\bibitem [{\citenamefont {Xi}\ \emph {et~al.}(2016)\citenamefont {Xi},
  \citenamefont {Wang}, \citenamefont {Zhao}, \citenamefont {Park},
  \citenamefont {Law}, \citenamefont {Berger}, \citenamefont {Forr\'o},
  \citenamefont {Shen},\ and\ \citenamefont {Mak}}]{NatPhys2016Xi}%
  \BibitemOpen
  \bibfield  {author} {\bibinfo {author} {\bibfnamefont {X.}~\bibnamefont
  {Xi}}, \bibinfo {author} {\bibfnamefont {Z.}~\bibnamefont {Wang}}, \bibinfo
  {author} {\bibfnamefont {W.}~\bibnamefont {Zhao}}, \bibinfo {author}
  {\bibfnamefont {J.-H.}\ \bibnamefont {Park}}, \bibinfo {author}
  {\bibfnamefont {K.~T.}\ \bibnamefont {Law}}, \bibinfo {author} {\bibfnamefont
  {H.}~\bibnamefont {Berger}}, \bibinfo {author} {\bibfnamefont
  {L.}~\bibnamefont {Forr\'o}}, \bibinfo {author} {\bibfnamefont
  {J.}~\bibnamefont {Shen}}, \ and\ \bibinfo {author} {\bibfnamefont {K.~F.}\
  \bibnamefont {Mak}},\ }\href@noop {} {\bibfield  {journal} {\bibinfo
  {journal} {Nat. Phys.}\ }\textbf {\bibinfo {volume} {12}},\ \bibinfo {pages}
  {139} (\bibinfo {year} {2016})}\BibitemShut {NoStop}%
\bibitem [{\citenamefont {Tsen}\ \emph {et~al.}(2016)\citenamefont {Tsen},
  \citenamefont {Hunt}, \citenamefont {Kim}, \citenamefont {Yuan},
  \citenamefont {Jia}, \citenamefont {Cava}, \citenamefont {Hone},
  \citenamefont {Kim}, \citenamefont {Dean},\ and\ \citenamefont
  {Pasupathy}}]{NatPhys2016Tsen}%
  \BibitemOpen
  \bibfield  {author} {\bibinfo {author} {\bibfnamefont {A.~W.}\ \bibnamefont
  {Tsen}}, \bibinfo {author} {\bibfnamefont {B.}~\bibnamefont {Hunt}}, \bibinfo
  {author} {\bibfnamefont {Y.~D.}\ \bibnamefont {Kim}}, \bibinfo {author}
  {\bibfnamefont {Z.~J.}\ \bibnamefont {Yuan}}, \bibinfo {author}
  {\bibfnamefont {S.}~\bibnamefont {Jia}}, \bibinfo {author} {\bibfnamefont
  {R.~J.}\ \bibnamefont {Cava}}, \bibinfo {author} {\bibfnamefont
  {J.}~\bibnamefont {Hone}}, \bibinfo {author} {\bibfnamefont {P.}~\bibnamefont
  {Kim}}, \bibinfo {author} {\bibfnamefont {C.~R.}\ \bibnamefont {Dean}}, \
  and\ \bibinfo {author} {\bibfnamefont {A.~N.}\ \bibnamefont {Pasupathy}},\
  }\href@noop {} {\bibfield  {journal} {\bibinfo  {journal} {Nat. Phys.}\
  }\textbf {\bibinfo {volume} {12}},\ \bibinfo {pages} {208} (\bibinfo {year}
  {2016})}\BibitemShut {NoStop}%
\bibitem [{\citenamefont {Cao}\ \emph {et~al.}(2015)\citenamefont {Cao},
  \citenamefont {Mishchenko}, \citenamefont {Yu}, \citenamefont {Khestanova},
  \citenamefont {Rooney}, \citenamefont {Prestat}, \citenamefont {Kretinin},
  \citenamefont {Blake}, \citenamefont {Shalom}, \citenamefont {Woods},
  \citenamefont {Chapman}, \citenamefont {Balakrishnan}, \citenamefont
  {Grigorieva}, \citenamefont {Novoselov}, \citenamefont {Piot}, \citenamefont
  {Potemski}, \citenamefont {Watanabe}, \citenamefont {Taniguchi},
  \citenamefont {Haigh}, \citenamefont {Geim},\ and\ \citenamefont
  {Gorbachev}}]{NanoLett2015Cao}%
  \BibitemOpen
  \bibfield  {author} {\bibinfo {author} {\bibfnamefont {Y.}~\bibnamefont
  {Cao}}, \bibinfo {author} {\bibfnamefont {A.}~\bibnamefont {Mishchenko}},
  \bibinfo {author} {\bibfnamefont {G.~L.}\ \bibnamefont {Yu}}, \bibinfo
  {author} {\bibfnamefont {E.}~\bibnamefont {Khestanova}}, \bibinfo {author}
  {\bibfnamefont {A.~P.}\ \bibnamefont {Rooney}}, \bibinfo {author}
  {\bibfnamefont {E.}~\bibnamefont {Prestat}}, \bibinfo {author} {\bibfnamefont
  {A.~V.}\ \bibnamefont {Kretinin}}, \bibinfo {author} {\bibfnamefont
  {P.}~\bibnamefont {Blake}}, \bibinfo {author} {\bibfnamefont {M.~B.}\
  \bibnamefont {Shalom}}, \bibinfo {author} {\bibfnamefont {C.}~\bibnamefont
  {Woods}}, \bibinfo {author} {\bibfnamefont {J.}~\bibnamefont {Chapman}},
  \bibinfo {author} {\bibfnamefont {G.}~\bibnamefont {Balakrishnan}}, \bibinfo
  {author} {\bibfnamefont {I.~V.}\ \bibnamefont {Grigorieva}}, \bibinfo
  {author} {\bibfnamefont {K.~S.}\ \bibnamefont {Novoselov}}, \bibinfo {author}
  {\bibfnamefont {B.~A.}\ \bibnamefont {Piot}}, \bibinfo {author}
  {\bibfnamefont {M.}~\bibnamefont {Potemski}}, \bibinfo {author}
  {\bibfnamefont {K.}~\bibnamefont {Watanabe}}, \bibinfo {author}
  {\bibfnamefont {T.}~\bibnamefont {Taniguchi}}, \bibinfo {author}
  {\bibfnamefont {S.~J.}\ \bibnamefont {Haigh}}, \bibinfo {author}
  {\bibfnamefont {A.~K.}\ \bibnamefont {Geim}}, \ and\ \bibinfo {author}
  {\bibfnamefont {R.~V.}\ \bibnamefont {Gorbachev}},\ }\href@noop {} {\bibfield
   {journal} {\bibinfo  {journal} {Nano Lett.}\ }\textbf {\bibinfo {volume}
  {15}},\ \bibinfo {pages} {4914} (\bibinfo {year} {2015})}\BibitemShut
  {NoStop}%
\bibitem [{\citenamefont {Xi}\ \emph {et~al.}(2015)\citenamefont {Xi},
  \citenamefont {Zhao}, \citenamefont {Wang}, \citenamefont {Berger},
  \citenamefont {Forr{\'o}}, \citenamefont {Shan},\ and\ \citenamefont
  {Mak}}]{NatNano2015Xi}%
  \BibitemOpen
  \bibfield  {author} {\bibinfo {author} {\bibfnamefont {X.}~\bibnamefont
  {Xi}}, \bibinfo {author} {\bibfnamefont {L.}~\bibnamefont {Zhao}}, \bibinfo
  {author} {\bibfnamefont {Z.}~\bibnamefont {Wang}}, \bibinfo {author}
  {\bibfnamefont {H.}~\bibnamefont {Berger}}, \bibinfo {author} {\bibfnamefont
  {L.}~\bibnamefont {Forr{\'o}}}, \bibinfo {author} {\bibfnamefont
  {J.}~\bibnamefont {Shan}}, \ and\ \bibinfo {author} {\bibfnamefont {K.~F.}\
  \bibnamefont {Mak}},\ }\href@noop {} {\bibfield  {journal} {\bibinfo
  {journal} {Nat. Nano.}\ }\textbf {\bibinfo {volume} {10}},\ \bibinfo {pages}
  {765} (\bibinfo {year} {2015})}\BibitemShut {NoStop}%
\bibitem [{\citenamefont {Ugeda}\ \emph {et~al.}(2016)\citenamefont {Ugeda},
  \citenamefont {Bradley}, \citenamefont {Zhang}, \citenamefont {Onishi},
  \citenamefont {Chen}, \citenamefont {Ruan}, \citenamefont
  {Ojeda-Aristizabal}, \citenamefont {Ryu}, \citenamefont {Edmonds},
  \citenamefont {Tsai}, \citenamefont {Riss}, \citenamefont {Mo}, \citenamefont
  {Lee}, \citenamefont {Zettl}, \citenamefont {Hussain}, \citenamefont {Shen},\
  and\ \citenamefont {Crommie}}]{NatPhys2016Ugeda}%
  \BibitemOpen
  \bibfield  {author} {\bibinfo {author} {\bibfnamefont {M.~M.}\ \bibnamefont
  {Ugeda}}, \bibinfo {author} {\bibfnamefont {A.~J.}\ \bibnamefont {Bradley}},
  \bibinfo {author} {\bibfnamefont {Y.}~\bibnamefont {Zhang}}, \bibinfo
  {author} {\bibfnamefont {S.}~\bibnamefont {Onishi}}, \bibinfo {author}
  {\bibfnamefont {Y.}~\bibnamefont {Chen}}, \bibinfo {author} {\bibfnamefont
  {W.}~\bibnamefont {Ruan}}, \bibinfo {author} {\bibfnamefont {C.}~\bibnamefont
  {Ojeda-Aristizabal}}, \bibinfo {author} {\bibfnamefont {H.}~\bibnamefont
  {Ryu}}, \bibinfo {author} {\bibfnamefont {M.~T.}\ \bibnamefont {Edmonds}},
  \bibinfo {author} {\bibfnamefont {H.-Z.}\ \bibnamefont {Tsai}}, \bibinfo
  {author} {\bibfnamefont {A.}~\bibnamefont {Riss}}, \bibinfo {author}
  {\bibfnamefont {S.-K.}\ \bibnamefont {Mo}}, \bibinfo {author} {\bibfnamefont
  {D.}~\bibnamefont {Lee}}, \bibinfo {author} {\bibfnamefont {A.}~\bibnamefont
  {Zettl}}, \bibinfo {author} {\bibfnamefont {Z.}~\bibnamefont {Hussain}},
  \bibinfo {author} {\bibfnamefont {Z.-X.}\ \bibnamefont {Shen}}, \ and\
  \bibinfo {author} {\bibfnamefont {M.~F.}\ \bibnamefont {Crommie}},\
  }\href@noop {} {\bibfield  {journal} {\bibinfo  {journal} {Nat Phys}\
  }\textbf {\bibinfo {volume} {12}},\ \bibinfo {pages} {92} (\bibinfo {year}
  {2016})}\BibitemShut {NoStop}%
\bibitem [{\citenamefont {Wilson}\ \emph {et~al.}(1975)\citenamefont {Wilson},
  \citenamefont {Salvo},\ and\ \citenamefont {Mahajan}}]{AdvPhys1975Wilson}%
  \BibitemOpen
  \bibfield  {author} {\bibinfo {author} {\bibfnamefont {J.~A.}\ \bibnamefont
  {Wilson}}, \bibinfo {author} {\bibfnamefont {F.~J.~D.}\ \bibnamefont
  {Salvo}}, \ and\ \bibinfo {author} {\bibfnamefont {S.}~\bibnamefont
  {Mahajan}},\ }\href@noop {} {\bibfield  {journal} {\bibinfo  {journal} {Adv.
  Phys.}\ }\textbf {\bibinfo {volume} {24}},\ \bibinfo {pages} {117} (\bibinfo
  {year} {1975})}\BibitemShut {NoStop}%
\bibitem [{\citenamefont {Rossnagel}(2011)}]{JPCM2011Rossnagel}%
  \BibitemOpen
  \bibfield  {author} {\bibinfo {author} {\bibfnamefont {K.}~\bibnamefont
  {Rossnagel}},\ }\href@noop {} {\bibfield  {journal} {\bibinfo  {journal} {J.
  Phys.: Cond. Matter}\ }\textbf {\bibinfo {volume} {23}},\ \bibinfo {pages}
  {213001} (\bibinfo {year} {2011})}\BibitemShut {NoStop}%
\bibitem [{\citenamefont {Revolinsky}\ \emph {et~al.}(1963)\citenamefont
  {Revolinsky}, \citenamefont {Lautenschlager},\ and\ \citenamefont
  {Armitage}}]{SSC1963Revolinsky}%
  \BibitemOpen
  \bibfield  {author} {\bibinfo {author} {\bibfnamefont {E.}~\bibnamefont
  {Revolinsky}}, \bibinfo {author} {\bibfnamefont {E.~P.}\ \bibnamefont
  {Lautenschlager}}, \ and\ \bibinfo {author} {\bibfnamefont {C.~H.}\
  \bibnamefont {Armitage}},\ }\href@noop {} {\bibfield  {journal} {\bibinfo
  {journal} {Solid State Commun.}\ }\textbf {\bibinfo {volume} {1}},\ \bibinfo
  {pages} {59} (\bibinfo {year} {1963})}\BibitemShut {NoStop}%
\bibitem [{\citenamefont {Revolinsky}\ \emph {et~al.}(1965)\citenamefont
  {Revolinsky}, \citenamefont {Spiering},\ and\ \citenamefont
  {Beerntsen}}]{JPCS1965Revolinsky}%
  \BibitemOpen
  \bibfield  {author} {\bibinfo {author} {\bibfnamefont {E.}~\bibnamefont
  {Revolinsky}}, \bibinfo {author} {\bibfnamefont {G.~A.}\ \bibnamefont
  {Spiering}}, \ and\ \bibinfo {author} {\bibfnamefont {D.~J.}\ \bibnamefont
  {Beerntsen}},\ }\href@noop {} {\bibfield  {journal} {\bibinfo  {journal} {J.
  Phys. Chem. Solids}\ }\textbf {\bibinfo {volume} {26}},\ \bibinfo {pages}
  {1029} (\bibinfo {year} {1965})}\BibitemShut {NoStop}%
\bibitem [{\citenamefont {Wilson}\ \emph {et~al.}(1974)\citenamefont {Wilson},
  \citenamefont {Di~Salvo},\ and\ \citenamefont {Mahajan}}]{PRL1974Wilson}%
  \BibitemOpen
  \bibfield  {author} {\bibinfo {author} {\bibfnamefont {J.~A.}\ \bibnamefont
  {Wilson}}, \bibinfo {author} {\bibfnamefont {F.~J.}\ \bibnamefont
  {Di~Salvo}}, \ and\ \bibinfo {author} {\bibfnamefont {S.}~\bibnamefont
  {Mahajan}},\ }\href@noop {} {\bibfield  {journal} {\bibinfo  {journal} {Phys.
  Rev. Lett.}\ }\textbf {\bibinfo {volume} {32}},\ \bibinfo {pages} {882}
  (\bibinfo {year} {1974})}\BibitemShut {NoStop}%
\bibitem [{\citenamefont {Straub}\ \emph {et~al.}(1999)\citenamefont {Straub},
  \citenamefont {Finteis}, \citenamefont {Claessen}, \citenamefont {Steiner},
  \citenamefont {H\"ufner}, \citenamefont {Blaha}, \citenamefont {Oglesby},\
  and\ \citenamefont {Bucher}}]{PRL1999Straub}%
  \BibitemOpen
  \bibfield  {author} {\bibinfo {author} {\bibfnamefont {T.}~\bibnamefont
  {Straub}}, \bibinfo {author} {\bibfnamefont {T.}~\bibnamefont {Finteis}},
  \bibinfo {author} {\bibfnamefont {R.}~\bibnamefont {Claessen}}, \bibinfo
  {author} {\bibfnamefont {P.}~\bibnamefont {Steiner}}, \bibinfo {author}
  {\bibfnamefont {S.}~\bibnamefont {H\"ufner}}, \bibinfo {author}
  {\bibfnamefont {P.}~\bibnamefont {Blaha}}, \bibinfo {author} {\bibfnamefont
  {C.~S.}\ \bibnamefont {Oglesby}}, \ and\ \bibinfo {author} {\bibfnamefont
  {E.}~\bibnamefont {Bucher}},\ }\href@noop {} {\bibfield  {journal} {\bibinfo
  {journal} {Phys. Rev. Lett.}\ }\textbf {\bibinfo {volume} {82}},\ \bibinfo
  {pages} {4504} (\bibinfo {year} {1999})}\BibitemShut {NoStop}%
\bibitem [{\citenamefont {Shen}\ \emph {et~al.}(2008)\citenamefont {Shen},
  \citenamefont {Zhang}, \citenamefont {Yang}, \citenamefont {Wei},
  \citenamefont {Ou}, \citenamefont {Dong}, \citenamefont {Xie}, \citenamefont
  {He}, \citenamefont {Zhao}, \citenamefont {Zhou}, \citenamefont {Arita},
  \citenamefont {Shimada}, \citenamefont {Namatame}, \citenamefont {Taniguchi},
  \citenamefont {Shi},\ and\ \citenamefont {Feng}}]{PRL2008Shen}%
  \BibitemOpen
  \bibfield  {author} {\bibinfo {author} {\bibfnamefont {D.~W.}\ \bibnamefont
  {Shen}}, \bibinfo {author} {\bibfnamefont {Y.}~\bibnamefont {Zhang}},
  \bibinfo {author} {\bibfnamefont {L.~X.}\ \bibnamefont {Yang}}, \bibinfo
  {author} {\bibfnamefont {J.}~\bibnamefont {Wei}}, \bibinfo {author}
  {\bibfnamefont {H.~W.}\ \bibnamefont {Ou}}, \bibinfo {author} {\bibfnamefont
  {J.~K.}\ \bibnamefont {Dong}}, \bibinfo {author} {\bibfnamefont {B.~P.}\
  \bibnamefont {Xie}}, \bibinfo {author} {\bibfnamefont {C.}~\bibnamefont
  {He}}, \bibinfo {author} {\bibfnamefont {J.~F.}\ \bibnamefont {Zhao}},
  \bibinfo {author} {\bibfnamefont {B.}~\bibnamefont {Zhou}}, \bibinfo {author}
  {\bibfnamefont {M.}~\bibnamefont {Arita}}, \bibinfo {author} {\bibfnamefont
  {K.}~\bibnamefont {Shimada}}, \bibinfo {author} {\bibfnamefont
  {H.}~\bibnamefont {Namatame}}, \bibinfo {author} {\bibfnamefont
  {M.}~\bibnamefont {Taniguchi}}, \bibinfo {author} {\bibfnamefont
  {J.}~\bibnamefont {Shi}}, \ and\ \bibinfo {author} {\bibfnamefont {D.~L.}\
  \bibnamefont {Feng}},\ }\href@noop {} {\bibfield  {journal} {\bibinfo
  {journal} {Phys. Rev. Lett.}\ }\textbf {\bibinfo {volume} {101}},\ \bibinfo
  {pages} {226406} (\bibinfo {year} {2008})}\BibitemShut {NoStop}%
\bibitem [{\citenamefont {Borisenko}\ \emph {et~al.}(2009)\citenamefont
  {Borisenko}, \citenamefont {Kordyuk}, \citenamefont {Zabolotnyy},
  \citenamefont {Inosov}, \citenamefont {Evtushinsky}, \citenamefont
  {B\"uchner}, \citenamefont {Yaresko}, \citenamefont {Varykhalov},
  \citenamefont {Follath}, \citenamefont {Eberhardt}, \citenamefont {Patthey},\
  and\ \citenamefont {Berger}}]{PRL2009Borisenko}%
  \BibitemOpen
  \bibfield  {author} {\bibinfo {author} {\bibfnamefont {S.~V.}\ \bibnamefont
  {Borisenko}}, \bibinfo {author} {\bibfnamefont {A.~A.}\ \bibnamefont
  {Kordyuk}}, \bibinfo {author} {\bibfnamefont {V.~B.}\ \bibnamefont
  {Zabolotnyy}}, \bibinfo {author} {\bibfnamefont {D.~S.}\ \bibnamefont
  {Inosov}}, \bibinfo {author} {\bibfnamefont {D.}~\bibnamefont {Evtushinsky}},
  \bibinfo {author} {\bibfnamefont {B.}~\bibnamefont {B\"uchner}}, \bibinfo
  {author} {\bibfnamefont {A.~N.}\ \bibnamefont {Yaresko}}, \bibinfo {author}
  {\bibfnamefont {A.}~\bibnamefont {Varykhalov}}, \bibinfo {author}
  {\bibfnamefont {R.}~\bibnamefont {Follath}}, \bibinfo {author} {\bibfnamefont
  {W.}~\bibnamefont {Eberhardt}}, \bibinfo {author} {\bibfnamefont
  {L.}~\bibnamefont {Patthey}}, \ and\ \bibinfo {author} {\bibfnamefont
  {H.}~\bibnamefont {Berger}},\ }\href@noop {} {\bibfield  {journal} {\bibinfo
  {journal} {Phys. Rev. Lett.}\ }\textbf {\bibinfo {volume} {102}},\ \bibinfo
  {pages} {166402} (\bibinfo {year} {2009})}\BibitemShut {NoStop}%
\bibitem [{\citenamefont {Varma}\ and\ \citenamefont
  {Simons}(1983)}]{PRL1983Varma}%
  \BibitemOpen
  \bibfield  {author} {\bibinfo {author} {\bibfnamefont {C.~M.}\ \bibnamefont
  {Varma}}\ and\ \bibinfo {author} {\bibfnamefont {A.~L.}\ \bibnamefont
  {Simons}},\ }\href@noop {} {\bibfield  {journal} {\bibinfo  {journal} {Phys.
  Rev. Lett.}\ }\textbf {\bibinfo {volume} {51}},\ \bibinfo {pages} {138}
  (\bibinfo {year} {1983})}\BibitemShut {NoStop}%
\bibitem [{\citenamefont {Valla}\ \emph {et~al.}(2004)\citenamefont {Valla},
  \citenamefont {Fedorov}, \citenamefont {Johnson}, \citenamefont {Glans},
  \citenamefont {McGuinness}, \citenamefont {Smith}, \citenamefont {Andrei},\
  and\ \citenamefont {Berger}}]{PRL2004Valla}%
  \BibitemOpen
  \bibfield  {author} {\bibinfo {author} {\bibfnamefont {T.}~\bibnamefont
  {Valla}}, \bibinfo {author} {\bibfnamefont {A.~V.}\ \bibnamefont {Fedorov}},
  \bibinfo {author} {\bibfnamefont {P.~D.}\ \bibnamefont {Johnson}}, \bibinfo
  {author} {\bibfnamefont {P.-A.}\ \bibnamefont {Glans}}, \bibinfo {author}
  {\bibfnamefont {C.}~\bibnamefont {McGuinness}}, \bibinfo {author}
  {\bibfnamefont {K.~E.}\ \bibnamefont {Smith}}, \bibinfo {author}
  {\bibfnamefont {E.~Y.}\ \bibnamefont {Andrei}}, \ and\ \bibinfo {author}
  {\bibfnamefont {H.}~\bibnamefont {Berger}},\ }\href@noop {} {\bibfield
  {journal} {\bibinfo  {journal} {Phys. Rev. Lett.}\ }\textbf {\bibinfo
  {volume} {92}},\ \bibinfo {pages} {086401} (\bibinfo {year}
  {2004})}\BibitemShut {NoStop}%
\bibitem [{\citenamefont {Weber}\ \emph {et~al.}(2011)\citenamefont {Weber},
  \citenamefont {Rosenkranz}, \citenamefont {Castellan}, \citenamefont
  {Osborn}, \citenamefont {Hott}, \citenamefont {Heid}, \citenamefont {Bohnen},
  \citenamefont {Egami}, \citenamefont {Said},\ and\ \citenamefont
  {Reznik}}]{PRL2011Weber}%
  \BibitemOpen
  \bibfield  {author} {\bibinfo {author} {\bibfnamefont {F.}~\bibnamefont
  {Weber}}, \bibinfo {author} {\bibfnamefont {S.}~\bibnamefont {Rosenkranz}},
  \bibinfo {author} {\bibfnamefont {J.-P.}\ \bibnamefont {Castellan}}, \bibinfo
  {author} {\bibfnamefont {R.}~\bibnamefont {Osborn}}, \bibinfo {author}
  {\bibfnamefont {R.}~\bibnamefont {Hott}}, \bibinfo {author} {\bibfnamefont
  {R.}~\bibnamefont {Heid}}, \bibinfo {author} {\bibfnamefont {K.-P.}\
  \bibnamefont {Bohnen}}, \bibinfo {author} {\bibfnamefont {T.}~\bibnamefont
  {Egami}}, \bibinfo {author} {\bibfnamefont {A.~H.}\ \bibnamefont {Said}}, \
  and\ \bibinfo {author} {\bibfnamefont {D.}~\bibnamefont {Reznik}},\
  }\href@noop {} {\bibfield  {journal} {\bibinfo  {journal} {Phys. Rev. Lett.}\
  }\textbf {\bibinfo {volume} {107}},\ \bibinfo {pages} {107403} (\bibinfo
  {year} {2011})}\BibitemShut {NoStop}%
\bibitem [{\citenamefont {Rahn}\ \emph {et~al.}(2012)\citenamefont {Rahn},
  \citenamefont {Hellmann}, \citenamefont {Kall\"ane}, \citenamefont {Sohrt},
  \citenamefont {Kim}, \citenamefont {Kipp},\ and\ \citenamefont
  {Rossnagel}}]{PRB2012Rahn}%
  \BibitemOpen
  \bibfield  {author} {\bibinfo {author} {\bibfnamefont {D.~J.}\ \bibnamefont
  {Rahn}}, \bibinfo {author} {\bibfnamefont {S.}~\bibnamefont {Hellmann}},
  \bibinfo {author} {\bibfnamefont {M.}~\bibnamefont {Kall\"ane}}, \bibinfo
  {author} {\bibfnamefont {C.}~\bibnamefont {Sohrt}}, \bibinfo {author}
  {\bibfnamefont {T.~K.}\ \bibnamefont {Kim}}, \bibinfo {author} {\bibfnamefont
  {L.}~\bibnamefont {Kipp}}, \ and\ \bibinfo {author} {\bibfnamefont
  {K.}~\bibnamefont {Rossnagel}},\ }\href@noop {} {\bibfield  {journal}
  {\bibinfo  {journal} {Phys. Rev. B}\ }\textbf {\bibinfo {volume} {85}},\
  \bibinfo {pages} {224532} (\bibinfo {year} {2012})}\BibitemShut {NoStop}%
\bibitem [{\citenamefont {Soumyanarayanan}\ \emph {et~al.}(2013)\citenamefont
  {Soumyanarayanan}, \citenamefont {Yee}, \citenamefont {He}, \citenamefont
  {van Wezel}, \citenamefont {Rahn}, \citenamefont {Rossnagel}, \citenamefont
  {Hudson}, \citenamefont {Norman},\ and\ \citenamefont
  {Hoffman}}]{PNAS2013Soumyanarayana}%
  \BibitemOpen
  \bibfield  {author} {\bibinfo {author} {\bibfnamefont {A.}~\bibnamefont
  {Soumyanarayanan}}, \bibinfo {author} {\bibfnamefont {M.~M.}\ \bibnamefont
  {Yee}}, \bibinfo {author} {\bibfnamefont {Y.}~\bibnamefont {He}}, \bibinfo
  {author} {\bibfnamefont {J.}~\bibnamefont {van Wezel}}, \bibinfo {author}
  {\bibfnamefont {D.~J.}\ \bibnamefont {Rahn}}, \bibinfo {author}
  {\bibfnamefont {K.}~\bibnamefont {Rossnagel}}, \bibinfo {author}
  {\bibfnamefont {E.~W.}\ \bibnamefont {Hudson}}, \bibinfo {author}
  {\bibfnamefont {M.~R.}\ \bibnamefont {Norman}}, \ and\ \bibinfo {author}
  {\bibfnamefont {J.~E.}\ \bibnamefont {Hoffman}},\ }\href@noop {} {\bibfield
  {journal} {\bibinfo  {journal} {Proc. Natl. Acad. Sci. USA}\ }\textbf
  {\bibinfo {volume} {110}},\ \bibinfo {pages} {1623} (\bibinfo {year}
  {2013})}\BibitemShut {NoStop}%
\bibitem [{\citenamefont {Arguello}\ \emph {et~al.}(2015)\citenamefont
  {Arguello}, \citenamefont {Rosenthal}, \citenamefont {Andrade}, \citenamefont
  {Jin}, \citenamefont {Yeh}, \citenamefont {Zaki}, \citenamefont {Jia},
  \citenamefont {Cava}, \citenamefont {Fernandes}, \citenamefont {Millis},
  \citenamefont {Valla}, \citenamefont {Osgood},\ and\ \citenamefont
  {Pasupathy}}]{PRL2015Arguello}%
  \BibitemOpen
  \bibfield  {author} {\bibinfo {author} {\bibfnamefont {C.~J.}\ \bibnamefont
  {Arguello}}, \bibinfo {author} {\bibfnamefont {E.~P.}\ \bibnamefont
  {Rosenthal}}, \bibinfo {author} {\bibfnamefont {E.~F.}\ \bibnamefont
  {Andrade}}, \bibinfo {author} {\bibfnamefont {W.}~\bibnamefont {Jin}},
  \bibinfo {author} {\bibfnamefont {P.~C.}\ \bibnamefont {Yeh}}, \bibinfo
  {author} {\bibfnamefont {N.}~\bibnamefont {Zaki}}, \bibinfo {author}
  {\bibfnamefont {S.}~\bibnamefont {Jia}}, \bibinfo {author} {\bibfnamefont
  {R.~J.}\ \bibnamefont {Cava}}, \bibinfo {author} {\bibfnamefont {R.~M.}\
  \bibnamefont {Fernandes}}, \bibinfo {author} {\bibfnamefont {A.~J.}\
  \bibnamefont {Millis}}, \bibinfo {author} {\bibfnamefont {T.}~\bibnamefont
  {Valla}}, \bibinfo {author} {\bibfnamefont {R.~M.}\ \bibnamefont {Osgood}}, \
  and\ \bibinfo {author} {\bibfnamefont {A.~N.}\ \bibnamefont {Pasupathy}},\
  }\href@noop {} {\bibfield  {journal} {\bibinfo  {journal} {Phys. Rev. Lett.}\
  }\textbf {\bibinfo {volume} {114}},\ \bibinfo {pages} {037001} (\bibinfo
  {year} {2015})}\BibitemShut {NoStop}%
\bibitem [{\citenamefont {Frindt}(1972)}]{PRL1972Frindt}%
  \BibitemOpen
  \bibfield  {author} {\bibinfo {author} {\bibfnamefont {R.~F.}\ \bibnamefont
  {Frindt}},\ }\href@noop {} {\bibfield  {journal} {\bibinfo  {journal} {Phys.
  Rev. Lett.}\ }\textbf {\bibinfo {volume} {28}},\ \bibinfo {pages} {299}
  (\bibinfo {year} {1972})}\BibitemShut {NoStop}%
\bibitem [{\citenamefont {Staley}\ \emph {et~al.}(2009)\citenamefont {Staley},
  \citenamefont {Wu}, \citenamefont {Eklund}, \citenamefont {Liu},
  \citenamefont {Li},\ and\ \citenamefont {Xu}}]{PRB2009Staley}%
  \BibitemOpen
  \bibfield  {author} {\bibinfo {author} {\bibfnamefont {N.~E.}\ \bibnamefont
  {Staley}}, \bibinfo {author} {\bibfnamefont {J.}~\bibnamefont {Wu}}, \bibinfo
  {author} {\bibfnamefont {P.}~\bibnamefont {Eklund}}, \bibinfo {author}
  {\bibfnamefont {Y.}~\bibnamefont {Liu}}, \bibinfo {author} {\bibfnamefont
  {L.}~\bibnamefont {Li}}, \ and\ \bibinfo {author} {\bibfnamefont
  {Z.}~\bibnamefont {Xu}},\ }\href@noop {} {\bibfield  {journal} {\bibinfo
  {journal} {Phys. Rev. B}\ }\textbf {\bibinfo {volume} {80}},\ \bibinfo
  {pages} {184505} (\bibinfo {year} {2009})}\BibitemShut {NoStop}%
\bibitem [{\citenamefont {Johannes}\ and\ \citenamefont
  {Mazin}(2008)}]{PRB2008Johaness}%
  \BibitemOpen
  \bibfield  {author} {\bibinfo {author} {\bibfnamefont {M.~D.}\ \bibnamefont
  {Johannes}}\ and\ \bibinfo {author} {\bibfnamefont {I.~I.}\ \bibnamefont
  {Mazin}},\ }\href@noop {} {\bibfield  {journal} {\bibinfo  {journal} {Phys.
  Rev. B}\ }\textbf {\bibinfo {volume} {77}},\ \bibinfo {pages} {165135}
  (\bibinfo {year} {2008})}\BibitemShut {NoStop}%
\bibitem [{\citenamefont {Doran}\ \emph {et~al.}(1978)\citenamefont {Doran},
  \citenamefont {Titterington}, \citenamefont {Ricco}, \citenamefont
  {Schreiber},\ and\ \citenamefont {Wexler}}]{JPhysC1978Doran}%
  \BibitemOpen
  \bibfield  {author} {\bibinfo {author} {\bibfnamefont {N.~J.}\ \bibnamefont
  {Doran}}, \bibinfo {author} {\bibfnamefont {D.}~\bibnamefont {Titterington}},
  \bibinfo {author} {\bibfnamefont {B.}~\bibnamefont {Ricco}}, \bibinfo
  {author} {\bibfnamefont {M.}~\bibnamefont {Schreiber}}, \ and\ \bibinfo
  {author} {\bibfnamefont {G.}~\bibnamefont {Wexler}},\ }\href@noop {}
  {\bibfield  {journal} {\bibinfo  {journal} {J. Phys. C: Solid State Phys.}\
  }\textbf {\bibinfo {volume} {11}},\ \bibinfo {pages} {699} (\bibinfo {year}
  {1978})}\BibitemShut {NoStop}%
\bibitem [{\citenamefont {Johannes}\ \emph {et~al.}(2006)\citenamefont
  {Johannes}, \citenamefont {Mazin},\ and\ \citenamefont
  {Howells}}]{PRB2006Johannes}%
  \BibitemOpen
  \bibfield  {author} {\bibinfo {author} {\bibfnamefont {M.~D.}\ \bibnamefont
  {Johannes}}, \bibinfo {author} {\bibfnamefont {I.~I.}\ \bibnamefont {Mazin}},
  \ and\ \bibinfo {author} {\bibfnamefont {C.~A.}\ \bibnamefont {Howells}},\
  }\href@noop {} {\bibfield  {journal} {\bibinfo  {journal} {Phys. Rev. B}\
  }\textbf {\bibinfo {volume} {73}},\ \bibinfo {pages} {205102} (\bibinfo
  {year} {2006})}\BibitemShut {NoStop}%
\bibitem [{\citenamefont {Rice}\ and\ \citenamefont
  {Scott}(1975)}]{PRL1975Rice}%
  \BibitemOpen
  \bibfield  {author} {\bibinfo {author} {\bibfnamefont {T.~M.}\ \bibnamefont
  {Rice}}\ and\ \bibinfo {author} {\bibfnamefont {G.~K.}\ \bibnamefont
  {Scott}},\ }\href@noop {} {\bibfield  {journal} {\bibinfo  {journal} {Phys.
  Rev. Lett.}\ }\textbf {\bibinfo {volume} {35}},\ \bibinfo {pages} {120}
  (\bibinfo {year} {1975})}\BibitemShut {NoStop}%
\bibitem [{\citenamefont {Kiss}\ \emph {et~al.}(2007)\citenamefont {Kiss},
  \citenamefont {Yokoya}, \citenamefont {Chainani}, \citenamefont {Shin},
  \citenamefont {Hanguri}, \citenamefont {Nohara},\ and\ \citenamefont
  {Takagi}}]{NatPhys2007Kiss}%
  \BibitemOpen
  \bibfield  {author} {\bibinfo {author} {\bibfnamefont {T.}~\bibnamefont
  {Kiss}}, \bibinfo {author} {\bibfnamefont {T.}~\bibnamefont {Yokoya}},
  \bibinfo {author} {\bibfnamefont {A.}~\bibnamefont {Chainani}}, \bibinfo
  {author} {\bibfnamefont {S.}~\bibnamefont {Shin}}, \bibinfo {author}
  {\bibfnamefont {T.}~\bibnamefont {Hanguri}}, \bibinfo {author} {\bibfnamefont
  {M.}~\bibnamefont {Nohara}}, \ and\ \bibinfo {author} {\bibfnamefont
  {H.}~\bibnamefont {Takagi}},\ }\href@noop {} {\bibfield  {journal} {\bibinfo
  {journal} {Nat. Phys.}\ }\textbf {\bibinfo {volume} {3}},\ \bibinfo {pages}
  {720} (\bibinfo {year} {2007})}\BibitemShut {NoStop}%
\bibitem [{\citenamefont {Calandra}\ \emph {et~al.}(2009)\citenamefont
  {Calandra}, \citenamefont {Mazin},\ and\ \citenamefont
  {Mauri}}]{PRB2009Calandra}%
  \BibitemOpen
  \bibfield  {author} {\bibinfo {author} {\bibfnamefont {M.}~\bibnamefont
  {Calandra}}, \bibinfo {author} {\bibfnamefont {I.~I.}\ \bibnamefont {Mazin}},
  \ and\ \bibinfo {author} {\bibfnamefont {F.}~\bibnamefont {Mauri}},\
  }\href@noop {} {\bibfield  {journal} {\bibinfo  {journal} {Phys. Rev. B}\
  }\textbf {\bibinfo {volume} {80}},\ \bibinfo {pages} {241108} (\bibinfo
  {year} {2009})}\BibitemShut {NoStop}%
\bibitem [{\citenamefont {Leb\`egue}\ and\ \citenamefont
  {Eriksson}(2009)}]{PRB2009Lebegue}%
  \BibitemOpen
  \bibfield  {author} {\bibinfo {author} {\bibfnamefont {S.}~\bibnamefont
  {Leb\`egue}}\ and\ \bibinfo {author} {\bibfnamefont {O.}~\bibnamefont
  {Eriksson}},\ }\href@noop {} {\bibfield  {journal} {\bibinfo  {journal}
  {Phys. Rev. B}\ }\textbf {\bibinfo {volume} {79}},\ \bibinfo {pages} {115409}
  (\bibinfo {year} {2009})}\BibitemShut {NoStop}%
\bibitem [{\citenamefont {Perdew}\ \emph {et~al.}(1996)\citenamefont {Perdew},
  \citenamefont {Burke},\ and\ \citenamefont {Ernzerhof}}]{PRL1996Perdew}%
  \BibitemOpen
  \bibfield  {author} {\bibinfo {author} {\bibfnamefont {J.~P.}\ \bibnamefont
  {Perdew}}, \bibinfo {author} {\bibfnamefont {K.}~\bibnamefont {Burke}}, \
  and\ \bibinfo {author} {\bibfnamefont {M.}~\bibnamefont {Ernzerhof}},\
  }\href@noop {} {\bibfield  {journal} {\bibinfo  {journal} {Phys. Rev. Lett.}\
  }\textbf {\bibinfo {volume} {77}},\ \bibinfo {pages} {3865} (\bibinfo {year}
  {1996})}\BibitemShut {NoStop}%
\bibitem [{\citenamefont {Troullier}\ and\ \citenamefont
  {Martins}(1991)}]{PRB1991Troullier}%
  \BibitemOpen
  \bibfield  {author} {\bibinfo {author} {\bibfnamefont {N.}~\bibnamefont
  {Troullier}}\ and\ \bibinfo {author} {\bibfnamefont {J.~L.}\ \bibnamefont
  {Martins}},\ }\href@noop {} {\bibfield  {journal} {\bibinfo  {journal} {Phys.
  Rev. B}\ }\textbf {\bibinfo {volume} {43}},\ \bibinfo {pages} {1993}
  (\bibinfo {year} {1991})}\BibitemShut {NoStop}%
\bibitem [{\citenamefont {Louie}\ \emph {et~al.}(1982)\citenamefont {Louie},
  \citenamefont {Froyen},\ and\ \citenamefont {Cohen}}]{PRB1982Louie}%
  \BibitemOpen
  \bibfield  {author} {\bibinfo {author} {\bibfnamefont {S.~G.}\ \bibnamefont
  {Louie}}, \bibinfo {author} {\bibfnamefont {S.}~\bibnamefont {Froyen}}, \
  and\ \bibinfo {author} {\bibfnamefont {M.~L.}\ \bibnamefont {Cohen}},\
  }\href@noop {} {\bibfield  {journal} {\bibinfo  {journal} {Phys. Rev. B}\
  }\textbf {\bibinfo {volume} {26}},\ \bibinfo {pages} {1738} (\bibinfo {year}
  {1982})}\BibitemShut {NoStop}%
\bibitem [{\citenamefont {Rohlfing}\ \emph {et~al.}(1995)\citenamefont
  {Rohlfing}, \citenamefont {Kr\"uger},\ and\ \citenamefont
  {Pollmann}}]{PRL1995Rohlfing}%
  \BibitemOpen
  \bibfield  {author} {\bibinfo {author} {\bibfnamefont {M.}~\bibnamefont
  {Rohlfing}}, \bibinfo {author} {\bibfnamefont {P.}~\bibnamefont {Kr\"uger}},
  \ and\ \bibinfo {author} {\bibfnamefont {J.}~\bibnamefont {Pollmann}},\
  }\href@noop {} {\bibfield  {journal} {\bibinfo  {journal} {Phys. Rev. Lett.}\
  }\textbf {\bibinfo {volume} {75}},\ \bibinfo {pages} {3489} (\bibinfo {year}
  {1995})}\BibitemShut {NoStop}%
\bibitem [{\citenamefont {Marini}\ \emph {et~al.}(2001)\citenamefont {Marini},
  \citenamefont {Onida},\ and\ \citenamefont {Del~Sole}}]{PRL2002Marini}%
  \BibitemOpen
  \bibfield  {author} {\bibinfo {author} {\bibfnamefont {A.}~\bibnamefont
  {Marini}}, \bibinfo {author} {\bibfnamefont {G.}~\bibnamefont {Onida}}, \
  and\ \bibinfo {author} {\bibfnamefont {R.}~\bibnamefont {Del~Sole}},\
  }\href@noop {} {\bibfield  {journal} {\bibinfo  {journal} {Phys. Rev. Lett.}\
  }\textbf {\bibinfo {volume} {88}},\ \bibinfo {pages} {016403} (\bibinfo
  {year} {2001})}\BibitemShut {NoStop}%
\bibitem [{\citenamefont {Giannozzi}\ \emph {et~al.}(2009)\citenamefont
  {Giannozzi}, \citenamefont {Baroni}, \citenamefont {Bonini}, \citenamefont
  {Calandra}, \citenamefont {Car}, \citenamefont {Cavazzoni}, \citenamefont
  {Ceresoli}, \citenamefont {Chiarotti}, \citenamefont {Cococcioni},
  \citenamefont {Dabo}, \citenamefont {Corso}, \citenamefont {de~Gironcoli},
  \citenamefont {Fabris}, \citenamefont {Fratesi}, \citenamefont {Gebauer},
  \citenamefont {Gerstmann}, \citenamefont {Gougoussis}, \citenamefont
  {Kokalj}, \citenamefont {Lazzeri}, \citenamefont {Martin-Samos},
  \citenamefont {Marzari}, \citenamefont {Mauri}, \citenamefont {Mazzarello},
  \citenamefont {Paolini}, \citenamefont {Pasquarello}, \citenamefont
  {Paulatto}, \citenamefont {Sbraccia}, \citenamefont {Scandolo}, \citenamefont
  {Sclauzero}, \citenamefont {Seitsonen}, \citenamefont {Smogunov},
  \citenamefont {Umari},\ and\ \citenamefont
  {Wentzcovitch}}]{JPhys2009Giannozzi}%
  \BibitemOpen
  \bibfield  {author} {\bibinfo {author} {\bibfnamefont {P.}~\bibnamefont
  {Giannozzi}}, \bibinfo {author} {\bibfnamefont {S.}~\bibnamefont {Baroni}},
  \bibinfo {author} {\bibfnamefont {N.}~\bibnamefont {Bonini}}, \bibinfo
  {author} {\bibfnamefont {M.}~\bibnamefont {Calandra}}, \bibinfo {author}
  {\bibfnamefont {R.}~\bibnamefont {Car}}, \bibinfo {author} {\bibfnamefont
  {C.}~\bibnamefont {Cavazzoni}}, \bibinfo {author} {\bibfnamefont
  {D.}~\bibnamefont {Ceresoli}}, \bibinfo {author} {\bibfnamefont {G.~L.}\
  \bibnamefont {Chiarotti}}, \bibinfo {author} {\bibfnamefont {M.}~\bibnamefont
  {Cococcioni}}, \bibinfo {author} {\bibfnamefont {I.}~\bibnamefont {Dabo}},
  \bibinfo {author} {\bibfnamefont {A.~D.}\ \bibnamefont {Corso}}, \bibinfo
  {author} {\bibfnamefont {S.}~\bibnamefont {de~Gironcoli}}, \bibinfo {author}
  {\bibfnamefont {S.}~\bibnamefont {Fabris}}, \bibinfo {author} {\bibfnamefont
  {G.}~\bibnamefont {Fratesi}}, \bibinfo {author} {\bibfnamefont
  {R.}~\bibnamefont {Gebauer}}, \bibinfo {author} {\bibfnamefont
  {U.}~\bibnamefont {Gerstmann}}, \bibinfo {author} {\bibfnamefont
  {C.}~\bibnamefont {Gougoussis}}, \bibinfo {author} {\bibfnamefont
  {A.}~\bibnamefont {Kokalj}}, \bibinfo {author} {\bibfnamefont
  {M.}~\bibnamefont {Lazzeri}}, \bibinfo {author} {\bibfnamefont
  {L.}~\bibnamefont {Martin-Samos}}, \bibinfo {author} {\bibfnamefont
  {N.}~\bibnamefont {Marzari}}, \bibinfo {author} {\bibfnamefont
  {F.}~\bibnamefont {Mauri}}, \bibinfo {author} {\bibfnamefont
  {R.}~\bibnamefont {Mazzarello}}, \bibinfo {author} {\bibfnamefont
  {S.}~\bibnamefont {Paolini}}, \bibinfo {author} {\bibfnamefont
  {A.}~\bibnamefont {Pasquarello}}, \bibinfo {author} {\bibfnamefont
  {L.}~\bibnamefont {Paulatto}}, \bibinfo {author} {\bibfnamefont
  {C.}~\bibnamefont {Sbraccia}}, \bibinfo {author} {\bibfnamefont
  {S.}~\bibnamefont {Scandolo}}, \bibinfo {author} {\bibfnamefont
  {G.}~\bibnamefont {Sclauzero}}, \bibinfo {author} {\bibfnamefont {A.~P.}\
  \bibnamefont {Seitsonen}}, \bibinfo {author} {\bibfnamefont {A.}~\bibnamefont
  {Smogunov}}, \bibinfo {author} {\bibfnamefont {P.}~\bibnamefont {Umari}}, \
  and\ \bibinfo {author} {\bibfnamefont {R.~M.}\ \bibnamefont {Wentzcovitch}},\
  }\href@noop {} {\bibfield  {journal} {\bibinfo  {journal} {J. of Phys.:
  Condens. Mat.}\ }\textbf {\bibinfo {volume} {21}},\ \bibinfo {pages} {395502}
  (\bibinfo {year} {2009})}\BibitemShut {NoStop}%
\bibitem [{\citenamefont {Hedin}(1965)}]{PR1965Hedin}%
  \BibitemOpen
  \bibfield  {author} {\bibinfo {author} {\bibfnamefont {L.}~\bibnamefont
  {Hedin}},\ }\href@noop {} {\bibfield  {journal} {\bibinfo  {journal} {Phys.
  Rev.}\ }\textbf {\bibinfo {volume} {139}},\ \bibinfo {pages} {A796} (\bibinfo
  {year} {1965})}\BibitemShut {NoStop}%
\bibitem [{\citenamefont {Hybertsen}\ and\ \citenamefont
  {Louie}(1986)}]{PRB1986Hybertsen}%
  \BibitemOpen
  \bibfield  {author} {\bibinfo {author} {\bibfnamefont {M.~S.}\ \bibnamefont
  {Hybertsen}}\ and\ \bibinfo {author} {\bibfnamefont {S.~G.}\ \bibnamefont
  {Louie}},\ }\href@noop {} {\bibfield  {journal} {\bibinfo  {journal} {Phys.
  Rev. B}\ }\textbf {\bibinfo {volume} {34}},\ \bibinfo {pages} {5390}
  (\bibinfo {year} {1986})}\BibitemShut {NoStop}%
\bibitem [{\citenamefont {Deslippe}\ \emph {et~al.}(2012)\citenamefont
  {Deslippe}, \citenamefont {Samsonidze}, \citenamefont {Strubbe},
  \citenamefont {Jain}, \citenamefont {Cohen},\ and\ \citenamefont
  {Louie}}]{CPC2012Deslippe}%
  \BibitemOpen
  \bibfield  {author} {\bibinfo {author} {\bibfnamefont {J.}~\bibnamefont
  {Deslippe}}, \bibinfo {author} {\bibfnamefont {G.}~\bibnamefont
  {Samsonidze}}, \bibinfo {author} {\bibfnamefont {D.~A.}\ \bibnamefont
  {Strubbe}}, \bibinfo {author} {\bibfnamefont {M.}~\bibnamefont {Jain}},
  \bibinfo {author} {\bibfnamefont {M.~L.}\ \bibnamefont {Cohen}}, \ and\
  \bibinfo {author} {\bibfnamefont {S.~G.}\ \bibnamefont {Louie}},\ }\href@noop
  {} {\bibfield  {journal} {\bibinfo  {journal} {Comput. Phys. Commun.}\
  }\textbf {\bibinfo {volume} {183}},\ \bibinfo {pages} {1269 } (\bibinfo
  {year} {2012})}\BibitemShut {NoStop}%
\bibitem [{\citenamefont {Cooke}\ \emph {et~al.}(1974)\citenamefont {Cooke},
  \citenamefont {Davis},\ and\ \citenamefont {Mostoller}}]{PRB1974Cooke}%
  \BibitemOpen
  \bibfield  {author} {\bibinfo {author} {\bibfnamefont {J.~F.}\ \bibnamefont
  {Cooke}}, \bibinfo {author} {\bibfnamefont {H.~L.}\ \bibnamefont {Davis}}, \
  and\ \bibinfo {author} {\bibfnamefont {M.}~\bibnamefont {Mostoller}},\
  }\href@noop {} {\bibfield  {journal} {\bibinfo  {journal} {Phys. Rev. B}\
  }\textbf {\bibinfo {volume} {9}},\ \bibinfo {pages} {2485} (\bibinfo {year}
  {1974})}\BibitemShut {NoStop}%
\bibitem [{\citenamefont {Rath}\ and\ \citenamefont
  {Freeman}(1975)}]{PRB1975Rath}%
  \BibitemOpen
  \bibfield  {author} {\bibinfo {author} {\bibfnamefont {J.}~\bibnamefont
  {Rath}}\ and\ \bibinfo {author} {\bibfnamefont {A.~J.}\ \bibnamefont
  {Freeman}},\ }\href@noop {} {\bibfield  {journal} {\bibinfo  {journal} {Phys.
  Rev. B}\ }\textbf {\bibinfo {volume} {11}},\ \bibinfo {pages} {2109}
  (\bibinfo {year} {1975})}\BibitemShut {NoStop}%
\bibitem [{\citenamefont {Gupta}\ and\ \citenamefont
  {Freeman}(1976)}]{PRB1976Gupta}%
  \BibitemOpen
  \bibfield  {author} {\bibinfo {author} {\bibfnamefont {R.~P.}\ \bibnamefont
  {Gupta}}\ and\ \bibinfo {author} {\bibfnamefont {A.~J.}\ \bibnamefont
  {Freeman}},\ }\href@noop {} {\bibfield  {journal} {\bibinfo  {journal} {Phys.
  Rev. B}\ }\textbf {\bibinfo {volume} {13}},\ \bibinfo {pages} {4376}
  (\bibinfo {year} {1976})}\BibitemShut {NoStop}%
\bibitem [{\citenamefont {Myron}\ and\ \citenamefont
  {Freeman}(1975)}]{PRB1975Myron}%
  \BibitemOpen
  \bibfield  {author} {\bibinfo {author} {\bibfnamefont {H.~W.}\ \bibnamefont
  {Myron}}\ and\ \bibinfo {author} {\bibfnamefont {A.~J.}\ \bibnamefont
  {Freeman}},\ }\href@noop {} {\bibfield  {journal} {\bibinfo  {journal} {Phys.
  Rev. B}\ }\textbf {\bibinfo {volume} {11}},\ \bibinfo {pages} {2735}
  (\bibinfo {year} {1975})}\BibitemShut {NoStop}%
\bibitem [{\citenamefont {Myron}\ \emph {et~al.}(1977)\citenamefont {Myron},
  \citenamefont {Rath},\ and\ \citenamefont {Freeman}}]{PRB1977Myron}%
  \BibitemOpen
  \bibfield  {author} {\bibinfo {author} {\bibfnamefont {H.~W.}\ \bibnamefont
  {Myron}}, \bibinfo {author} {\bibfnamefont {J.}~\bibnamefont {Rath}}, \ and\
  \bibinfo {author} {\bibfnamefont {A.~J.}\ \bibnamefont {Freeman}},\
  }\href@noop {} {\bibfield  {journal} {\bibinfo  {journal} {Phys. Rev. B}\
  }\textbf {\bibinfo {volume} {15}},\ \bibinfo {pages} {885} (\bibinfo {year}
  {1977})}\BibitemShut {NoStop}%
\bibitem [{\citenamefont {Ge}\ and\ \citenamefont {Liu}(2012)}]{PRB2012Ge}%
  \BibitemOpen
  \bibfield  {author} {\bibinfo {author} {\bibfnamefont {Y.}~\bibnamefont
  {Ge}}\ and\ \bibinfo {author} {\bibfnamefont {A.~Y.}\ \bibnamefont {Liu}},\
  }\href@noop {} {\bibfield  {journal} {\bibinfo  {journal} {Phys. Rev. B}\
  }\textbf {\bibinfo {volume} {86}},\ \bibinfo {pages} {104101} (\bibinfo
  {year} {2012})}\BibitemShut {NoStop}%
\bibitem [{\citenamefont {Yazyev}\ \emph {et~al.}(2012)\citenamefont {Yazyev},
  \citenamefont {Kioupakis}, \citenamefont {Moore},\ and\ \citenamefont
  {Louie}}]{PRB2012Yazyev}%
  \BibitemOpen
  \bibfield  {author} {\bibinfo {author} {\bibfnamefont {O.~V.}\ \bibnamefont
  {Yazyev}}, \bibinfo {author} {\bibfnamefont {E.}~\bibnamefont {Kioupakis}},
  \bibinfo {author} {\bibfnamefont {J.~E.}\ \bibnamefont {Moore}}, \ and\
  \bibinfo {author} {\bibfnamefont {S.~G.}\ \bibnamefont {Louie}},\ }\href@noop
  {} {\bibfield  {journal} {\bibinfo  {journal} {Phys. Rev. B}\ }\textbf
  {\bibinfo {volume} {85}},\ \bibinfo {pages} {161101} (\bibinfo {year}
  {2012})}\BibitemShut {NoStop}%
\bibitem [{\citenamefont {Miasek}(1957)}]{PR1957Miasek}%
  \BibitemOpen
  \bibfield  {author} {\bibinfo {author} {\bibfnamefont {M.}~\bibnamefont
  {Miasek}},\ }\href@noop {} {\bibfield  {journal} {\bibinfo  {journal} {Phys.
  Rev.}\ }\textbf {\bibinfo {volume} {107}},\ \bibinfo {pages} {92} (\bibinfo
  {year} {1957})}\BibitemShut {NoStop}%
\bibitem [{\citenamefont {Egorov}\ \emph {et~al.}(1968)\citenamefont {Egorov},
  \citenamefont {Reser},\ and\ \citenamefont
  {Shirokovskii}}]{PhysSS1968Egorov}%
  \BibitemOpen
  \bibfield  {author} {\bibinfo {author} {\bibfnamefont {R.~F.}\ \bibnamefont
  {Egorov}}, \bibinfo {author} {\bibfnamefont {B.~I.}\ \bibnamefont {Reser}}, \
  and\ \bibinfo {author} {\bibfnamefont {V.~P.}\ \bibnamefont {Shirokovskii}},\
  }\href@noop {} {\bibfield  {journal} {\bibinfo  {journal} {Phys. Status
  Solidi B}\ }\textbf {\bibinfo {volume} {26}},\ \bibinfo {pages} {391}
  (\bibinfo {year} {1968})}\BibitemShut {NoStop}%
\bibitem [{\citenamefont {Kuznetsov}\ and\ \citenamefont
  {Men}(1978)}]{PhysSS1978Kuznetsov}%
  \BibitemOpen
  \bibfield  {author} {\bibinfo {author} {\bibfnamefont {V.~N.}\ \bibnamefont
  {Kuznetsov}}\ and\ \bibinfo {author} {\bibfnamefont {A.~N.}\ \bibnamefont
  {Men}},\ }\href@noop {} {\bibfield  {journal} {\bibinfo  {journal} {Phys.
  Status Solidi B}\ }\textbf {\bibinfo {volume} {85}},\ \bibinfo {pages} {95}
  (\bibinfo {year} {1978})}\BibitemShut {NoStop}%
\bibitem [{\citenamefont {Mattheiss}(1973)}]{PRL1973Mattheiss}%
  \BibitemOpen
  \bibfield  {author} {\bibinfo {author} {\bibfnamefont {L.~F.}\ \bibnamefont
  {Mattheiss}},\ }\href@noop {} {\bibfield  {journal} {\bibinfo  {journal}
  {Phys. Rev. Lett.}\ }\textbf {\bibinfo {volume} {30}},\ \bibinfo {pages}
  {784} (\bibinfo {year} {1973})}\BibitemShut {NoStop}%
\bibitem [{\citenamefont {Liu}\ \emph {et~al.}(2013)\citenamefont {Liu},
  \citenamefont {Shan}, \citenamefont {Yao}, \citenamefont {Yao},\ and\
  \citenamefont {Xiao}}]{PRB2013Liu}%
  \BibitemOpen
  \bibfield  {author} {\bibinfo {author} {\bibfnamefont {G.-B.}\ \bibnamefont
  {Liu}}, \bibinfo {author} {\bibfnamefont {W.-Y.}\ \bibnamefont {Shan}},
  \bibinfo {author} {\bibfnamefont {Y.}~\bibnamefont {Yao}}, \bibinfo {author}
  {\bibfnamefont {W.}~\bibnamefont {Yao}}, \ and\ \bibinfo {author}
  {\bibfnamefont {D.}~\bibnamefont {Xiao}},\ }\href@noop {} {\bibfield
  {journal} {\bibinfo  {journal} {Phys. Rev. B}\ }\textbf {\bibinfo {volume}
  {88}},\ \bibinfo {pages} {085433} (\bibinfo {year} {2013})}\BibitemShut
  {NoStop}%
\bibitem [{\citenamefont {Gray}\ \emph {et~al.}(1984)\citenamefont {Gray},
  \citenamefont {Hurst}, \citenamefont {Lewis},\ and\ \citenamefont
  {Meyer}}]{Gray}%
  \BibitemOpen
  \bibfield  {author} {\bibinfo {author} {\bibfnamefont {P.~R.}\ \bibnamefont
  {Gray}}, \bibinfo {author} {\bibfnamefont {P.~J.}\ \bibnamefont {Hurst}},
  \bibinfo {author} {\bibfnamefont {S.~H.}\ \bibnamefont {Lewis}}, \ and\
  \bibinfo {author} {\bibfnamefont {R.~G.}\ \bibnamefont {Meyer}},\ }\href@noop
  {} {\emph {\bibinfo {title} {Analysis and Design of Analog Integrated
  Circuits}}}\ (\bibinfo  {publisher} {Wiley},\ \bibinfo {address} {New York},\
  \bibinfo {year} {1984})\BibitemShut {NoStop}%
\bibitem [{\citenamefont {Ohta}\ \emph {et~al.}(2007)\citenamefont {Ohta},
  \citenamefont {Bostwick}, \citenamefont {McChesney}, \citenamefont {Seyller},
  \citenamefont {Horn},\ and\ \citenamefont {Rotenberg}}]{PRL2007Otha}%
  \BibitemOpen
  \bibfield  {author} {\bibinfo {author} {\bibfnamefont {T.}~\bibnamefont
  {Ohta}}, \bibinfo {author} {\bibfnamefont {A.}~\bibnamefont {Bostwick}},
  \bibinfo {author} {\bibfnamefont {J.~L.}\ \bibnamefont {McChesney}}, \bibinfo
  {author} {\bibfnamefont {T.}~\bibnamefont {Seyller}}, \bibinfo {author}
  {\bibfnamefont {K.}~\bibnamefont {Horn}}, \ and\ \bibinfo {author}
  {\bibfnamefont {E.}~\bibnamefont {Rotenberg}},\ }\href@noop {} {\bibfield
  {journal} {\bibinfo  {journal} {Phys. Rev. Lett.}\ }\textbf {\bibinfo
  {volume} {98}},\ \bibinfo {pages} {206802} (\bibinfo {year}
  {2007})}\BibitemShut {NoStop}%
\bibitem [{\citenamefont {Zhou}\ \emph {et~al.}(2007)\citenamefont {Zhou},
  \citenamefont {Gweon}, \citenamefont {Fedorov}, \citenamefont {First},
  \citenamefont {de~Heer}, \citenamefont {Lee}, \citenamefont {Guinea},
  \citenamefont {Neto},\ and\ \citenamefont {Lanzara}}]{NatMat2007Zhou}%
  \BibitemOpen
  \bibfield  {author} {\bibinfo {author} {\bibfnamefont {S.~Y.}\ \bibnamefont
  {Zhou}}, \bibinfo {author} {\bibfnamefont {G.-H.}\ \bibnamefont {Gweon}},
  \bibinfo {author} {\bibfnamefont {A.~V.}\ \bibnamefont {Fedorov}}, \bibinfo
  {author} {\bibfnamefont {P.~N.}\ \bibnamefont {First}}, \bibinfo {author}
  {\bibfnamefont {W.~A.}\ \bibnamefont {de~Heer}}, \bibinfo {author}
  {\bibfnamefont {D.-H.}\ \bibnamefont {Lee}}, \bibinfo {author} {\bibfnamefont
  {F.}~\bibnamefont {Guinea}}, \bibinfo {author} {\bibfnamefont {A.~H.~C.}\
  \bibnamefont {Neto}}, \ and\ \bibinfo {author} {\bibfnamefont
  {A.}~\bibnamefont {Lanzara}},\ }\href@noop {} {\bibfield  {journal} {\bibinfo
   {journal} {Nat. Mat.}\ }\textbf {\bibinfo {volume} {6}},\ \bibinfo {pages}
  {770 } (\bibinfo {year} {2007})}\BibitemShut {NoStop}%
\bibitem [{\citenamefont {Kim}\ \emph {et~al.}(2013)\citenamefont {Kim},
  \citenamefont {Ihm}, \citenamefont {Choi},\ and\ \citenamefont
  {Son}}]{SSC2013Kim}%
  \BibitemOpen
  \bibfield  {author} {\bibinfo {author} {\bibfnamefont {S.}~\bibnamefont
  {Kim}}, \bibinfo {author} {\bibfnamefont {J.}~\bibnamefont {Ihm}}, \bibinfo
  {author} {\bibfnamefont {H.~J.}\ \bibnamefont {Choi}}, \ and\ \bibinfo
  {author} {\bibfnamefont {Y.-W.}\ \bibnamefont {Son}},\ }\href@noop {}
  {\bibfield  {journal} {\bibinfo  {journal} {Solid State Comm.}\ }\textbf
  {\bibinfo {volume} {175-176}},\ \bibinfo {pages} {83} (\bibinfo {year}
  {2013})}\BibitemShut {NoStop}%
\bibitem [{\citenamefont {Silva-Guill\'en}\ \emph {et~al.}(2016)\citenamefont
  {Silva-Guill\'en}, \citenamefont {Ordej\'on}, \citenamefont {Guinea},\ and\
  \citenamefont {Canadell}}]{Canadell}%
  \BibitemOpen
  \bibfield  {author} {\bibinfo {author} {\bibfnamefont {J.~A.}\ \bibnamefont
  {Silva-Guill\'en}}, \bibinfo {author} {\bibfnamefont {P.}~\bibnamefont
  {Ordej\'on}}, \bibinfo {author} {\bibfnamefont {F.}~\bibnamefont {Guinea}}, \
  and\ \bibinfo {author} {\bibfnamefont {E.}~\bibnamefont {Canadell}},\
  }\href@noop {} {\bibfield  {journal} {\bibinfo  {journal} {2D Mater.}\
  }\textbf {\bibinfo {volume} {3}},\ \bibinfo {pages} {035028} (\bibinfo {year}
  {2016})}\BibitemShut {NoStop}%
\end{thebibliography}%

 \end{document}